\documentclass[journal,comsoc,twocolumn,10pt,twoside]{IEEEtran}
\usepackage{cite}

   \usepackage[pdftex]{graphicx}
\ifCLASSOPTIONcompsoc
  \usepackage[caption=false,font=normalsize,labelfont=sf,textfont=sf]{subfig}
\else
  \usepackage[caption=false,font=footnotesize,labelformat=simple]{subfig}
\fi

\usepackage{stfloats}
\usepackage{url}

\usepackage{multirow}
\usepackage{diagbox}
\usepackage{color}
\usepackage{amsmath,amsthm,amsfonts,amssymb}
\usepackage{comment}

\begin{document}
%
\title{Will Emerging Millimeter-Wave Cellular Networks Cause Harmful Interference to Weather Satellites?}

\author{Andreea Palade, Andra M. Voicu, Petri M\"ah\"onen and Ljiljana Simi\'c
\thanks{A.~Palade, A.~M.~Voicu, P.~M\"ah\"onen, and L.~Simi\'c  are with the Institute for Networked Systems, RWTH Aachen University, Aachen, Germany (e-mail: andreea.palade@rwth-aachen.de; avo@inets.rwth-aachen.de; pma@inets.rwth-aachen.de; lsi@inets.rwth-aachen.de).}
}

\maketitle

\begin{abstract}

We study whether realistic 5G mm-wave cellular networks would cause harmful out-of-band interference to weather satellites sensing in the 23.8 GHz band. We estimate uplink and downlink interference from a single interferer and a network of interferers in New York City, using real 3D building data and realistic antenna patterns. We perform detailed ray-tracing propagation simulations, for locations of the MetOp-B weather satellite and its scanning orientations and ground interferer antenna orientations for representative urban cell sites. In addition to the ITU-R threshold of \mbox{--136}~dBm/200~MHz, we propose an alternative set of harmful interference thresholds directly related to the sensitivity of the satellite sensor. Our results show that the 3GPP power leakage limits are sufficient to ensure that interference from a \emph{single} 5G device is not harmful if considering the ITU-R threshold, but not if the weather prediction software can tolerate only very low interference levels. Importantly, \emph{aggregate} interference resulting in practice from a 5G \emph{network} with realistic network densities is often harmful, even considering the least conservative ITU-R threshold. Overall, our comprehensive coexistence study thus strongly suggests that additional engineering and/or regulatory solutions will be necessary to protect weather satellite passive sensing from mm-wave cellular network interference.

\end{abstract}


%
\IEEEpeerreviewmaketitle

\section{Introduction}


The development of  5G-and-beyond cellular technology has lead to an increasing number of frequency bands being opened by ITU-R for International Mobile Telecommunications (IMT) services. However, cellular networks in several of these bands would have to coexist with incumbent satellite services in co- or adjacent bands.  
Notably, the IMT 24.25--27.5~GHz band is adjacent to the 23.6--24.0~GHz band (i.e.  23.8~GHz band), where Earth exploration-satellite service (EESS) systems employ passive sensing radiometers to measure water vapour and predict weather phenomena~\cite{Marcus2019}.
Major concerns have been raised by weather scientists regarding this coexistence case~\cite{Witze2019, Liu2021, Palmer2020}, sounding the alarm that the EESS passive systems would suffer from strong out-of-band interference from 5G mm-wave systems operating in the adjacent band, i.e. the 3GPP n258 band~\cite{ETSI_bs}, which would severely degrade weather forecasting. This has subsequently turned into a lively spectrum policy debate~\cite{Marcus2019, Shields2019, Chamberlain2019, Benish2020}, despite the power leakage limits of \mbox{--3}~dBm/200~MHz and 1~dBm/200~MHz  imposed by 3GPP for base stations (BSs) and users (UEs), respectively, to protect weather satellites in the 23.8~GHz band~\cite{ETSI_bs, ETSI_ue}.

Although this issue has received considerable attention from the weather science and spectrum policy communities, there is still only sparse engineering analysis evaluating the impact of 5G mm-wave out-of-band interference on weather satellites in the 23.8~GHz band~\cite{Eichen2019, Eichen2021,Yousefvand2020, Yao2019, Cho2019, Cho2020,  Caillet2021, Murakami2021}.
The authors in~\cite{Eichen2019, Eichen2021} proposed coexistence methods where the cellular networks pause their transmissions or vary their transmit power and traffic characteristics to protect the weather satellites; however, they did not study whether and under which conditions such methods would be needed in practice.  
The authors in~\cite{Yousefvand2020} took a different perspective and evaluated the impact of interference on the precision of numerical weather prediction models, assuming as interference levels the maximum leakage powers allowed by the regulations. However, their first-order analysis used simplistic propagation models without any atmospheric attenuation, did not model directional antenna patterns of the satellite and ground interferer, and neglected to consider the effect of \emph{aggregate interference} from multiple ground devices, as expected in cellular networks. 


The authors in~\cite{Yao2019, Cho2019, Cho2020} estimated the aggregate interference from ground deployments to weather satellites in the 23.8~GHz band, finding that it sometimes exceeds the ITU-R harmful interference criteria for the satellites, i.e. the threshold of \mbox{--136}~dBm/200~MHz for 0.01\% of the time or of the area~\cite{ITU2017}.    
However,~\cite{Yao2019, Cho2019, Cho2020} assumed a simplistic circular ground area, using at most terrain data, where the satellite pointed at the center of this area, thus failing to capture the range of propagation conditions due to realistic building profiles and different satellite antenna orientations as the satellite scans different ground pixels.  
The authors in~\cite{Caillet2021} described studies conducted by European countries during \mbox{WRC-19} to estimate the aggregate interference and protect the weather satellites.  Although these preliminary studies confirmed potential interference issues, they also relied on simplistic ground deployment assumptions, as per ITU-R recommendations.    
Finally,  \cite{Murakami2021} evaluated the impact of element spacing in the ground device antenna array on the aggregate interference at the satellite, but also assumed hypothetical simplistic ground deployments with hexagonal cells. 

Consequently, it is not yet clear from the existing engineering literature whether and under which circumstances realistic 5G mm-wave cellular network deployments would cause harmful out-of-band interference to the weather satellites sensing in the 23.8~GHz band. In this context, this paper is the first to comprehensively address two major coexistence questions: \textbf{(i)~whether \emph{aggregate} interference from 5G mm-wave ground deployments can become harmful according to the ITU-R criteria}, given the per-device leakage power limit imposed by 3GPP; and \textbf{(ii)~in which cases interference may actually become harmful and degrade the whether prediction}, given the high sensitivity of the weather satellite sensor and the way the measurements are post-processed to forecast weather. 


We address the \textbf{first question} by presenting the most comprehensive and detailed study to date of 5G mm-wave network and weather satellite coexistence. We model the out-of-band uplink and downlink interference from a single interferer as well as a \emph{network} of interferers consisting of UEs and BSs located in New York City (NYC).
We study in detail the levels and spatial structure of the estimated interference, comprehensively modelling the impact of the urban propagation environment and the geometry of the ground/satellite coexistence scenario. We use real 3D building data from NYC and realistic directional antenna patterns -- thus modelling interference leaking from both the main lobe and sidelobes of the antenna -- to perform detailed ray-tracing propagation simulations, also including atmospheric attenuation. We consider real locations of the MetOp-B weather satellite above NYC and its scanning orientations, and realistic ground interferer antenna beam orientations from several representative local urban cell environments, and estimate the aggregate interference for a range of ground network densities. 


We address the \textbf{second question} by comparing the estimated interference not only against the ITU-R threshold, but also against a set of alternative harmful interference thresholds \{\mbox{--161}, \mbox{--151}, \mbox{--141}\}~dBm that correspond to a fraction \{0.01, 0.1, 1\}\% of the noise equivalent delta temperature of the weather satellite sensor. 
These thresholds constitute a proxy for different capabilities of the weather prediction software to remove by post-processing the impact of the interference on the resolution of the satellite passive sensor.
By directly relating the harmful interference threshold to the sensitivity of the weather satellite sensor, we take an important first step towards answering the question of when the weather prediction degradation from 5G mm-wave network interference would be unacceptably large. 

Our extensive simulation results show that the 3GPP power leakage limits are sufficient to ensure that interference from a \emph{single} 5G-NR device is not harmful from the ITU-R perspective.  
However, our new set of harmful interference thresholds suggests that interference even from a single ground device could sometimes become harmful, if the weather prediction software can tolerate only very low interference levels. 
Importantly, \emph{aggregate} interference resulting in practice from a 5G \emph{network} of uplink or downlink interferers with realistic network densities is often harmful, even considering the ITU-R threshold which is the least conservative
Overall, our comprehensive coexistence study strongly suggests that the concern expressed by weather scientists~\cite{Witze2019, Liu2021, Palmer2020} is well-founded and that ensuring the protection of weather satellites from 5G network interference will likely require additional solutions to be developed and implemented.
Such solutions for harmonious coexistence could include stricter regulatory requirements, standardizing explicit coexistence mechanisms~\cite{Eichen2019, Eichen2021} for 5G transmissions, as well as more sophisticated post-processing algorithms for the measured data, as proposed for e.g. coexistence between the satellite passive soil moisture and ocean salinity (SMOS) sensor in the L-band and terrestrial radar, TV, and radio transmissions~\cite{Daganzo2017,  Camps2011}.
Finally, we emphasize that our detailed approach to studying the impact of novel directional mm-wave cellular deployments is generally timely and relevant also in the methodological sense for other emerging coexistence cases between broadband wireless networks and passive sensing applications in the higher frequency bands~\cite{polese2021coexistence, Weiss2021}.



The remainder of this paper is organized as follows. Section~\ref{section:sysmodel} presents the system model, Section~\ref{section:scenarios_and_criteria} details the interference scenarios and harmful interference criteria, Section~\ref{section:Results} presents the results, and Section~\ref{section:conclusions} concludes the paper.

\section{System Model}
\label{section:sysmodel}

\begin{figure}
\begin{center}
  \includegraphics[width=1\linewidth]{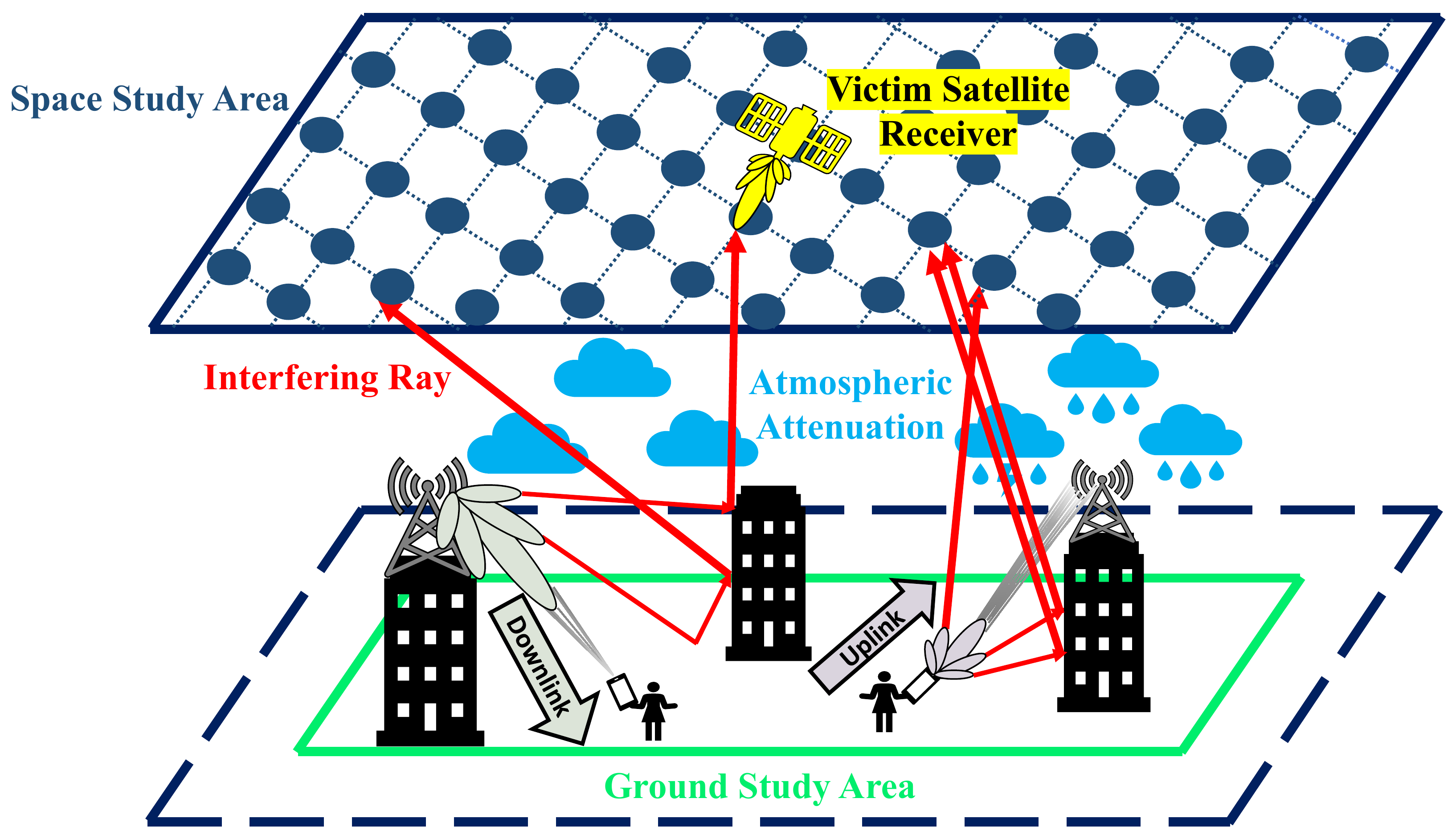} 
  \caption{System model: uplink/downlink transmissions from the ground cellular network cause interference to the victim weather satellite receiver passing above.}
  \label{fig:sys_model}
\end{center}
\end{figure}

In this section we present the system model for our coexistence study, as illustrated overall in Fig.~\ref{fig:sys_model}. We estimate the interference caused by a ground mm-wave cellular deployment to a victim weather satellite receiver, considering realistic antenna patterns and detailed propagation modelling based on 3D ray-tracing and realistic atmospheric attenuation values. We use this system model to study four distinct interference scenarios -- corresponding to downlink or uplink ground transmissions from a single interferer or a network of interferers -- as detailed in Section~\ref{section:inter_scenario}.

\subsection{Ground Cellular Network Model}
\label{section:cell_net_model}

We study the impact of out-of-band interference on the weather satellite in the $23.8$~GHz band where it takes water vapour measurements, from both uplink and downlink ground cellular transmissions in the $24$~GHz band (5G-NR FR2 band n258 as defined by 3GPP)~\cite{ETSI_bs}. We consider a $200$~MHz reference bandwidth\cite{ITU2017} and a transmit power of the ground interferer P$_{TX}$ equal to the leakage power limits established by 3GPP in order to protect satellite measurements, i.e. $P_{TX}=-3$~dBm for the BS in the downlink and $P_{TX} =  1$~dBm for the UE in the uplink~\cite{ETSI_ue,ETSI_bs}.

\begin{figure}[t!]
\centering
\captionsetup[subfigure]{width=0.48\linewidth}
\subfloat[]
          {\includegraphics[width=1\linewidth]{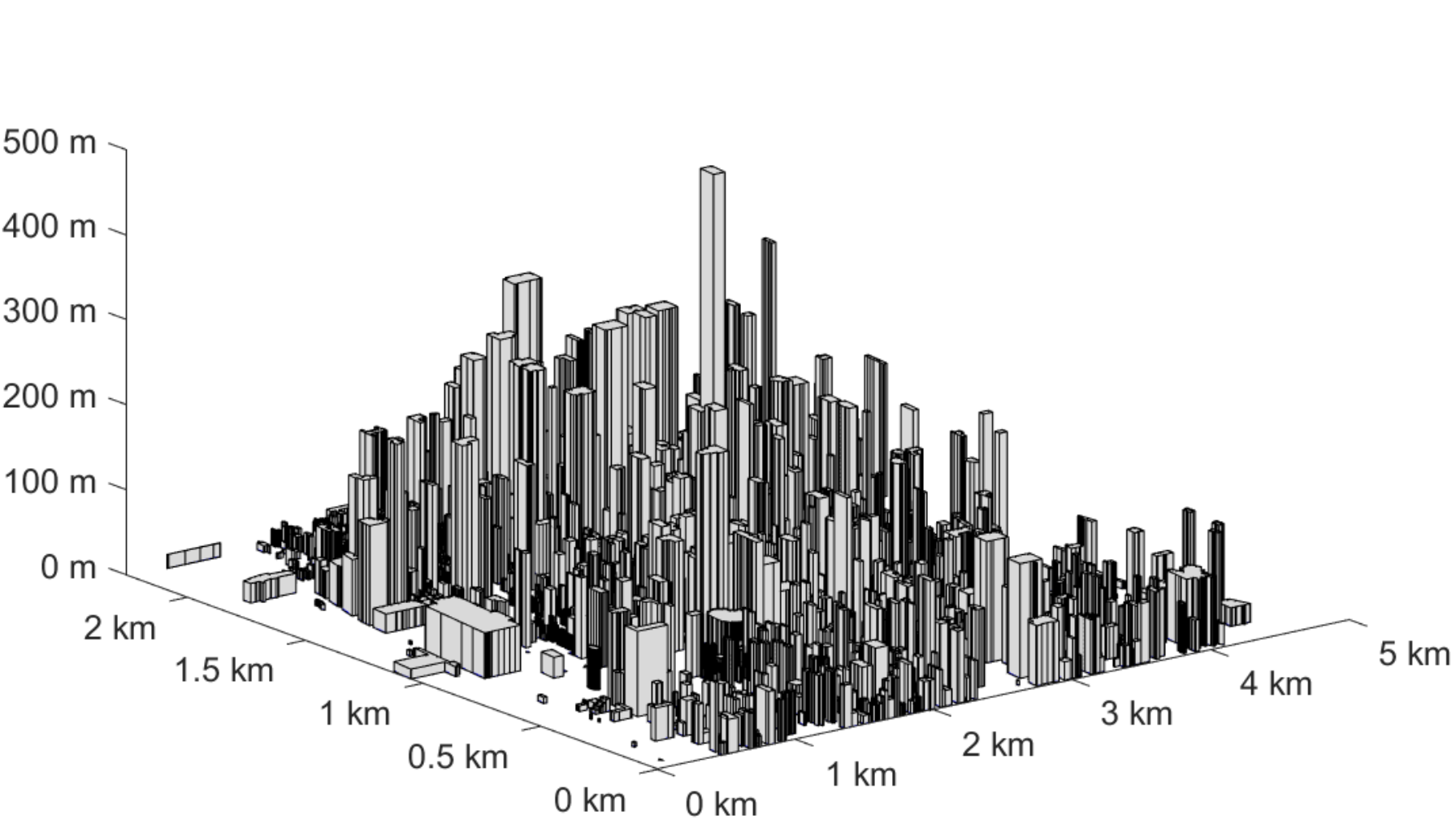} \label{ground_area_3d}}
\hfill
\subfloat[]
          {\includegraphics[width=1\linewidth]{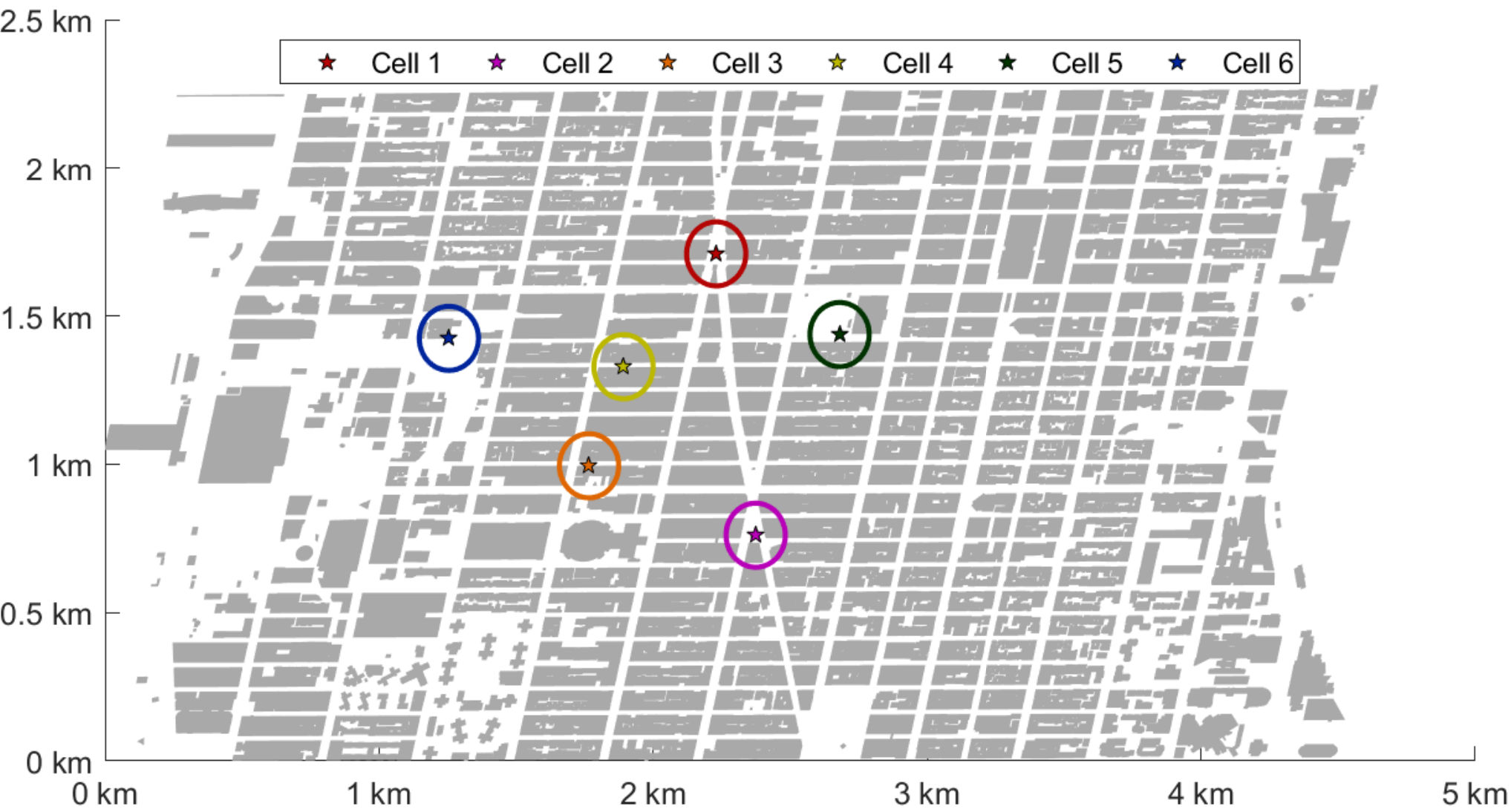} \label{bs_ground}}
\caption{Ground study area in Manhattan, NYC (a) in 3D and (b) in 2D showing the locations of six example cells (the star represents the BS and the surrounding circle represents the cell radius within which the UE is located).}
\label{fig:ground_area}
\end{figure}

We model the interferer as a ground mm-wave cell consisting of a BS and a UE, respectively transmitting in the downlink or uplink. We assume that the mm-wave BS is mounted on a building corner at a height of $h_{BS}=6$~m and that the UE is randomly placed in a line-of-sight (LOS) location within a nominal cell radius $r_{cell}$ from the BS, at a height of $h_{UE}=1.5$~m. In order to realistically model the propagation conditions of an urban environment where such mm-wave cellular networks are likely to be extensively deployed, we consider six example locations for the cell within an $8$~km$^2$ \emph{ground study area} in Manhattan, NYC using real 3D building data from~\cite{NYC_buildings}, as illustrated in Fig.~\ref{ground_area_3d}. Fig.~\ref{bs_ground} shows the six cell locations, which are selected to be representative of three distinct types of urban propagation environment: (i) Cell~$1$ and Cell~$2$ at a major intersection; (ii) Cell~$3$ and Cell~$4$ on a narrow street; and (iii) Cell~$5$ and Cell~$6$ bordering an open area. 


\subsection{Victim Satellite Receiver Model}
\label{section:space_area}
We consider the currently operational MetOp-B weather satellite, with its microwave passive sensor Advanced Microwave Sounding Unit-A (AMSU-A) which measures radiation around the spectral lines of water vapor and oxygen in the frequency range of $23.8$~GHz to $89$~GHz~\cite{Patel1993, Intent2017}.
We define a \emph{space study area} above our Manhattan ground study area, of square shape and length equal to the $2343$~km swath width of AMSU-A, and consider all possible positions -- and for each position, all scanning antenna orientations -- of the Metop-B victim satellite receiver within this area as it passes over the Manhattan ground cellular network, as follows.

\begin{figure}[t!]
\centering
\captionsetup[subfigure]{width=0.48\linewidth}
\subfloat[]
           {\includegraphics[width=0.48\linewidth]{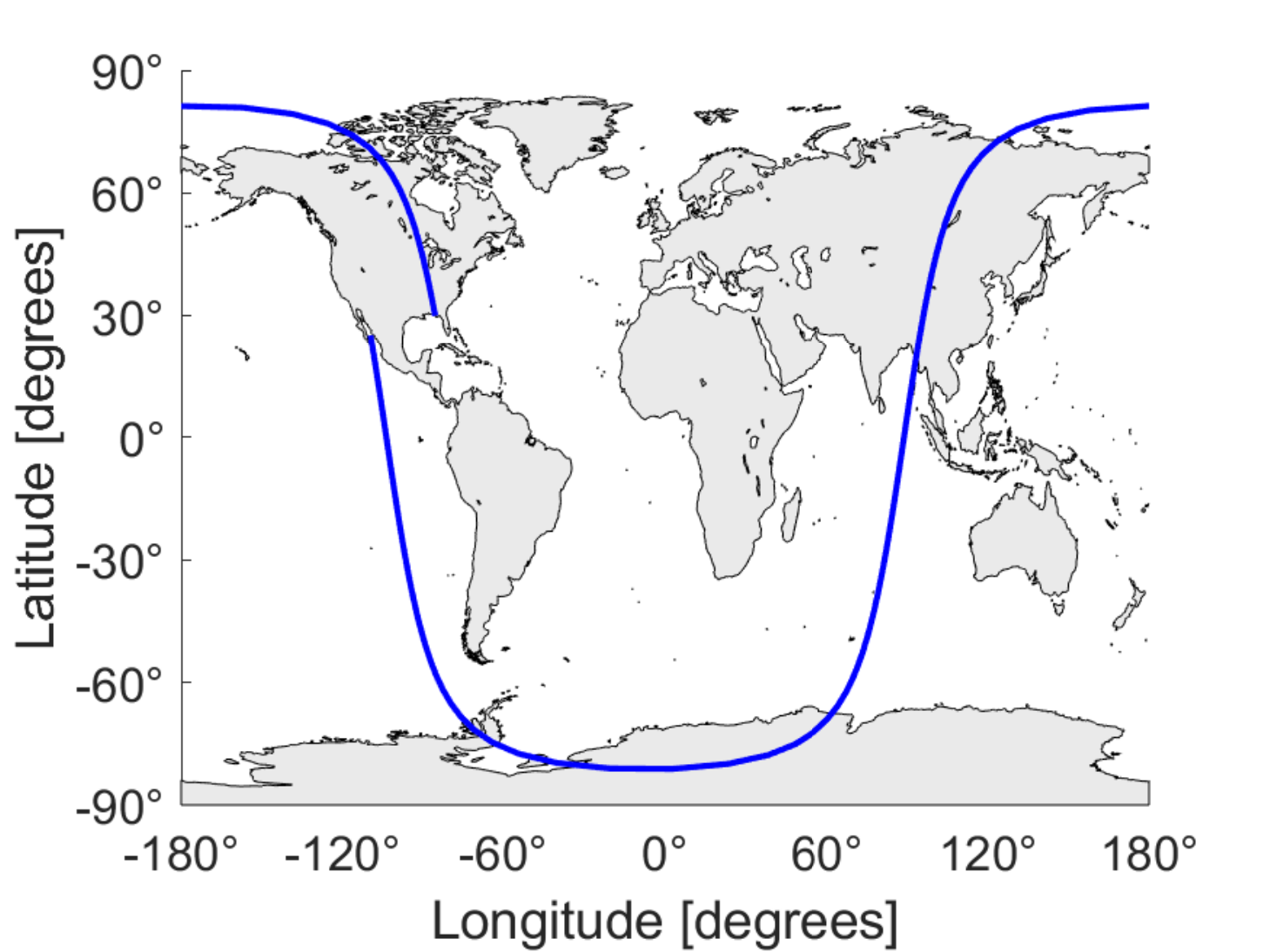} \label{fig:1_orbit}}
\hfill
\subfloat[]
          {\includegraphics[width=0.48\linewidth]{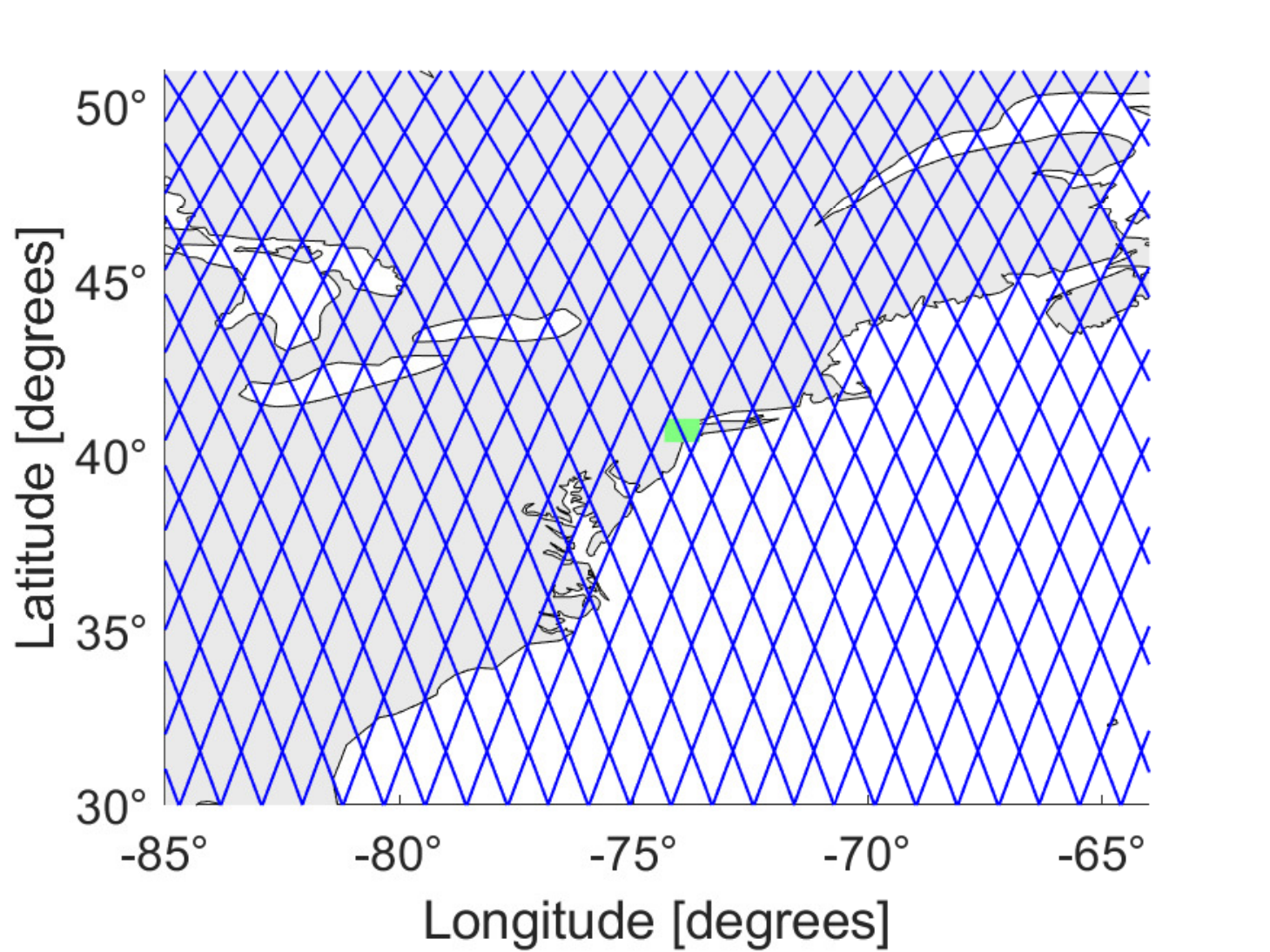} \label{fig:412_orbits}}

\caption{Ground track of the weather satellite MetOp-B (a)~over one orbital period and (b)~over one repeat cycle inside the space study area, above the NYC ground study area shown in green.}
\label{fig:ground_track}
\end{figure}

The satellite traces its \emph{ground track} path over the Earth's surface as it moves across the sky and takes an \emph{orbital period} to complete one orbit around the Earth; as the Earth is rotating, the satellite traces out a different ground track each time it completes one orbital period. The number of orbital periods after which it starts to repeat its ground track is the \emph{repeat cycle} and is equal to 29 days for the Metop-B satellite. We use the Simplified General Perturbations model 4 (SGP4) algorithm~\cite{Vallado2008, Vallado} to compute the ground track of the MetOp-B satellite over the repeat cycle, thus obtaining all its possible positions above the Earth. The ground track over one orbital period and the repeat cycle are illustrated in Figs.~\ref{fig:1_orbit} and~\ref{fig:412_orbits}, respectively.

\begin{figure}[tb!]
\centering
\captionsetup[subfigure]{width=1\linewidth}
\subfloat[]
           {\includegraphics[width=0.7\linewidth]{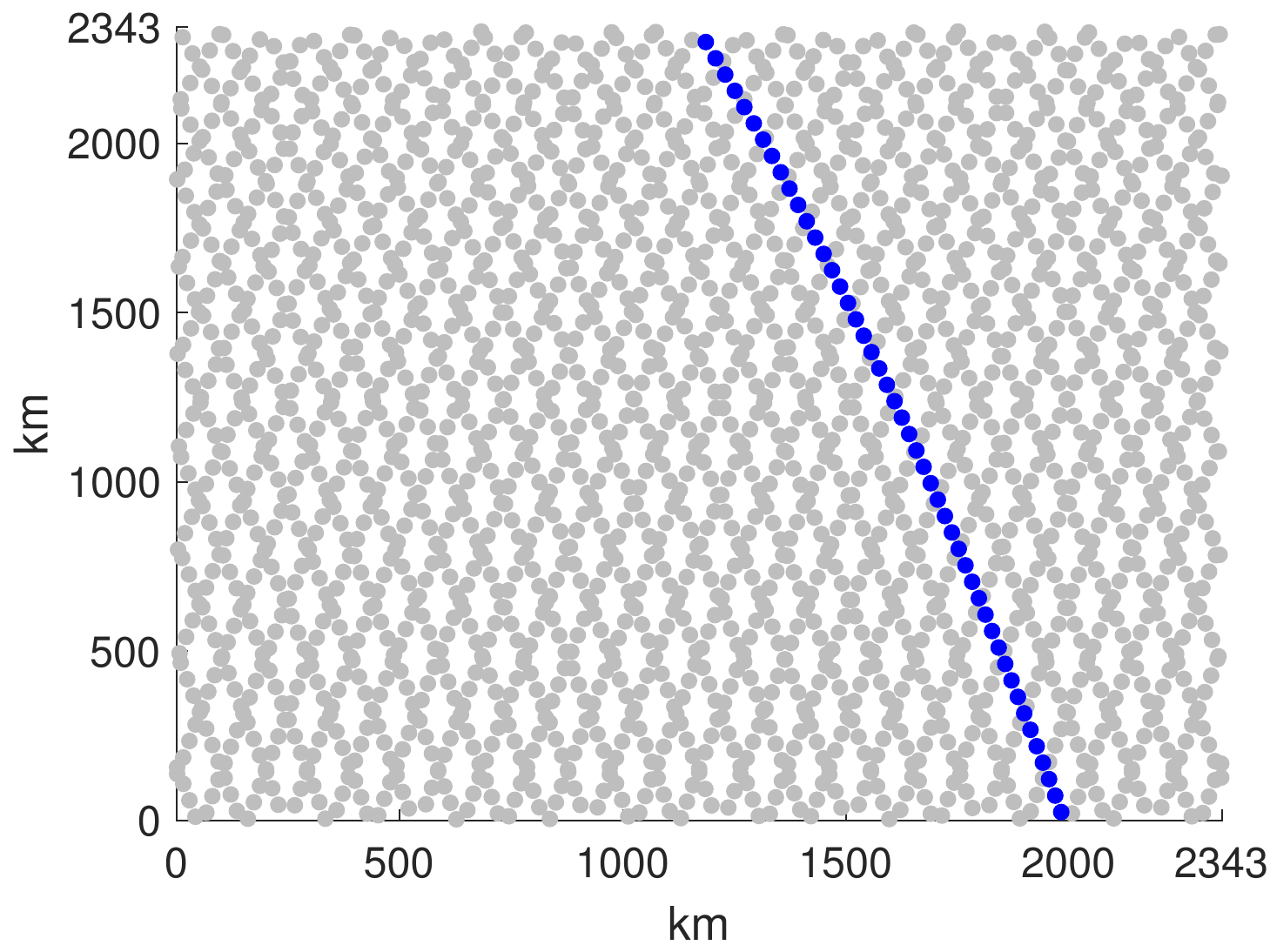} \label{fig:sat_pos}}
\hfill
\subfloat[]
          {\includegraphics[width=1\linewidth]{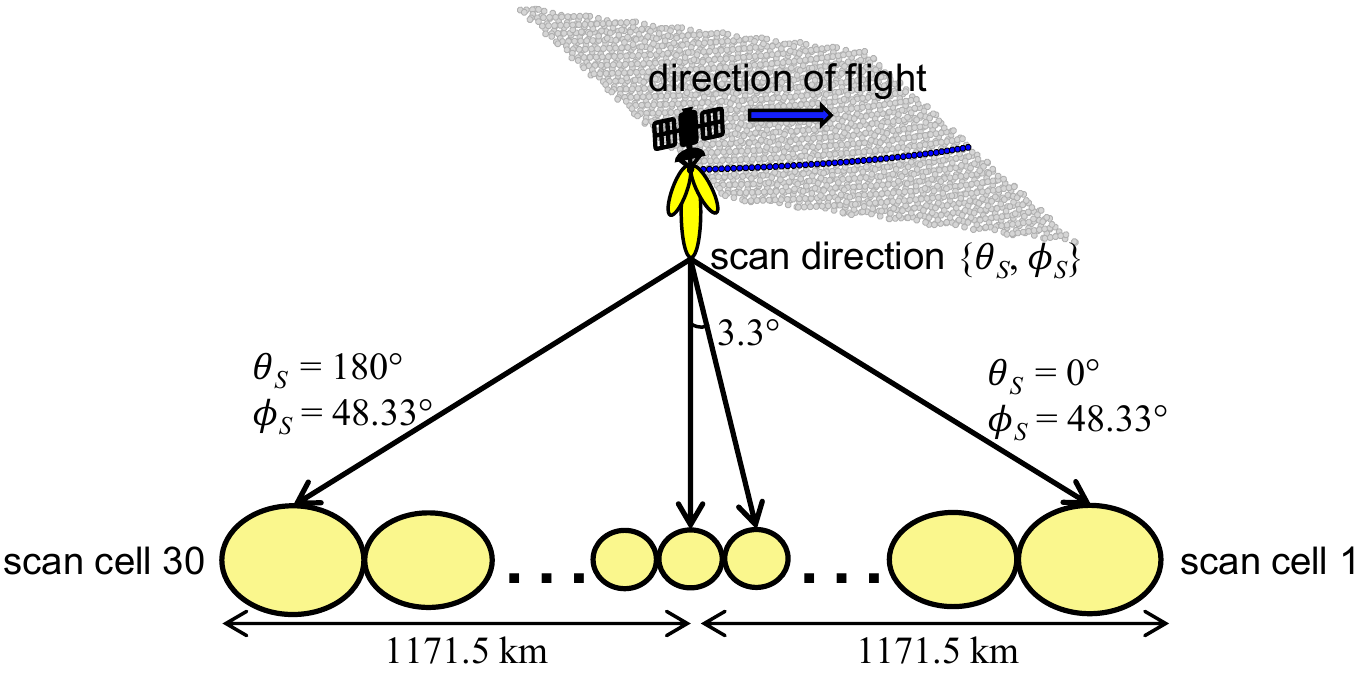} \label{fig:amsu_geometry}}

\caption{Space study area above NYC showing (a) all MetOp-B satellite positions (with an example of consecutive positions represented in blue) and (b)~AMSU-A passive sensor scanning geometry (adapted from~\cite{Patel1993, Intent2017}).}
\label{fig:scan_geo}
\end{figure}

The AMSU-A sensor is a cross track scanner, meaning that it measures radiation in an area perpendicular to the direction of flight of the satellite. The satellite antenna reflectors rotate $48.33^{\circ}$ on each side from the nadir in a step-and-stare sequence, scanning a total of 30 scan pixels in one revolution. We express the orientation of the satellite antenna as $\{\theta_{S}, \phi_{S} \} $, where $\theta_{S}=[-180^{\circ},180^{\circ}]$ is the azimuth angle and $\phi_{S}=[-90^{\circ}, 90^{\circ}]$ is the elevation angle. The field of view at each satellite antenna position is approximately $3.3^{\circ}$, leading to a $50$~km scan pixel diameter at the nadir and a total of $2343$~km swath width for a nominal satellite altitude of $870$~km ~\cite{Patel1993}. 
In our interference analysis, we thus consider 30 distinct orientations $\{\theta_{S}, \phi_{S} \} $ of the satellite receiver antenna for each satellite position sampled at $50$~km intervals along the ground track within the space study area. Fig.~\ref{fig:sat_pos} illustrates the sampled satellite positions inside the space study area, showing in blue an example of consecutive positions along the ground track within one orbital period. Fig.~\ref{fig:amsu_geometry} shows the corresponding scanning geometry of the AMSU-A sensor, also indicating the satellite's direction of flight.

\subsection{Antenna Models}
\label{section:antennas}

\begin{figure}[t!]
\centering
\captionsetup[subfigure]{width=0.5\linewidth}

\subfloat[$4\times 4$ URA UE antenna]
          {\includegraphics[width=0.47\linewidth]{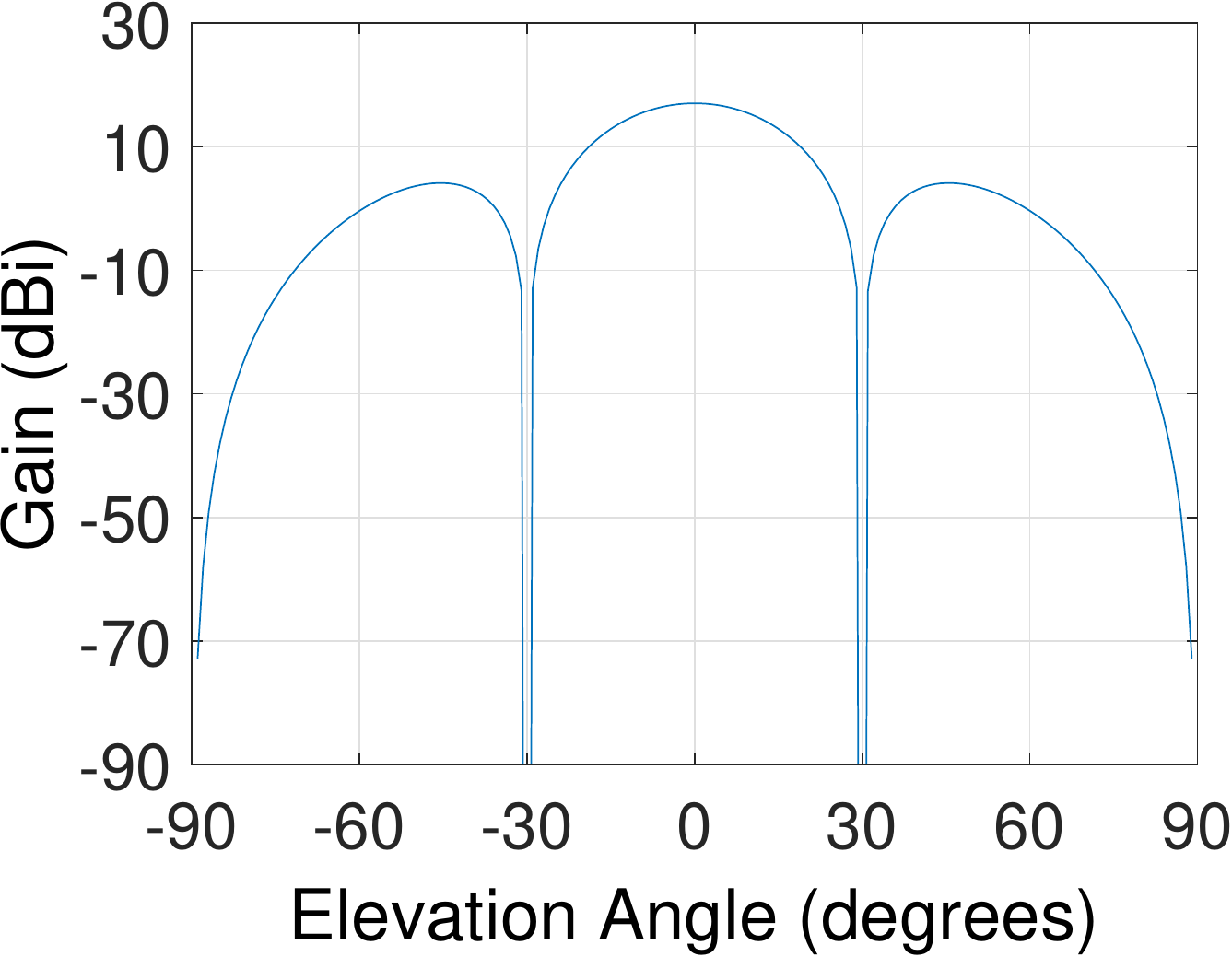} \label{4_4_2D}}
~            
\subfloat[$16\times 16$ BS antenna]
          {\includegraphics[width=0.47\linewidth]{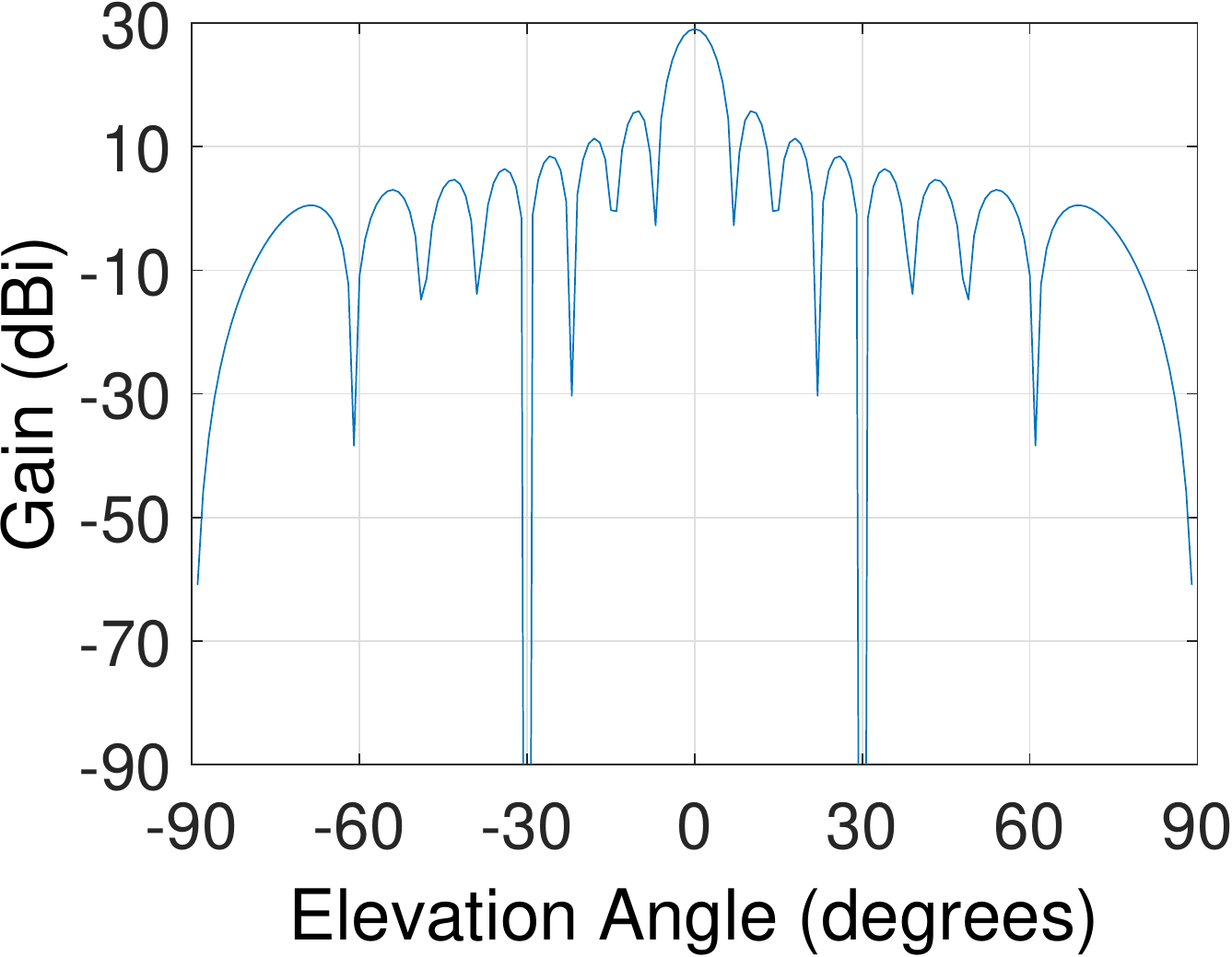} \label{16_16_2D}}          
\caption{Ground cellular interferer antenna pattern in the azimuth plane, for transmissions in the (a)~uplink from the UE or (b)~downlink from the BS.}

\label{antennas}
\end{figure}

We model the BS and UE antennas as uniform rectangular arrays (URAs) with patch antenna elements,\footnote{Specifically, we approximate the patch element with a cosine element with the radiation pattern given by $\cos^m(\phi)\cos^n(\theta)$ with $m=n=0.5$.} using the Matlab Toolbox ``Phased Array System''~\cite{MathWorks}.  We assume $16\times 16$ and $4\times 4$ URAs with maximum gains of $29$~dBi and $17$~dBi and half power beamwidths of $6.4^{\circ}$ and $25.8^{\circ}$ for the BS and UE, respectively. Fig.~\ref{antennas} illustrates the corresponding antenna patterns in the azimuth plane, $G_{BS}(\theta=0^{\circ},\phi)$ and $G_{UE}(\theta=0^{\circ},\phi)$, where $\theta$ and $\phi$ are the azimuth and elevation angles, respectively. We define the orientation of the BS/UE antenna as $ \{ \theta_{BS}, \phi_{BS} \}$ and $ \{ \theta_{UE}, \phi_{UE}  \}$, respectively. For an arbitrary angle-of-departure (AoD) direction from the transmitting ground interferer $ \{ \theta_j, \phi_j \} $, the antenna gain is given by $G_{BS}(\theta_j-\theta_{BS},\phi_j-\phi_{BS})$ or  $G_{UE}(\theta_j-\theta_{UE},\phi_j-\phi_{UE})$, when the BS/UE antenna antenna main lobe is oriented in the direction $\{\theta_{BS}, \phi_{BS} \} $ or $\{\theta_{UE}, \phi_{UE} \} $, respectively.

We model the satellite antenna pattern  using real measured AMSU-U data and characteristics reported in~\cite{Mo1999, ITU2010}. The AMSU-A antenna is an offset Cassegrain reflector with a maximum gain of $34.4$~dB~\cite{ITU2010} and a half power beamwidth of $3.3^{\circ}$. Fig.~\ref{amsu_ant_2D} illustrates the  satellite antenna pattern in the azimuth plane, $G_{RX}(\theta=0^{\circ},\phi)$, as given in \cite{Mo1999}. We note that we assumed for simplicity that the overall pattern is symmetrical in the azimuth and elevation planes, i.e. we created the 3D pattern from $G_{RX}(\theta=0^{\circ},\phi)$ by interpolating over the range of $\theta=[-180^{\circ}, 180^{\circ}]$. For an arbitrary angle-of-arrival (AoA) direction at the victim satellite receiver $ \{ \theta_k, \phi_k \} $ and when the satellite antenna is oriented in the direction $\{\theta_{S}, \phi_{S} \} $, the satellite antenna gain is given by $G_{RX}(\theta_k-\theta_{S},\phi_k-\phi_{S})$.

\begin{figure}[t!]
\centering
\captionsetup[subfigure]{width=0.5\linewidth}
%
          {\includegraphics[width=0.5\linewidth]{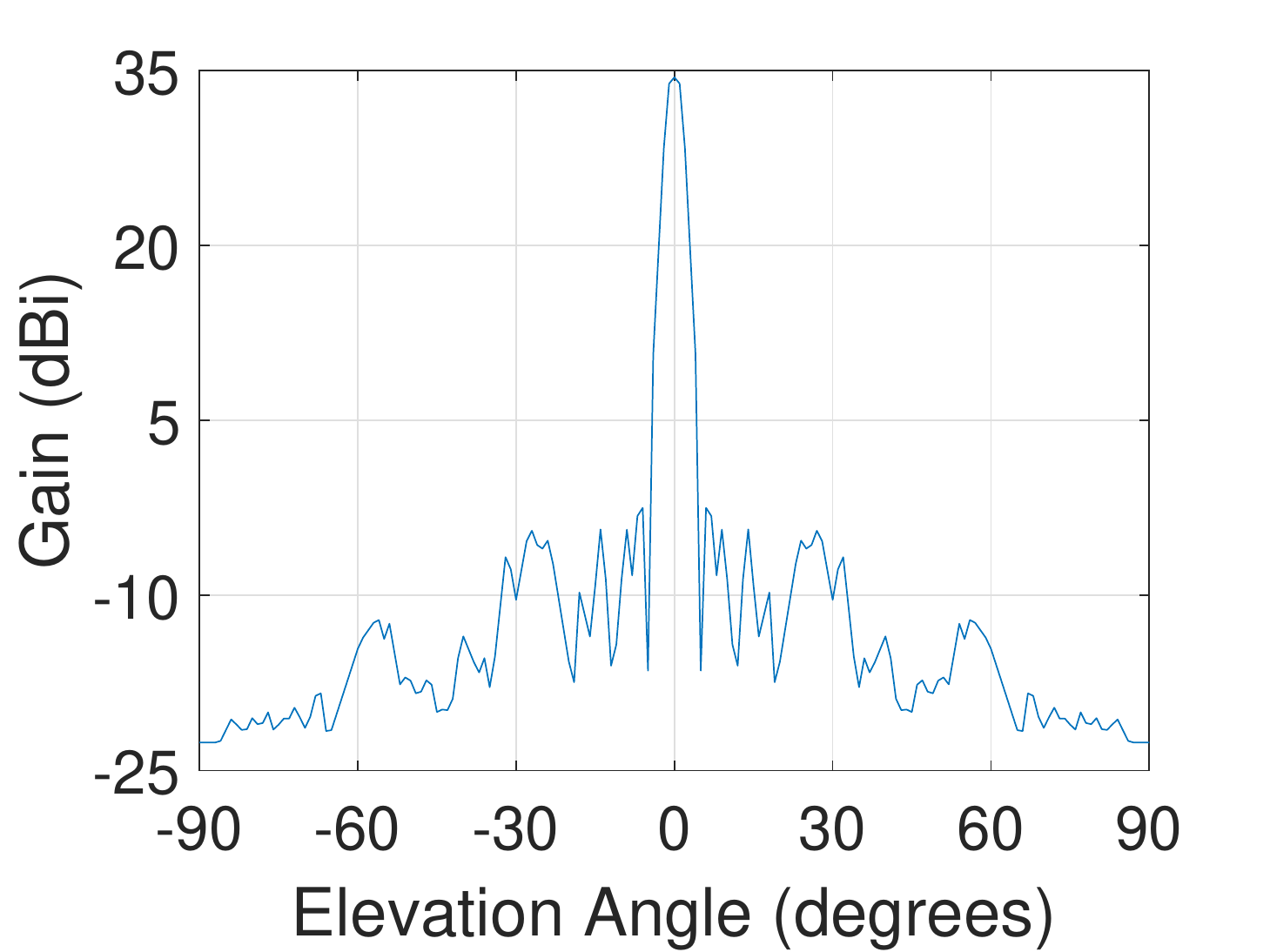} \label{amsu_ant_2D}}

\caption{AMSU-A victim satellite receiver antenna pattern in the azimuth plane.}
\label{amsu_ant_2D}
\end{figure}

\subsection{Propagation Modelling}
\label{section:prop_model}

We obtain site-specific propagation data using the open-source mm-wave ray-tracing tool~\cite{Simic2017} and real 3D building data of our Manhattan ground study area (\emph{cf.}~Section~\ref{section:cell_net_model}). We perform a dedicated high-resolution ray-tracing simulation for each considered BS/UE ground interferer location, launching rays omnidirectionally with granularity of $0.5^{\circ}$ and $0.1^{\circ}$ in the elevation and azimuth, respectively. We consider a receiver sphere of $50$~km diameter for collecting the incident interfering rays at the victim satellite at its orbiting height, which ranges from $815$~km to $820$~km above Earth, for each considered satellite position in the satellite study area specified in Section~\ref{section:space_area}. The ray-tracing simulation considers free-space propagation and reflections of up to six bounces.\footnote{Diffraction is neglected as it is not a significant propagation mechanism at mm-wave frequencies~\cite{Molisch2014}. We also note that the accuracy of our ray-tracer has been validated against real mm-wave antenna-array measurements in an urban environment~\cite{Ichkov2020}.} We assume a reflection loss of $4.7$~dB and $3$~dB for the ground and buildings, respectively~\cite{GonzalezDominguez2016,Ichkov2020}.

For each considered satellite position, the output of the omnidirectional ray-tracing simulation is the set of received interfering rays and the corresponding computed path loss -- consisting of free-space plus ground/building reflection loss -- for each ray. We post-process this omnidirectional ray-tracing output by applying the directional antenna gains given the orientations of satellite receiver antenna $\{\theta_{S}, \phi_{S} \} $ and BS/UE ground interferer antenna $\{\theta_{BS}, \phi_{BS} \} $ or $\{\theta_{UE}, \phi_{UE} \} $, respectively, as corresponding to the interference scenarios defined in Section~\ref{section:inter_scenario}. The final interference calculation additionally takes into account the atmospheric attenuation and sums the interference power of all incident rays, as detailed in Section~\ref{section:inter_calc}.

We compute the atmospheric attenuation using the MATLAB tool ``ITU-R Propagation Models Software Library'' from the Centre National d'\'Etudes Spatiales (CNES)~\cite{cnes}. Specifically, the atmospheric attenuation is defined as attenuation due to rain, gas, clouds, and scintillation, and may be calculated using~\cite{ITU2015}:
\begin{eqnarray}
L_{atm}(p)&=& A_{G}(p)+\sqrt{(A_{R}(p)+A_{C}(p))^2+A_{S}(p)^3} , \label{eq:atm_att}
\end{eqnarray}
where $A_{G}(p)$ is the gaseous attenuation due to water vapour and oxygen, $A_{R}(p)$ is the rain attenuation, $A_{C}(p)$ is the attenuation due to clouds and $A_{S}(p)$ is the attenuation due to tropospheric scintillation, and $p$ is the unavailability probability, which is defined in the range from $0.001\%$ to $50\%$, so that $L_{atm}(p)$ is minimum for $p = 50\%$  and maximum for  $p = 0.001\%$.

\begin{figure}[t!]
\centering
\captionsetup[subfigure]{width=9\linewidth}
\subfloat[downlink interference scenario geometry]
          {\includegraphics[width=0.8\linewidth]{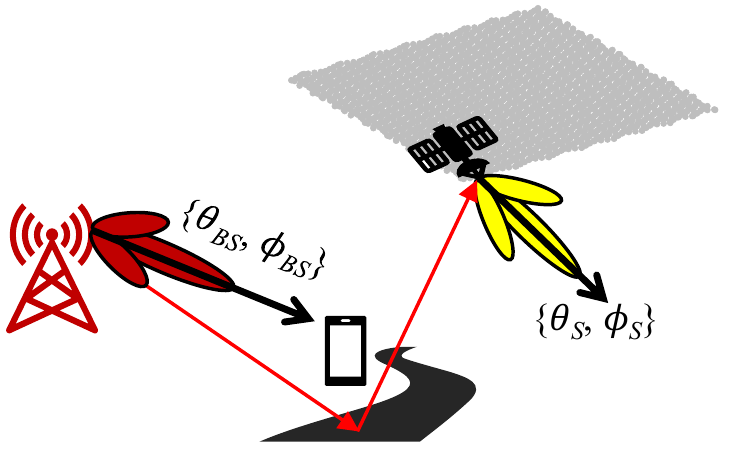} \label{downlink_scenario}}
\hfill
\subfloat[BS (star) and LOS UE positions (shaded) considered inside each cell]
          {\includegraphics[width=1\linewidth]{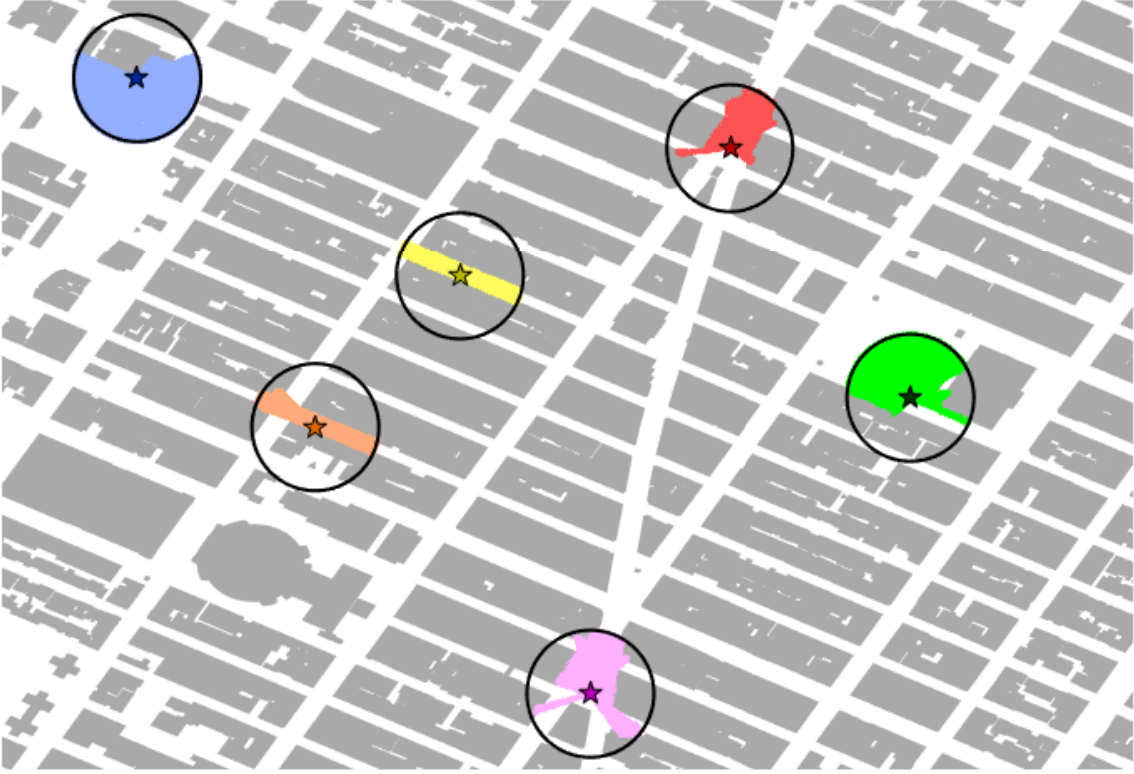} \label{downlink_ground}}
\hfill
\subfloat[CDF of the elevation angles of the BS $\phi_ {BS}$ for each cell]
          {\includegraphics[width=1\linewidth]{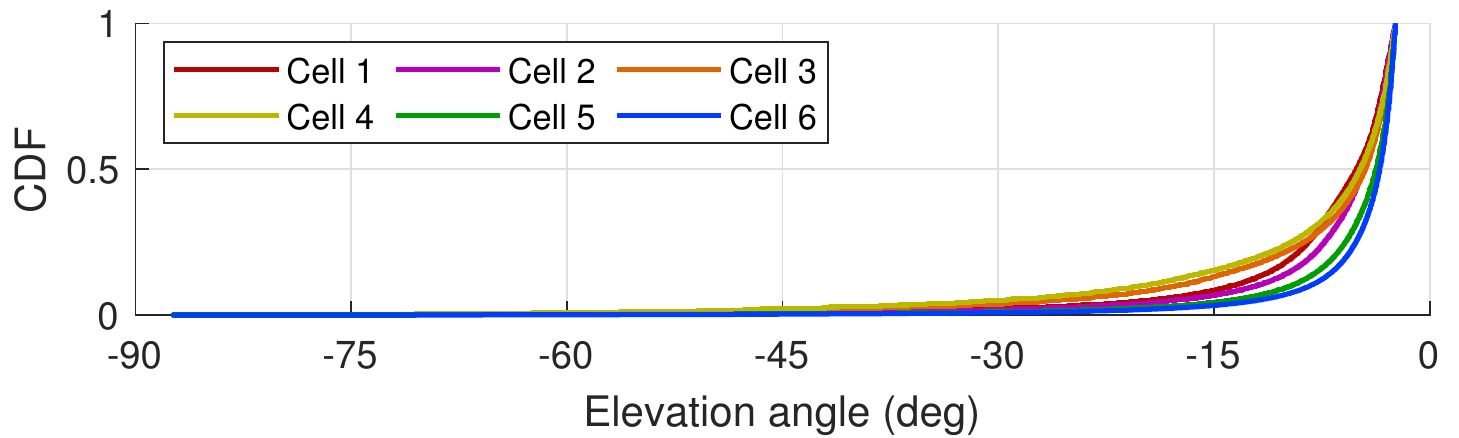} \label{downlink_cdf_angles}}
\caption{Single downlink interferer scenario.}
\label{downlink_trans}
\end{figure}

\begin{figure}[t!]
\centering
\captionsetup[subfigure]{width=9\linewidth}
\subfloat[uplink interference scenario geometry]
          {\includegraphics[width=0.8\linewidth]{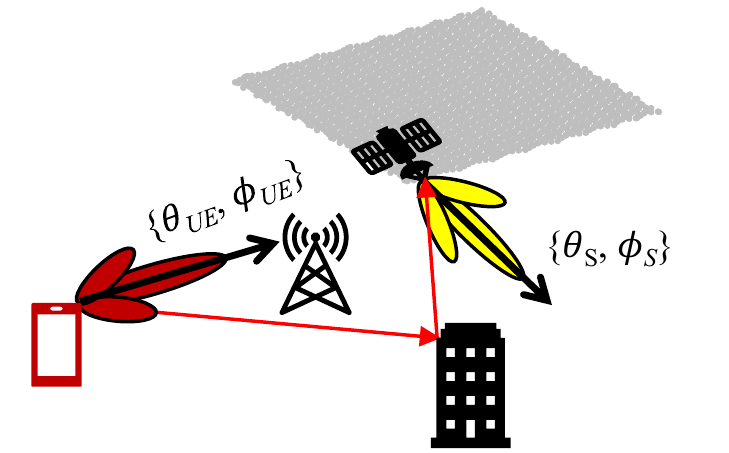} \label{uplink_scenario}}
\hfill
\subfloat[BS (star) and UE (cross) positions considered inside each cell]
          {\includegraphics[width=1\linewidth]{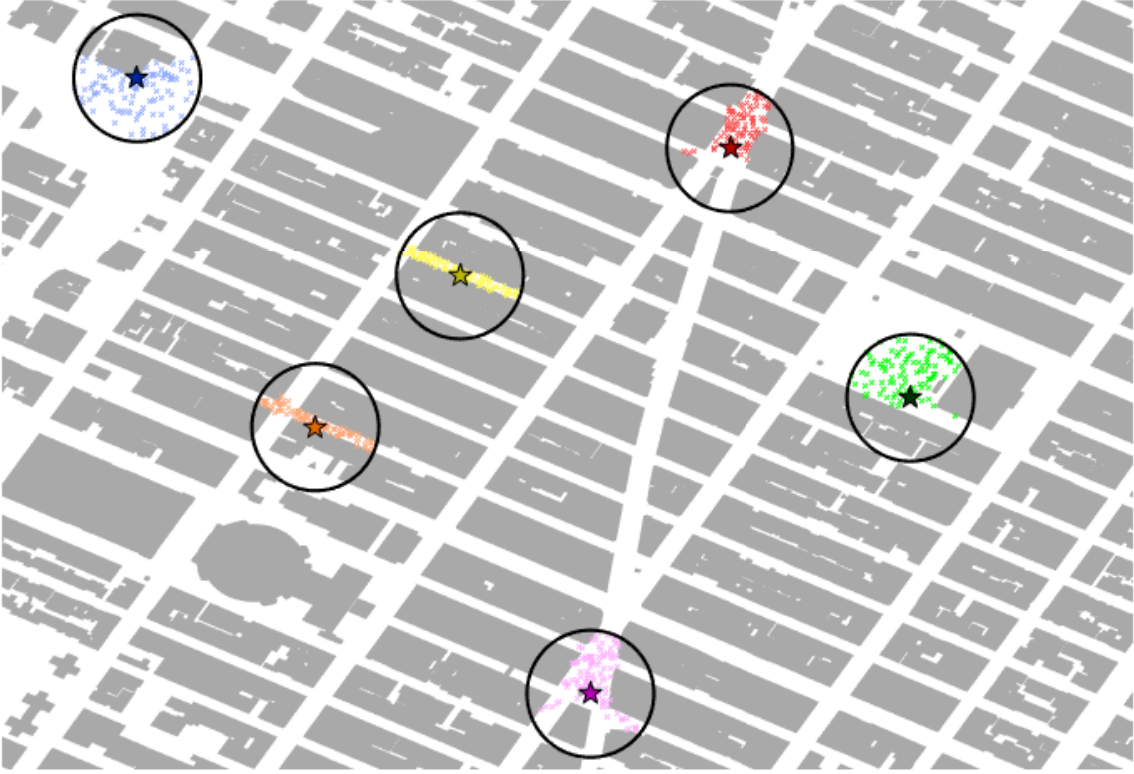} \label{uplink_ground}}
\hfill
\subfloat[CDF of the elevation angles of the UEs  $\phi_ {UE}$ for each cell]
          {\includegraphics[width=1\linewidth]{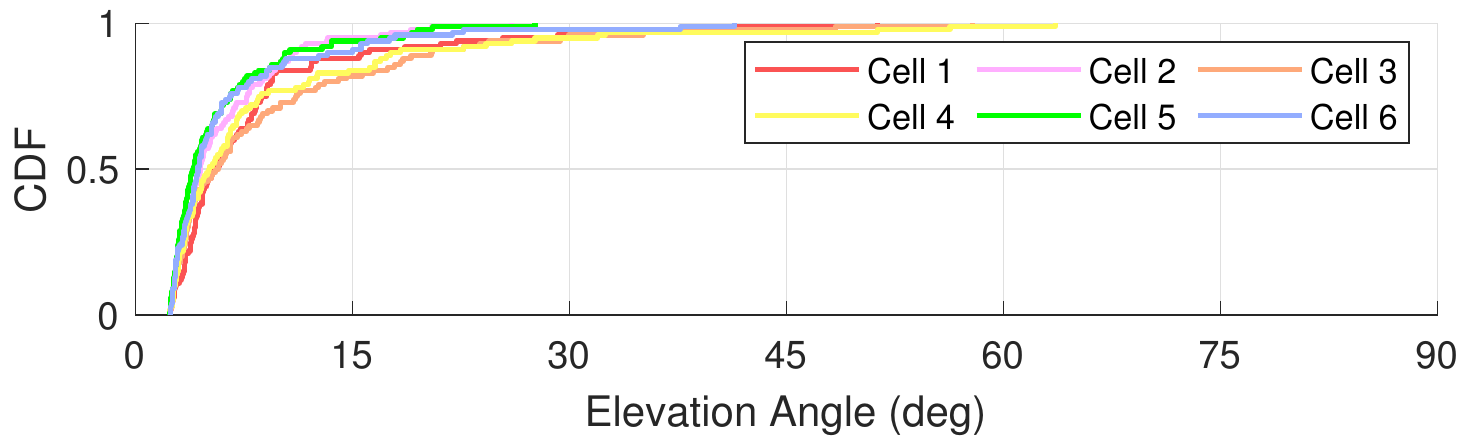} \label{uplink_cdf_angles}}
\caption{Single uplink interferer scenario.}
\label{downlink_trans}
\end{figure}

\section{Interference Scenarios \& \\Harmful Interference Criteria}
\label{section:scenarios_and_criteria}
\subsection{Interference Scenarios}
\label{section:inter_scenario}

In this section we present the four distinct interference scenarios in our coexistence study in Section~\ref{section:Results}: downlink or uplink ground transmissions from a single interferer or a network of interferers. Throughout, we assume that only one ground transmitter is active at a time per cell; this corresponds to a time division multiple access (TDMA) system, such that only one UE is actively served by the BS at a time and the BS/UE antennas are then beamformed towards each other, i.e. oriented along the LOS direction between the BS and the given UE. The corresponding range of considered BS/UE ground interferer antenna orientations $\{\theta_{BS}, \phi_{BS} \} $ or $\{\theta_{UE}, \phi_{UE} \} $ is defined specific to each scenario, for each of our six example cells. For each considered BS/UE orientation, we then compute the interference at each of the victim satellite receiver positions in the space study area and for each of the $30$  satellite sensor orientations $\{\theta_{S}, \phi_{S} \} $ (\emph{cf.}~Section~\ref{section:space_area}).

\subsubsection{Single Downlink Interferer Scenario}
\label{section:scenario_DLsingle}

This scenario considers the impact of interference on the victim satellite receiver from a single BS transmitting in the downlink, as illustrated in Fig.~\ref{downlink_scenario}.  In order to capture the distribution of realistic beam orientations of the ground interferer, we consider for each cell the range of BS antenna orientations $\{\theta_{BS}, \phi_{BS} \} $ corresponding to all LOS UE positions sampled on a $1$~m $\times 1$~m grid within the cell boundary defined by $r_{cell}=108$~m, as illustrated in Fig.~\ref{downlink_ground}. The resulting distribution of the elevation angle $\phi_{BS}$ is plotted in Fig.~\ref{downlink_cdf_angles} for each of our six example cells, where smaller negative angles correspond to UEs closer to the BS. We note that, as illustrated in Fig.~\ref{downlink_scenario}, even though the BS antenna is always directed downwards towards the UE, reflections -- in particular off the ground -- of rays transmitted through the BS antenna main lobe or sidelobes can nonetheless reach the weather satellite above, causing interference.

\subsubsection{Single Uplink Interferer Scenario}
\label{section:scenario_ULsingle}
This scenario considers the impact of interference on the victim satellite receiver from a single UE transmitting in the uplink, as illustrated in Fig.~\ref{uplink_scenario}. We consider a distribution of realistic beam orientations of the ground interferer, given for each cell by the range of UE antenna orientations $\{\theta_{UE}, \phi_{UE} \} $ corresponding to $100$ uniformly randomly sampled LOS UE positions within the cell boundary defined by $r_{cell}=108$~m, as illustrated in Fig.~\ref{uplink_ground}. We note that  since the interference calculation requires a dedicated ray-tracing simulation for each ground transmitter location (\emph{cf.} Section~\ref{section:prop_model}), it was computationally infeasible to consider as many UE locations as for the downlink scenario. Nonetheless, the smaller number of UE locations considered for the uplink still allow us to obtain a representative distribution of $\{\theta_{UE}, \phi_{UE} \} $ for each cell, as evident from comparing the distributions in Figs.~\ref{downlink_cdf_angles}~and~\ref{uplink_cdf_angles}  for the opposite angles $\phi_{BS}$ and $\phi_{UE}$, respectively. We note that since the UE antenna is pointed upwards towards the BS, uplink ground transmissions can interfere at the weather satellite via rays transmitted through the UE antenna main lobe or sidelobes that either travel directly or are reflected off buildings, as illustrated in Fig.~\ref{uplink_scenario}.


\subsubsection{Network of Downlink Interferers Scenario}
\label{section:scenario_DLnetwork}

\begin{figure}
\begin{center}
  \includegraphics[width=1\linewidth]{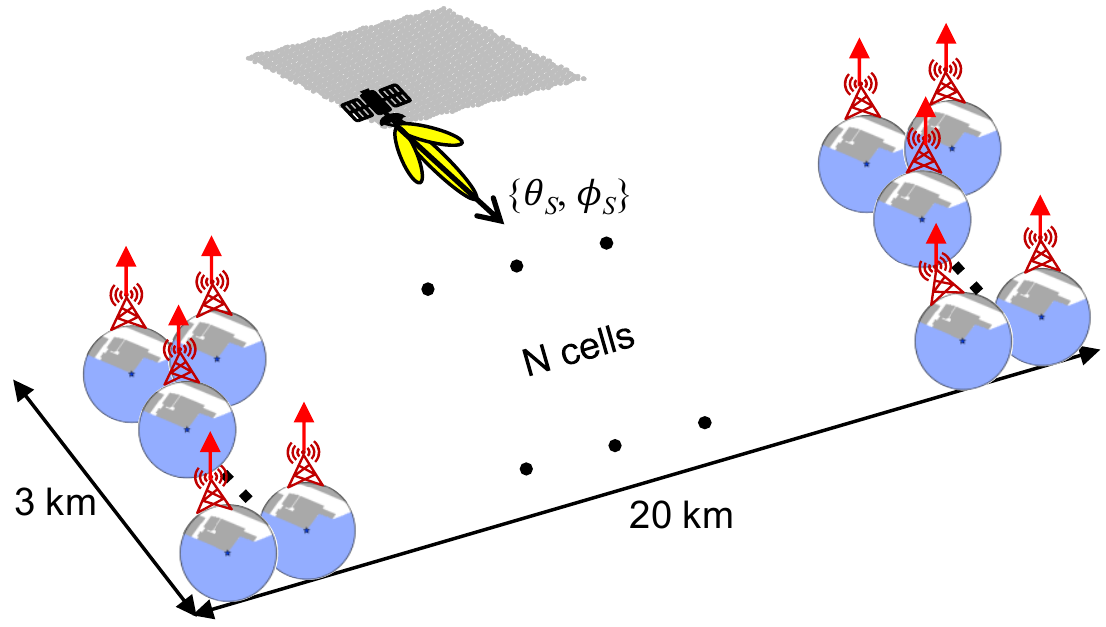} 
  \caption{Network of downlink interferers scenario, illustrating the simplified aggregate interference model of a mm-wave Manhattan network assuming homogeneous cells (of example Cell 6 type).}
  \label{fig:cells_manhattan_bs}
\end{center}
\end{figure}

This scenario considers the impact of interference on the victim satellite receiver from a \emph{network} of BSs transmitting in the downlink, representative of a real multi-cell ground mm-wave network deployment. We consider a range of realistic network densities $\lambda_{cell}=\{25, 50, 100, 200\}$~BSs/km$^2$, ranging from sparse hotspot mm-wave coverage for offloading in early non-standalone 5G networks to dense coverage in mature standalone 5G-NR deployment~\cite{LSImmW-coverage}. In order to keep our overall simulations computationally feasible,
we assume a simplified aggregate network interference model of a city-wide mm-wave network covering the $A_{M}=60$~km$^2$ area of Manhattan, consisting of $N=\lambda_{cell} \times A_{M}$ \emph{homogeneous} cells, i.e. all of cell type 1-6. We assume a frequency reuse factor 1 for the network, such that the $N$ cells operate simultaneously, resulting in aggregate interference from $N$ ground BSs at the satellite, equal simply to $N$ times the interference level caused by a single BS of the same cell type in the single-interferer scenario in Section~\ref{section:scenario_DLsingle}. Fig.~\ref{fig:cells_manhattan_bs} illustrates the scenario and underlying aggregate interference model. We note that the distribution of BS antenna orientations $\{\theta_{BS}, \phi_{BS} \} $ considered for each cell type is defined as in the single downlink interfer scenario in Section~\ref{section:scenario_DLsingle}, except that the UE positions are sampled within the the cell boundary defined by $r_{cell}=\{108, 74, 52, 36\}$~m, corresponding to the network densities of $\lambda_{cell}=\{25, 50, 100, 200\}$~BSs/km$^2$, respectively.\footnote{We map the network density to nominal cell radius as half of the inter-BS distance, assuming uniformly distributed cells over the ground study area.}

\subsubsection{Network of Uplink Interferers Scenario}
\label{section:scenario_ULnetwork}

\begin{figure}
\begin{center}
  \includegraphics[width=1\linewidth]{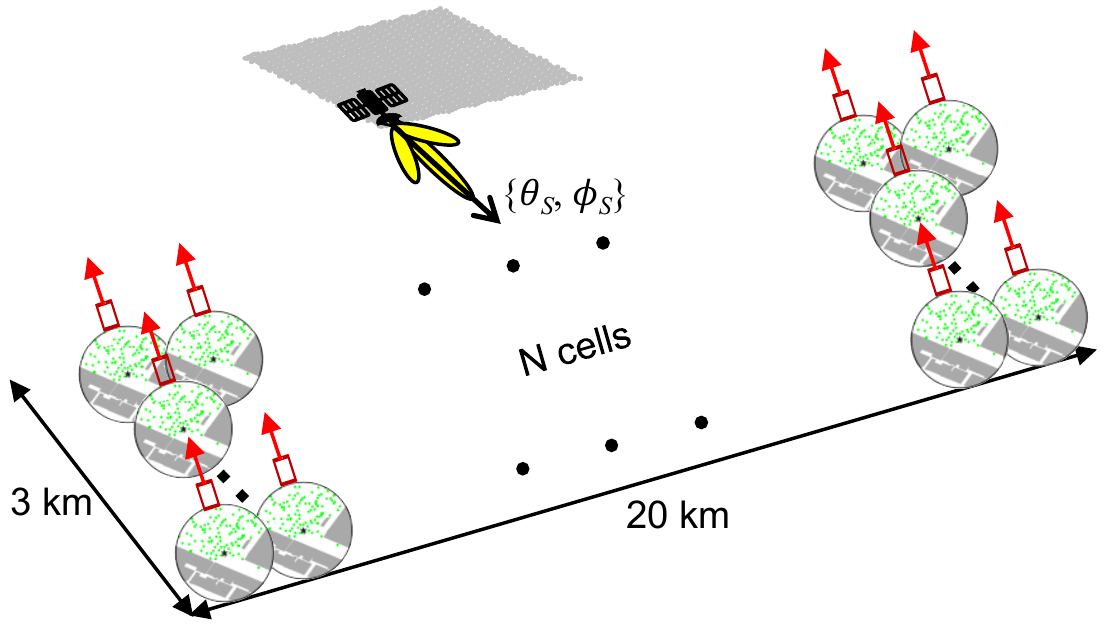} 
  \caption{Network of uplink interferers scenario, 
 illustrating the simplified aggregate interference model of a mm-wave Manhattan network assuming homogeneous cells (of example Cell 5 type).}
  \label{fig:cells_manhattan_ues}
\end{center}
\end{figure}

This scenario considers the impact of interference on the victim satellite receiver from a \emph{network} of UEs transmitting in the uplink, representative of a real multi-cell ground mm-wave network deployment. We consider the same range of network densities $\lambda_{cell}$ and same aggregate interference model of the city-wide Manhattan mm-wave ground network as defined for the downlink in Section~\ref{section:scenario_DLnetwork}. Fig.~\ref{fig:cells_manhattan_ues} illustrates the uplink network scenario and underlying aggregate interference model. We note that the distribution of UE antenna orientations $\{\theta_{UE}, \phi_{UE} \} $ considered for each cell type is defined equivalently to the downlink network case, i.e. as in the single uplink interfer scenario in Section~\ref{section:scenario_ULsingle}, except only considering the sampled UE positions within the cell boundary defined by $r_{cell}$ corresponding to the given $\lambda_{cell}$.

\subsection{Interference Calculation}
\label{section:inter_calc}

Let us assume that there are $M$ incident  rays from the ground interferer received at a given victim satellite receiver position. The total attenuation for the $m^{th}$ incident interfering ray is defined as
\begin{eqnarray}
L_{m}&=& L_{FS}L_{GL}L_{BL}^{Q}L_{atm}(p), \label{eq:total_loss}
\end{eqnarray}
where $L_{FS}$ is the free space path loss calculated using the Friis transmission formula~\cite{Rappaport2002}, $L_{GL}$ is the ground reflection loss of the ray, $L_{BL}^Q$ is the reflection loss due to building materials after the ray has been reflected in $Q$ building interactions, and $L_{atm}(p)$ is the atmospheric attenuation as defined in~(\ref{eq:atm_att}). We note that we compute $L_{FS}$, $L_{GL}$, and $L_{BL}^Q$ for the $m^{th}$ ray using ray-tracing simulations with $Q\leq6$, as detailed in Section~\ref{section:prop_model}.

The total interference power received at the victim satellite position for a given satellite sensor orientation $\{\theta_{S}, \phi_{S} \} $ and ground interferer antenna orientation $\{\theta_{G}, \phi_{G} \} $ is given by
\begin{eqnarray}
\resizebox{.9\hsize}{!}{$I(\theta_{S},\phi_{S}, \theta_{G},\phi_{G})= P_{TX}\sum_{m=1}^{M}\frac{ G_{TX}(\theta_{j}-\theta_{G},\phi_{j}-\phi_{G})G_{RX}(\theta_{k}-\theta_{S},\phi_{k}-\phi_{S})}{L_{m}}$}, \label{eq:inter}
\end{eqnarray}
where $\{\theta_{j}, \phi_{j} \}$ and $\{\theta_{k}, \phi_{k} \}$ are the AoD and AoA of the $m^{th}$ ray at the ground interferer and victim satellite receiver, respectively; $\{\theta_{G}, \phi_{G} \}$ are given by $\{\theta_{BS}, \phi_{BS} \}$ or $\{\theta_{UE}, \phi_{UE} \}$, and $G_{TX}(\theta, \phi)$ is given by $G_{BS}(\theta, \phi)$ or $G_{UE}(\theta, \phi)$, as defined in Section~\ref{section:antennas}, for the single downlink and uplink interferer scenarios defined in Sections~\ref{section:scenario_DLsingle}~and~\ref{section:scenario_ULsingle}, respectively.

For the network of downlink or uplink interferers scenarios defined in Sections~\ref{section:scenario_DLnetwork}~and~\ref{section:scenario_ULnetwork}, the aggregate interference power received at the victim satellite from the network of interferers is given by
\begin{eqnarray}
I_{agg}(\theta_{S},\phi_{S}, \theta_{G},\phi_{G})&=& N \times I(\theta_{S},\phi_{S}, \theta_{G},\phi_{G}), \label{eq:agg_int}
\end{eqnarray}
where $I(\theta_{S},\phi_{S}, \theta_{G},\phi_{G})$ is the interference received from a single BS/UE ground interferer in the downlink or uplink, respectively, as defined by~(\ref{eq:inter}).

\subsection{Harmful Interference Criteria}
\label{section:Evaluation}


We evaluate whether the interference given by~(\ref{eq:inter}), (\ref{eq:agg_int}) is harmful to the weather satellite by comparing it against two types of interference thresholds: $\gamma_1$ defined by the ITU-R as a function of the noise fluctuations of the passive sensor (\emph{cf.} Section~\ref{Subsection:gamma_one}); and $\{\gamma_2, \gamma_3, \gamma_4\}$ which we define as the maximum interference that would not considerably degrade the AMSU-A radiometer sensitivity (\emph{cf.} Section~\ref{Subsection:gamma_two}). We estimate the likelihood of exceeding the given interference threshold in terms of the percentage of time, as detailed in Section~\ref{Subsection:harmful_likelihood}.
\subsubsection{Noise Power Fluctuation Threshold $\gamma_1$}
\label{Subsection:gamma_one}

ITU-R~\cite{ITU2017} represents the minimum discernible power change at the satellite passive sensor as a function of the noise fluctuation $\Delta T_{N}$: 
\begin{eqnarray}
\Delta P&=& k\Delta T_{N}B  , \label{eq:deltaP}
\end{eqnarray}
where $k$ is the Boltzmann constant and $B$ is the reference bandwith of $200$~MHz. For the purpose of protecting satellite passive sensing, ITU-R defines the maximum tolerable interference threshold in the frequency band $23.6-24$~GHz as $20\%$ of $\Delta P$: 
\begin{eqnarray}
\gamma_1 &=& 0.2\Delta P = -136~\textrm{dBm}   . \label{eq:gamma_1}
\end{eqnarray}

\subsubsection{Radiometer Sensitivity Thresholds $\gamma_2$, $\gamma_3$, and $\gamma_4$}
\label{Subsection:gamma_two}

In order to better model the impact of the interference on the resolution of the satellite radiometer, we propose a set of alternative interference thresholds, corresponding to different candidate maximum interference levels that would not significantly degrade the sensitivity of the AMSU-A sensor. 
We thus take a first step towards directly mapping the interference from cellular ground deployments to degradation in weather prediction. 
In general, more sophisticated metrics can be defined to map more explicitly the interference to weather prediction degradation; however, defining such metrics is non-trivial due to the complexity of the weather prediction algorithms implemented in the software used by weather scientists.  
Thus,  our considered set of radiometer sensitivity thresholds is a proxy for different capabilities of the software algorithm to tolerate interference and predict weather with marginal degradation.     

To this end we consider the value of the noise equivalent delta temperature $NE\Delta T$=0.3~K, as specified by NOAA for AMSU-A sensors in the 23.8~GHz band~\cite{NOAA2014}. 
Since it is not clear how much additional temperature degradation the AMSU-A sensors could tolerate due to interference, we then assume that this temperature degradation is a fraction $x$=\{0.01, 0.1, 1\}\% of $NE\Delta T$, resulting in $x\times NE\Delta T$=\{0.00003, 0.0003, 0.003\}~K.  We thus cover a range of small temperature degradation values due to interference, that could be tolerated depending on the sensor and measurement post processing capabilities.
The equivalent maximum tolerable interference thresholds are then estimated as
\begin{eqnarray}
\gamma_2 = 1\% \times (k\times B\times NE\Delta T) = -141~\text{dBm}, \\
\gamma_3 = 0.1\% \times (k\times B\times NE\Delta T) = -151~\text{dBm}, \\
\gamma_4 = 0.01\% \times (k\times B\times NE\Delta T) = -161~\text{dBm}.
\end{eqnarray}

Finally, we note that the ITU-R interference threshold $\gamma_1$=--136~dBm is equivalent to a temperature degradation of $x\times NE\Delta T$=0.0091~K, where $x$=3.03\%.  Thus, we select sensitivity thresholds $\gamma_2$, $\gamma_3$, and $\gamma_4$ that are lower than the ITU-R threshold and thus relevant to study in practice, since interference levels exceeding the ITU-R threshold would be considered harmful in any case.

\subsubsection{Likelihood to Exceed Interference Thresholds}
\label{Subsection:harmful_likelihood}
The \mbox{ITU-R} specifies that the interference threshold may be exceeded without harming the satellite in $0.01\%$ of the area or time~\cite{ITU2017}. In line with this, we study whether the interference is deemed harmful by estimating the percentage of time for which a given interference threshold is exceeded. For this purpose, we consider the distribution of the estimated interference levels over all satellite positions and scanning orientations and over all orientations of the ground transmitter antenna for a given interference scenario (\emph{cf.}~Section~\ref{section:inter_scenario}) to be equivalent to the interference distribution over the total amount of scanning time. Consequently, we present our results in Section~\ref{section:Results} as CCDFs (complementary cumulative distribution functions) of the estimated interference, from which we read off the likelihood to exceed the interference thresholds $\gamma_1$, $\gamma_2$, $\gamma_3$, and $\gamma_4$ as the percentage of time. We also note that this representation allows us to interpret the results as a \emph{risk assessment chart}, which has been proposed as a useful tool for analysing spectrum coexistence in terms of the likelihood that a hazard occurs versus a consequence metric~\cite{Vries2017}.  We can thus meaningfully quantify the impact of a range of software algorithm capabilities on weather prediction when the satellite is coexisting with a 5G mm-wave network: the interference thresholds serve as a proxy for these capabilities, the consequence metric is the interference, and the hazard (i.e.  weather prediction degradation) occurs when the interference exceeds a given candidate threshold.

\section{Results}
\label{section:Results}

This section presents and discusses the results of our coexistence study.
Section~\ref{section:single_inter} presents the results for the \emph{single uplink} and \emph{downlink interferer scenarios}, Section~\ref{section:network_inter} focuses on the \emph{network of uplink} and \emph{downlink interferers scenarios}, and Section~\ref{section:atm_att_eff} discusses the impact of atmospheric attenuation on the interference level.

\subsection{Results for the Single Uplink and Downlink Interferer Scenarios}
\label{section:single_inter}

\begin{figure}[t!]
\centering
\captionsetup[subfigure]{width=0.48\linewidth}
          {\includegraphics[width=0.6\linewidth]{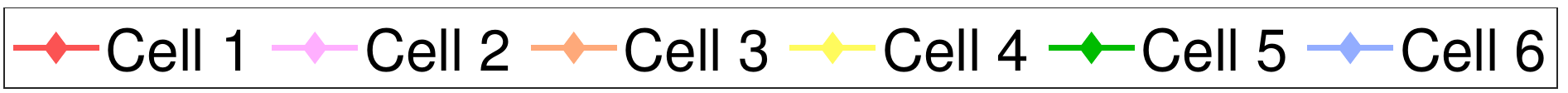}}
\subfloat
          {\includegraphics[width=1\linewidth]{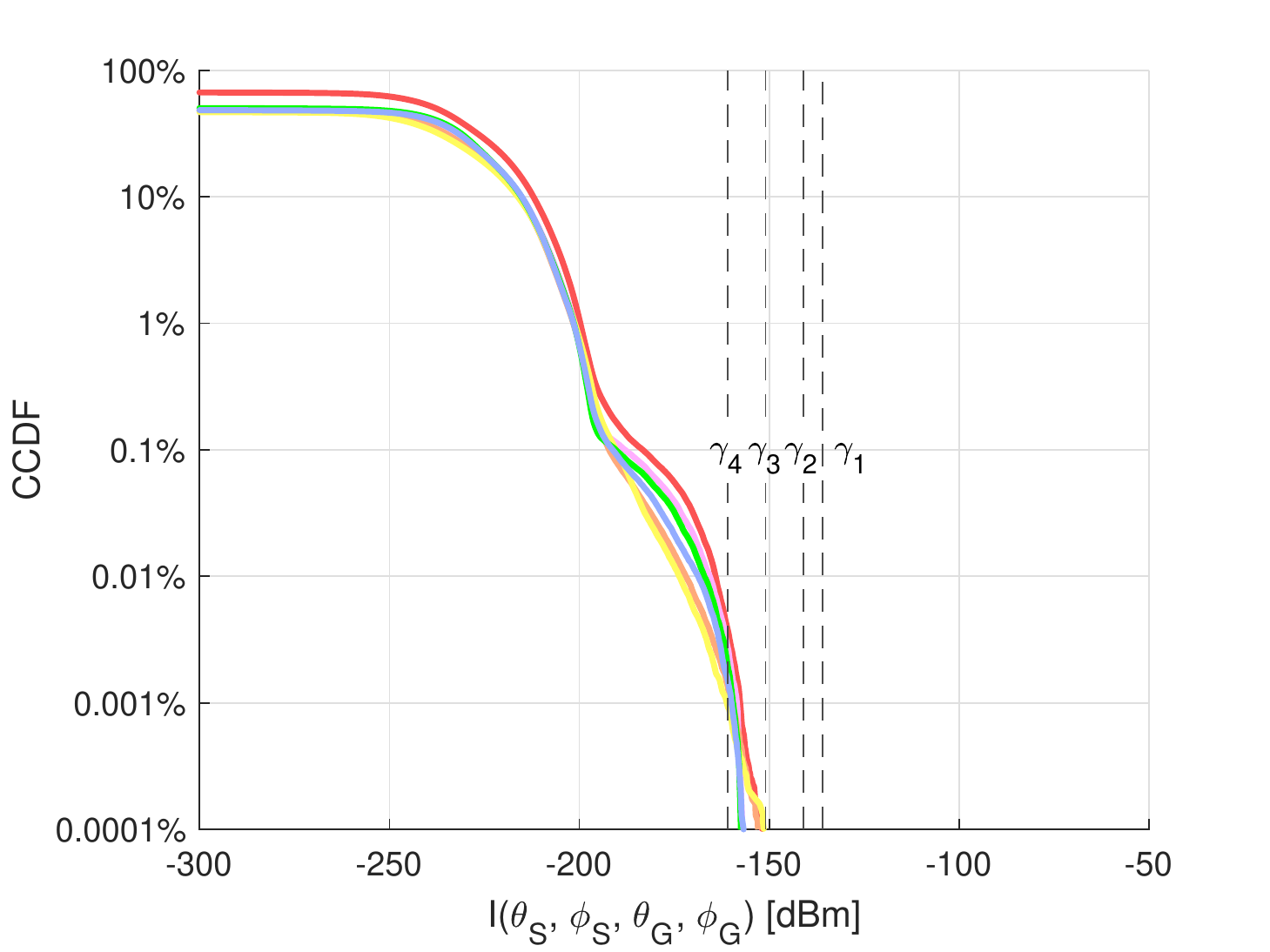} \label{res:uplink_single_interferer_23}}
\caption{Distribution of the interference $I(\theta_S, \phi_S, \theta_G, \phi_G)$, for the \emph{single uplink interferer scenario},  over all 100 UEs considered in each Cell~1--6 and all satellite positions and satellite antenna orientations,  for $p$=50\%.}
\label{res:uplink_single_interferer}
\end{figure}

\begin{figure}[t!]
\centering
\captionsetup[subfigure]{width=0.48\linewidth}
{\includegraphics[width=0.6\linewidth]{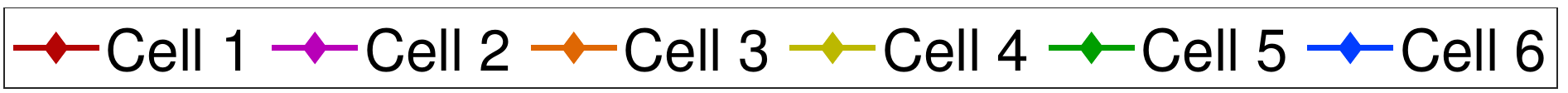}}
\subfloat
          {\includegraphics[width=1\linewidth]{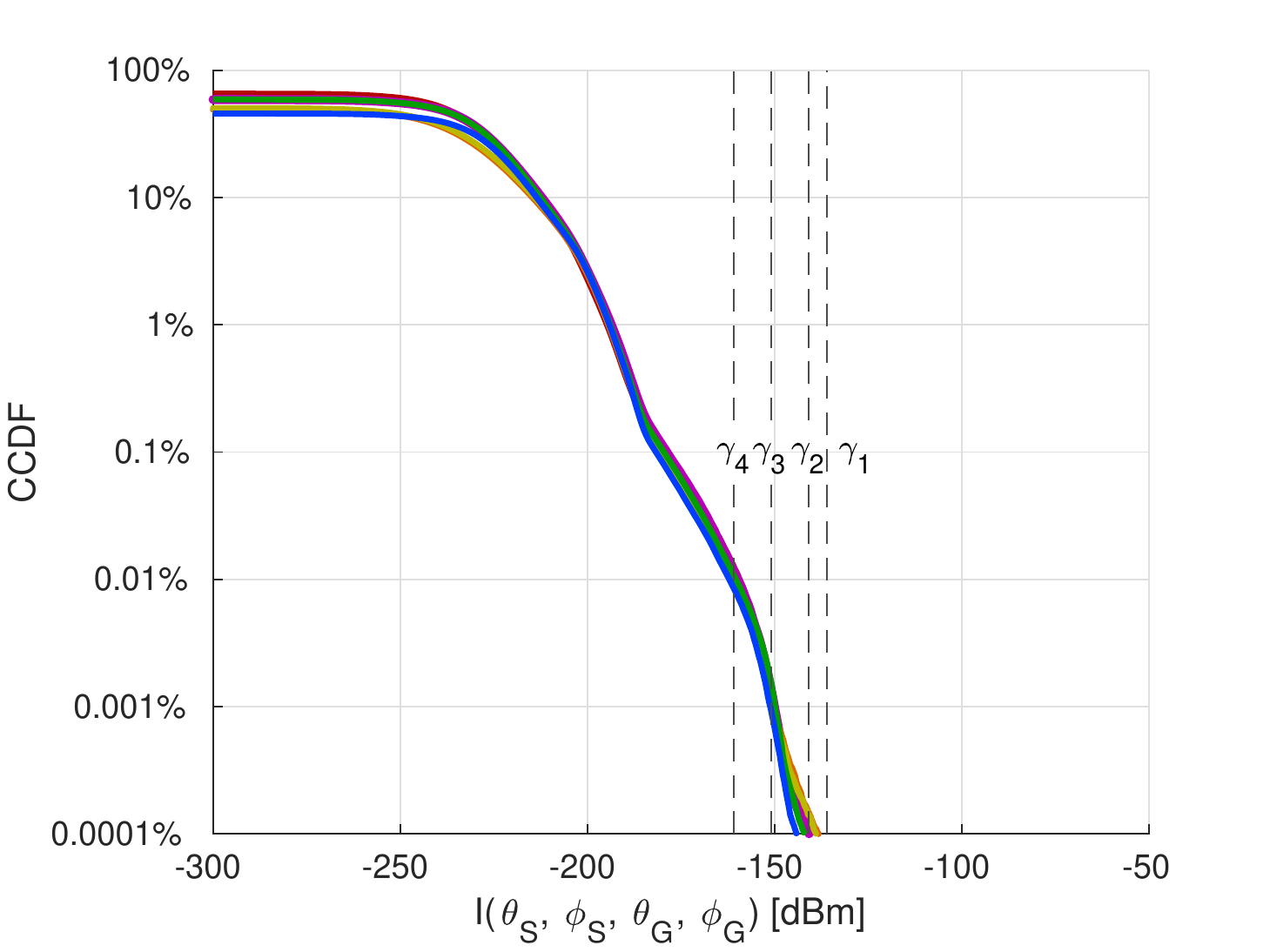} \label{res:downlink_single_interferer_23}}
\caption{Distribution of the interference $I(\theta_S, \phi_S, \theta_G, \phi_G)$,  for the \emph{single downlink interferer scenario}, over all possible BS orientations in each Cell~1--6 and all satellite positions and satellite antenna orientations, for $p$=50\%.}
\label{res:downlink_single_interferer}
\end{figure}

Figs.~\ref{res:uplink_single_interferer} and \ref{res:downlink_single_interferer} show the distributions of the interference at the satellite, for the \emph{single uplink} and \emph{downlink interferer scenarios}, respectively, over all satellite positions and satellite antenna orientations and all considered UE and BS orientations, for an unavailability probability $p$=50\%. 
We note that we selected $p$=50\% since this value corresponds to the lowest atmospheric attenuation and thus the worst-case interference at the weather satellite,  resulting in the most favourable conditions for the interference to become harmful; results for other values of $p$ are presented subsequently in Section~\ref{section:atm_att_eff}.
We discuss first Fig.~\ref{res:uplink_single_interferer} in Section~\ref{section:single_inter_ul} and then Fig.~\ref{res:downlink_single_interferer} in Section~\ref{section:single_inter_dl}. 

\subsubsection{Single Uplink Interferer Scenario}
\label{section:single_inter_ul}

\begin{table}[t!]
\centering
\caption{Likelihood that $I(\theta_S, \phi_S, \theta_G, \phi_G)$ exceeds $\gamma_1$ and $\gamma_4$, for the single uplink and downlink interferer scenarios, as obtained from Figs.~\ref{res:uplink_single_interferer} and \ref{res:downlink_single_interferer}}
\begin{tabular}{|c|p{1.3cm}|p{1.3cm}|p{1.3cm}|p{1.3cm}|}
\hline
\textbf{Cell} & \multicolumn{2}{c|}{\textbf{Uplink}} & \multicolumn{2}{c|}{\textbf{Downlink}} \\
\cline{2-5}
 & Pr($I> \gamma_1$) & Pr($I> \gamma_4$) & Pr($I> \gamma_1$) & Pr($I> \gamma_4$) \\
\hline
 1 & 0\%  & 0.0039\% &  0.00006\% & 0.0126\%  \\
\hline
 2 &  0\% & 0.0027\%  &  0.00005\% & 0.0118\%  \\
 \hline
 3 & 0\%  & 0.0014\%  & 0.00007\%  & 0.0101\% \\
 \hline
 4 &  0\% &  0.0009\% &  0.00007\% & 0.0089\%  \\
 \hline
 5 &  0\% & 0.0020\%  &  0.00004\% & 0.0108\%  \\
 \hline
 6 &  0\% & 0.0015\%  & 0.00003\%  & 0.0086\%  \\
\hline
\end{tabular}
\label{tab:likelihood_singleInterferer}
\end{table}

Let us consider the results in Fig.~\ref{res:uplink_single_interferer}. 
The estimated interference varies between \mbox{--300}~dBm and \mbox{--146}~dBm with a median of at most \mbox{--236}~dBm for all Cells~1--6. 
Moreover, the interference does not exceed the ITU-R threshold $\gamma_1$=\mbox{--136}~dBm for any of the cells, as also summarized in Table~\ref{tab:likelihood_singleInterferer}. This shows that the 3GPP leaked power limit imposed per interfering UE is suitable to protect the weather satellite according to the ITU-R criteria in the \emph{single uplink interferer scenario}. 
Similarly, the harmful interference thresholds $\gamma_2$=\mbox{--141}~dBm, $\gamma_3$=\mbox{--151}~dBm, and $\gamma_4$=\mbox{--161}~dBm are not exceeded with likelihoods higher than 0.01\% for any of the cells (\emph{cf.} Table~\ref{tab:likelihood_singleInterferer}).  
This shows that interference from a \emph{single} UE in the uplink is not considered as harmful, regardless of whether the satellite sensitivity or the ITU-R noise fluctuation threshold are adopted as the criterion.

Let us now explore further insights from Fig.~\ref{res:uplink_single_interferer}, in order to identify which types of cells are the strongest interferers. This is useful for evaluating the spatial structure of interference and the impact of different urban propagation environments on the interference level experienced at the weather satellite.  As we will show in Section~\ref{section:network_inter}, these aspects have important consequences for the \emph{aggregate} interference from multiple simultaneous transmissions in the \emph{network of interferers scenarios}, where the harmful interference thresholds can be exceeded. 
We observe that UEs in Cell~1 cause overall higher interference levels (e.g. highest median value) than the other cells, despite having a similar interference distribution. This is due to the local propagation environment, where Cell~1 is placed at an intersection in Times Square, as shown in Fig.~\ref{uplink_ground}. The buildings in this area are much higher than in the areas where the other cells are located, so the transmitted rays undergo more reflections off building walls and more of them are thus able to reach the satellite. 
Nonetheless, the highest interference levels (i.e.  tail of the distributions in Fig.~\ref{res:uplink_single_interferer}) are caused by UEs in Cells~3 and 4, which are placed on narrow streets, where the transmitted rays are reflected mostly by buildings located close to the ground transmitter.

\begin{figure}[tb!]
\centering
\captionsetup[subfigure]{width=0.48\linewidth}
{\includegraphics[scale=0.5]{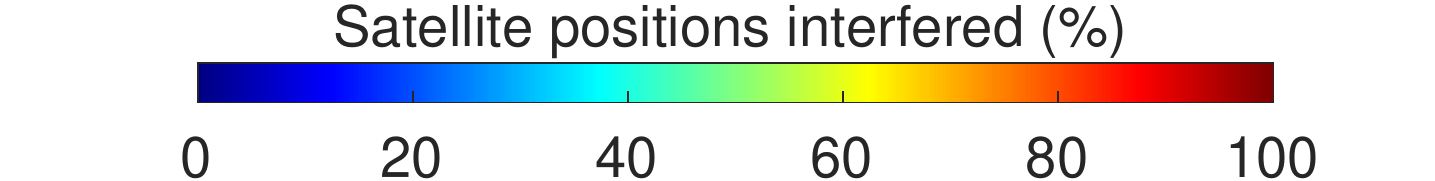}}
\subfloat[uplink]
          {\includegraphics[trim = {1.1cm 0.5cm 0.8cm 0.5cm}, width=0.5\linewidth]{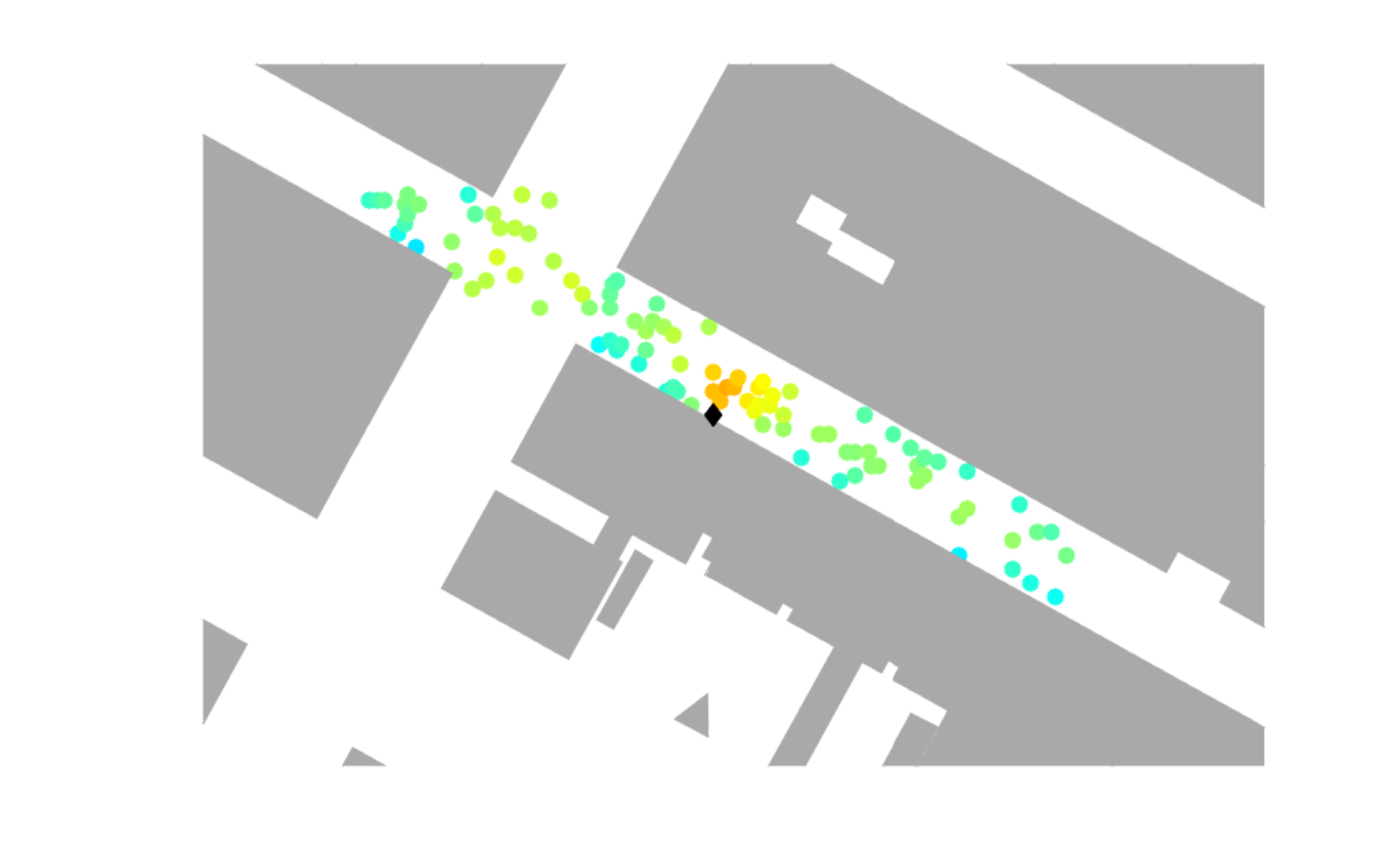} \label{res:uplink_ue3_heatmap}}
\subfloat[downlink]
          {\includegraphics[trim = {1.6cm 0.5cm 0.4cm 0.5cm}, width=0.5\linewidth]{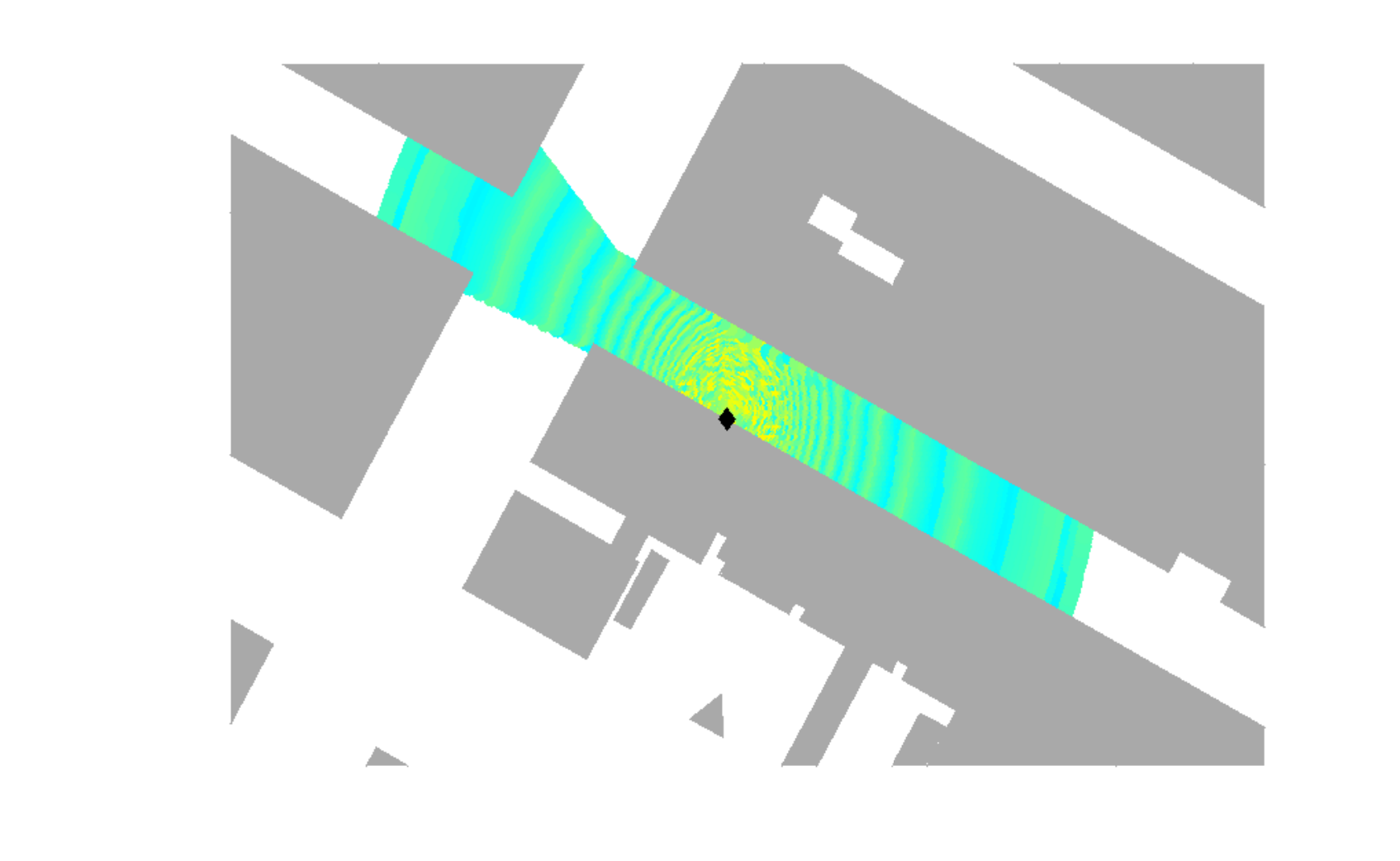} \label{res:downlink_ue3_heatmap}}
\caption{Heatmap of the percentage of satellite positions in the space study area interfered by (a)~each of the 100 served UEs in the uplink and (b)~each BS orientation in the downlink, for Cell~3 and $p$=50\%; BS position indicated by black diamond.}
\label{res:ues_heatmaps}
\end{figure}

\begin{figure}[tb!]
\centering
\captionsetup[subfigure]{width=0.48\linewidth}
\subfloat[uplink]
          {\includegraphics[width=0.5\linewidth]{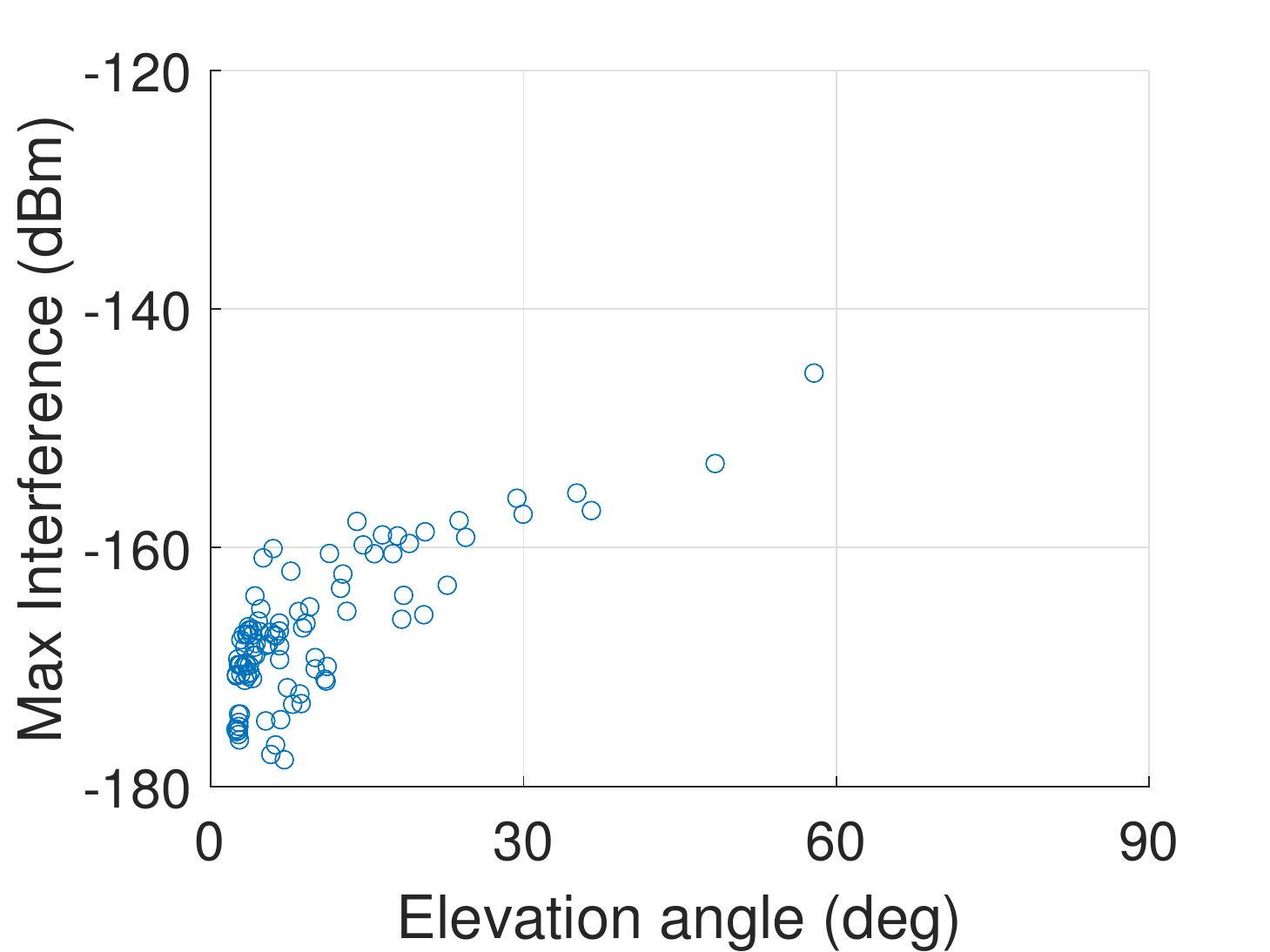} \label{res:uplink_elevation}}
\subfloat[downlink]
          {\includegraphics[width=0.5\linewidth]{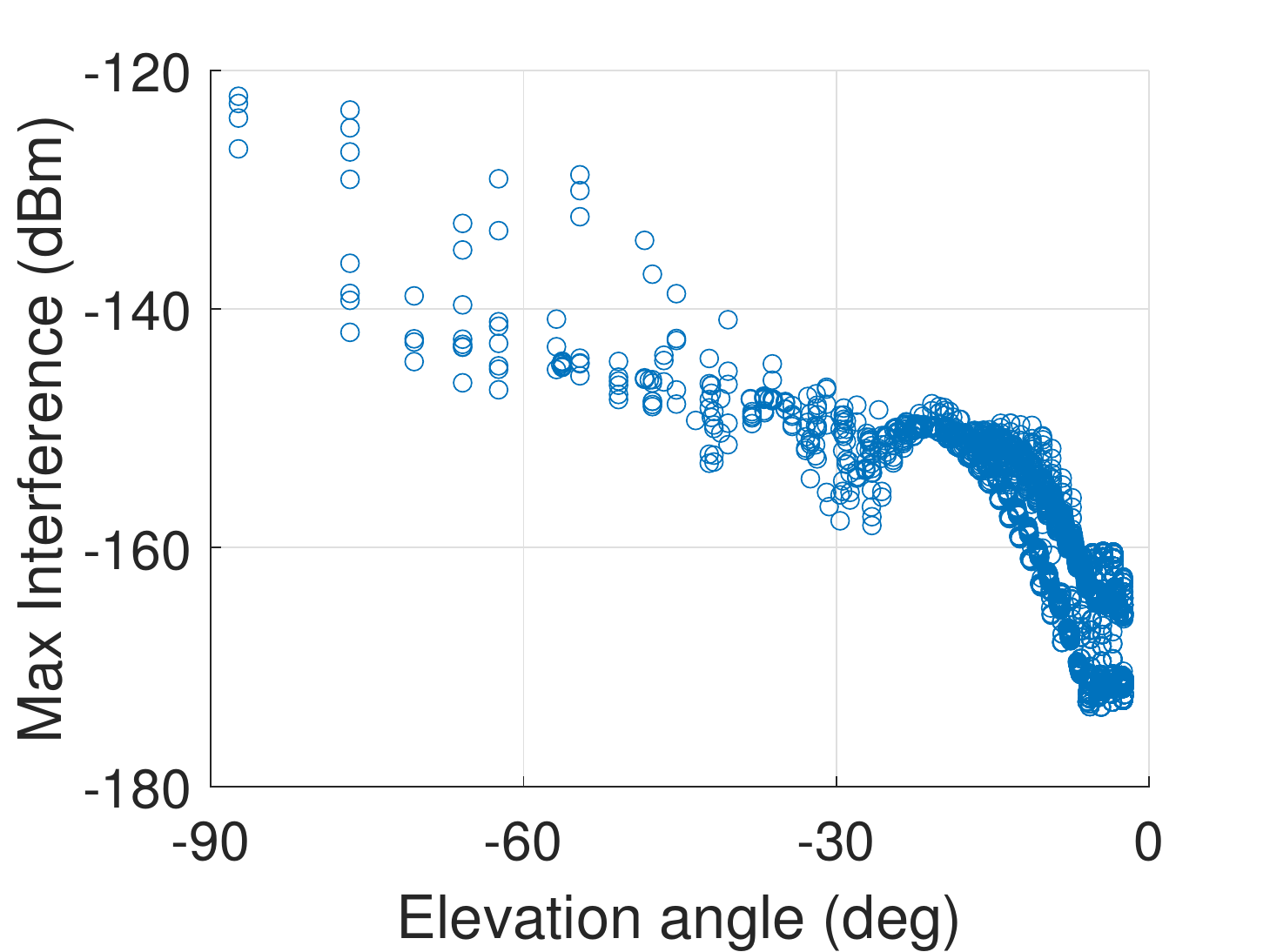} \label{res:downlink_elevation}}
\caption{Maximum interference $I(\theta_S, \phi_S, \theta_G, \phi_G)$ at each interfered satellite position (over all satellite scan angles $\{\theta_S, \phi_S\}$) versus the corresponding elevation angle of the interfering (a)~UE in the uplink ($\phi_G=\phi_{UE}$) and (b)~BS in the downlink ($\phi_G=\phi_{BS}$), for Cell~3 and $p$=50\%.}
\label{res:ues_elevation_dist}
\end{figure}

We next study the spatial structure of the interference in terms of which of the considered 100~UEs within a cell cause most interference at the satellite. To this end, we focus on the results for the example Cell~3, since its interference distribution is overall representative of most cell types and it also comprises UEs that cause the maximum interference levels.  Fig.~\ref{res:uplink_ue3_heatmap} shows the percentage of satellite positions that receive at least one interfering ray from each of the UEs in this cell. 
These results highlight overall two groups of UEs that interfere at many satellite positions: (i)~those placed directly in front of the BS, due to their high antenna elevation angles $\phi_{UE}$, consistent with their proximity to the BS; and (ii)~those located on the street intersection, due to the many surrounding buildings, which cause many ray reflections and thus enable the interfering rays to reach more satellite positions.
Alongside identifying the UEs that interfere at many satellite positions as in Fig.~\ref{res:uplink_ue3_heatmap}, another relevant aspect is identifying the UEs that cause the highest \emph{levels} of interference at each given position, thus being representative of the distribution tails in Fig.~\ref{res:uplink_single_interferer} where interference may be deemed as harmful.
To this end, Fig.~\ref{res:uplink_elevation} shows the maximum interference at each satellite position versus the antenna elevation angle $\phi_{UE}$ of the respective UE that causes this maximum interference.
Importantly, the high levels of interference are caused by the UEs with a high $\phi_{UE}$, which shows that such UEs not only interfere at many satellite positions (\emph{cf.} Fig.~\ref{res:uplink_ue3_heatmap}),  but are also the strongest interferers.

\begin{figure*}[b!]
\centering
\captionsetup[subfigure]{width=0.55\linewidth}

\subfloat[uplink, one example UE]
          {\includegraphics[width=0.25\linewidth]{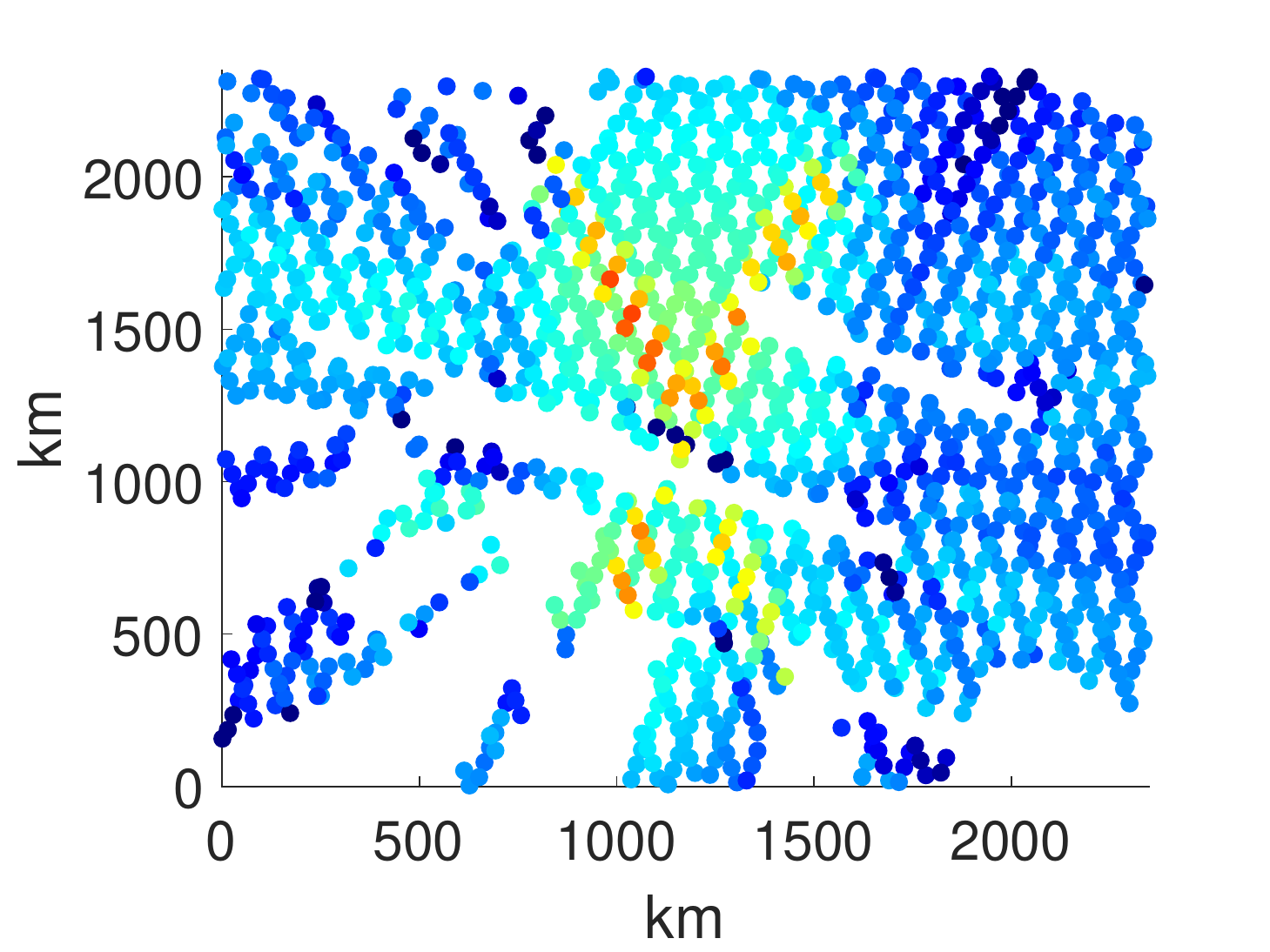} \label{res:uplink_sat_heatmap_1ue}}          
\subfloat[uplink, 100 UEs]
          {\includegraphics[width=0.25\linewidth]{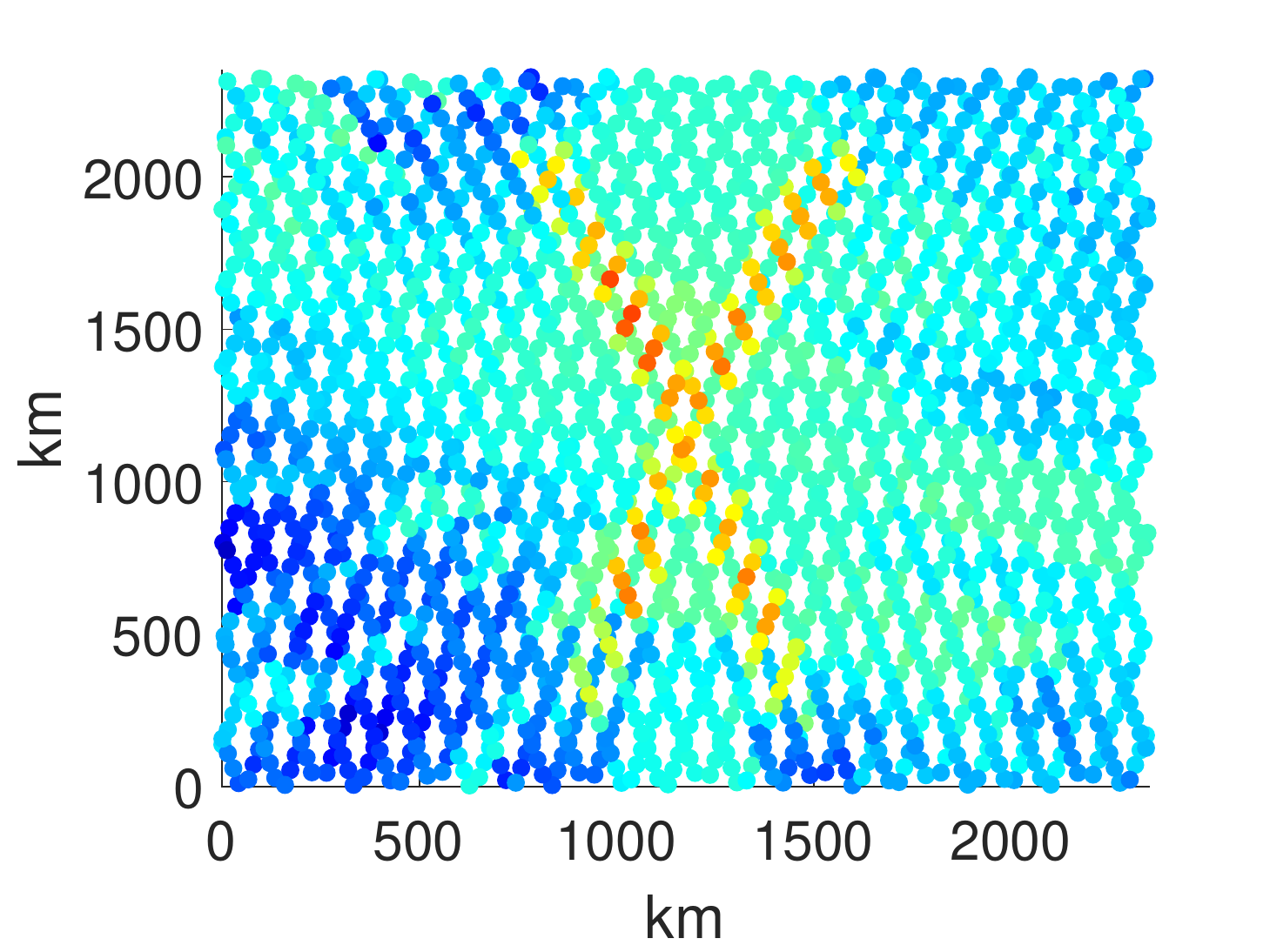} \label{res:uplink_sat_heatmap_100ue}}
\subfloat[downlink, serving any LOS UE]
          {\includegraphics[width=0.25\linewidth]{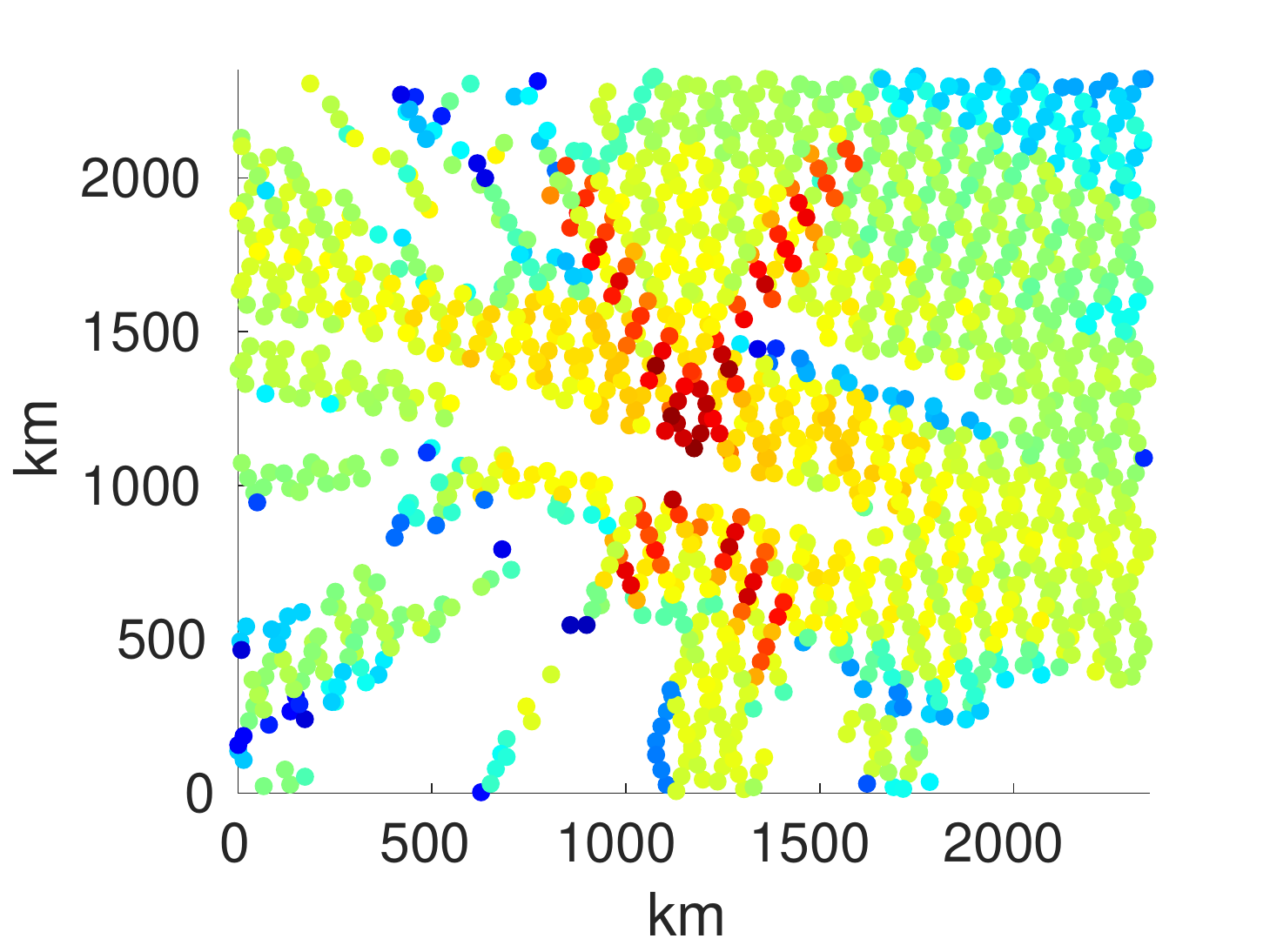} \label{res:downlink_sat_heatmap}}
\hspace{0.5cm}{\includegraphics[angle=90,scale=0.25]{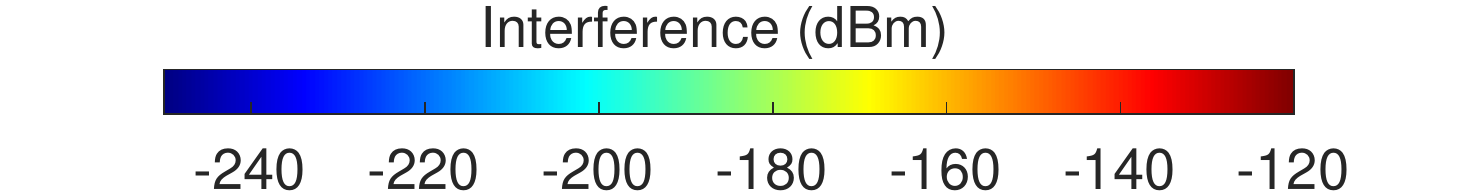}}
\caption{Heatmap of the maximum interference $I(\theta_S, \phi_S, \theta_G, \phi_G)$ at each satellite position (over all satellite scan angles $\{\theta_S, \phi_S\}$),  when the ground interferer is (a)~one example UE in the uplink,  (b)~any one of the 100~UEs in the uplink, and (c) the BS in the downlink over all its possible orientations, for Cell~3 and $p$=50\%.}
\label{res:sat_heatmaps}
\end{figure*}

\begin{figure*}[tb!]
\centering
\captionsetup[subfigure]{width=0.55\linewidth}
\subfloat[without antenna patterns]
          {\includegraphics[width=0.25\linewidth]{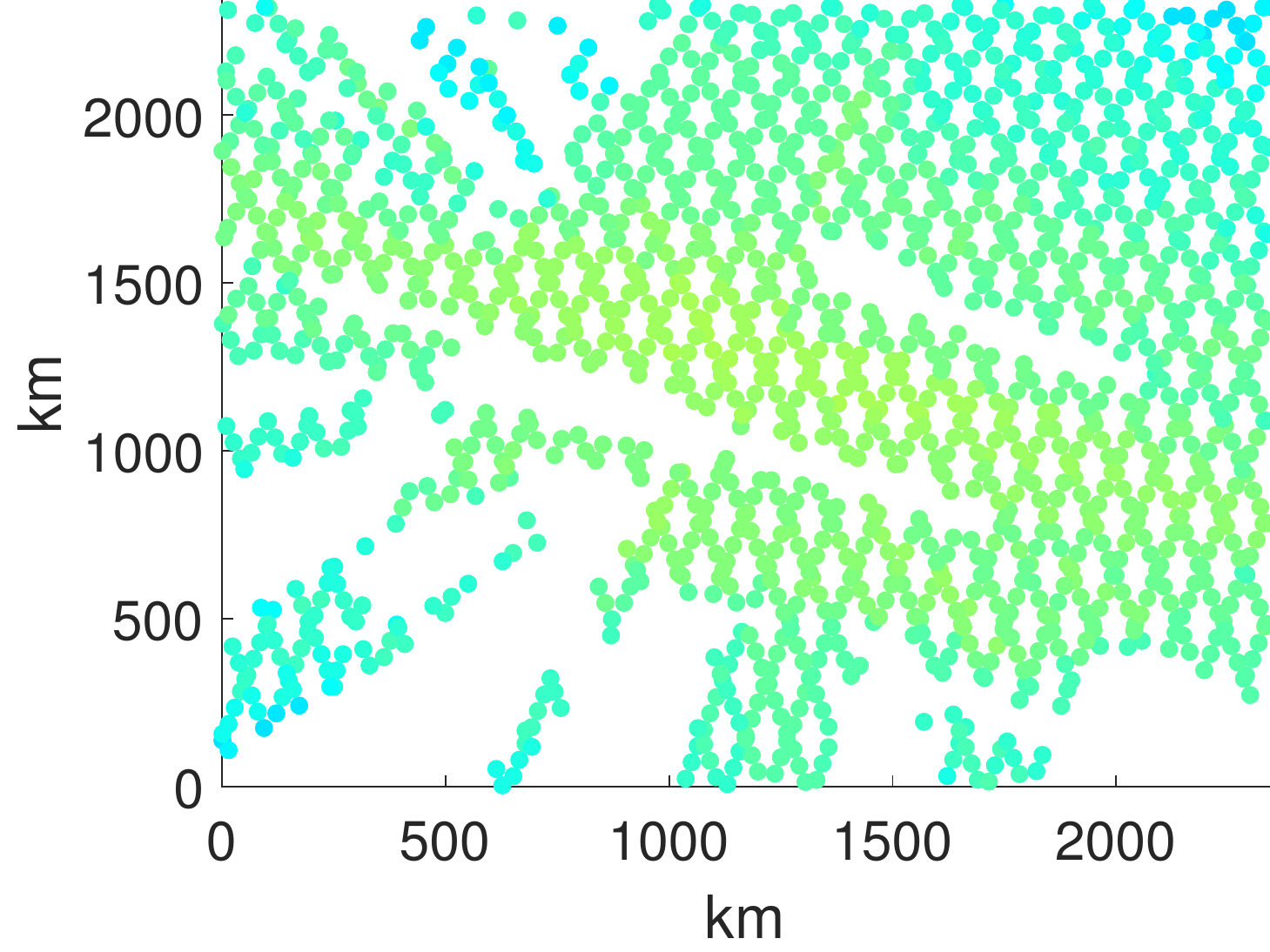} \label{res:scatter_heatmap_no_patterns}}         
\subfloat[with UE antenna pattern]
          {\includegraphics[width=0.25\linewidth]{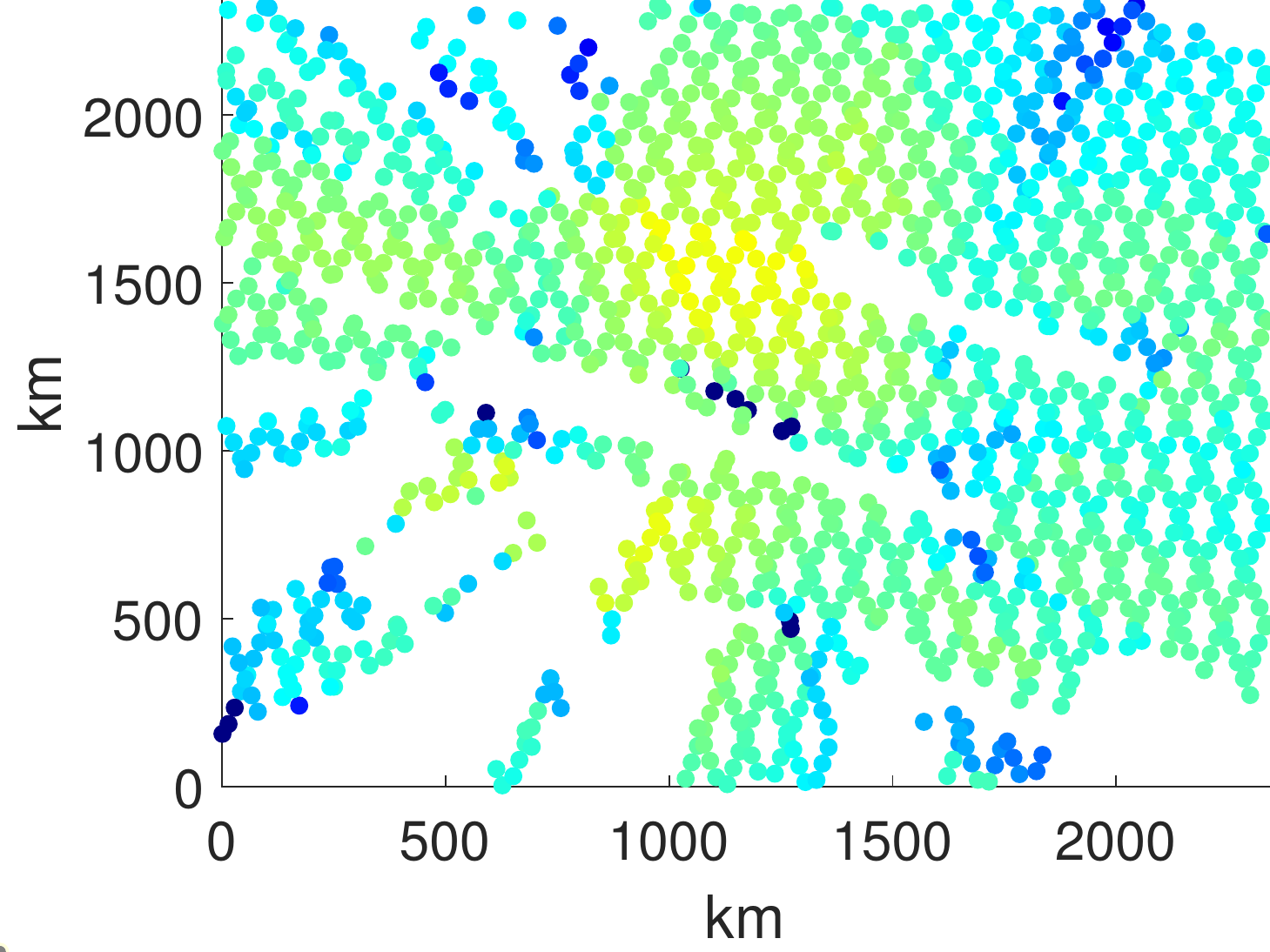} \label{res:scatter_heatmap_just_tx}}
\subfloat[with UE and satellite antenna patterns]
          {\includegraphics[width=0.25\linewidth]{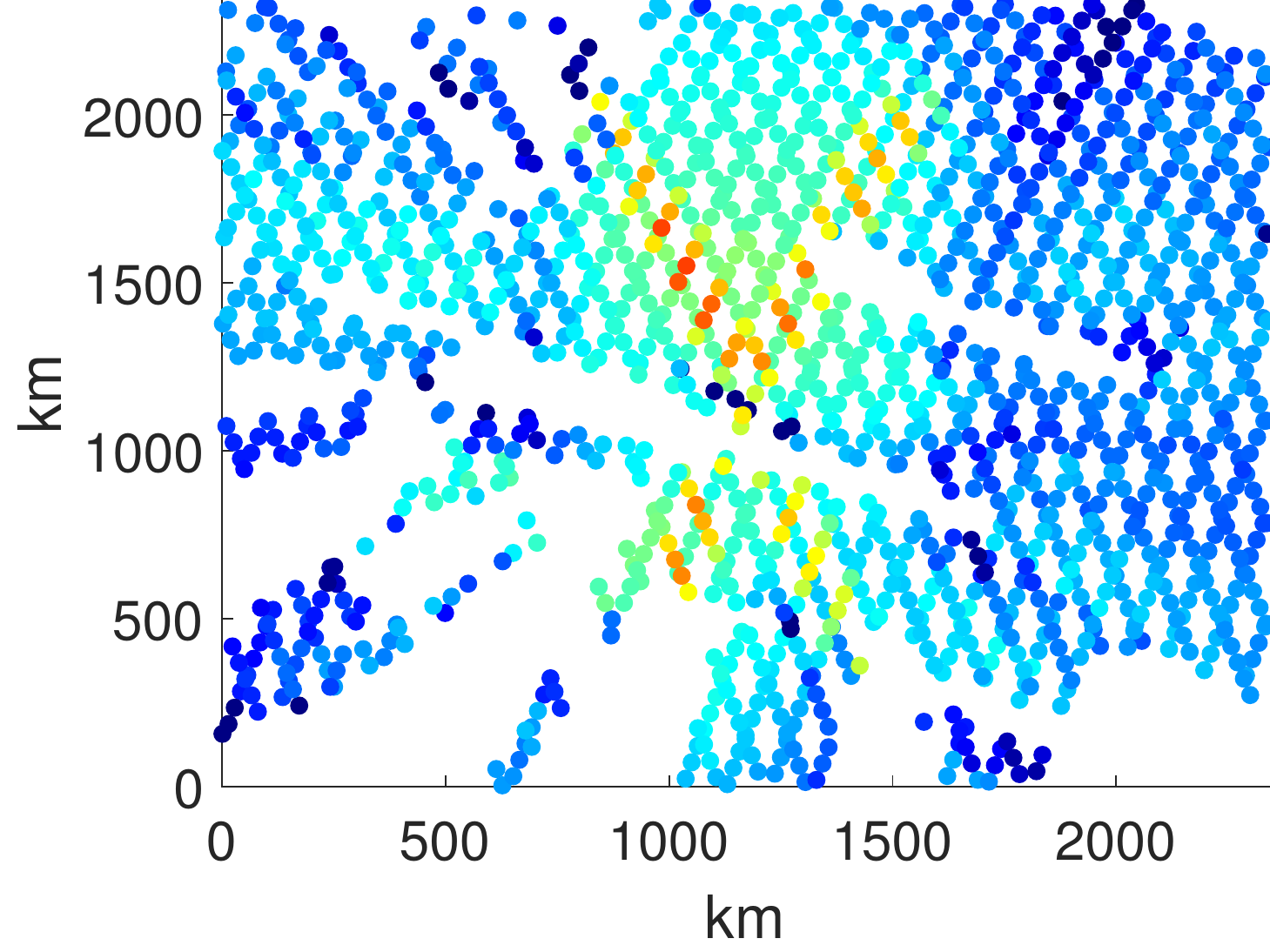} \label{res:scatter_heatmap_tx_rx}}
\hspace{0.5cm}{\includegraphics[angle=90,scale=0.25]{colorbar_powers_OLD.pdf}}
\caption{Heatmap of the maximum interference $I(\theta_S, \phi_S, \theta_G, \phi_G)$ at each satellite position (over all satellite scan angles $\{\theta_S, \phi_S\}$) caused by the single example Cell~3 UE as in Fig.~\ref{res:uplink_sat_heatmap_1ue}, as calculated (a)~without applying any directional antenna gains, (b)~with the UE antenna pattern, and (c)~with both UE and satellite antenna patterns.}
\label{res:scatter_heatmaps}
\end{figure*}

\begin{figure}
\begin{center}
  \includegraphics[width=0.6\linewidth]{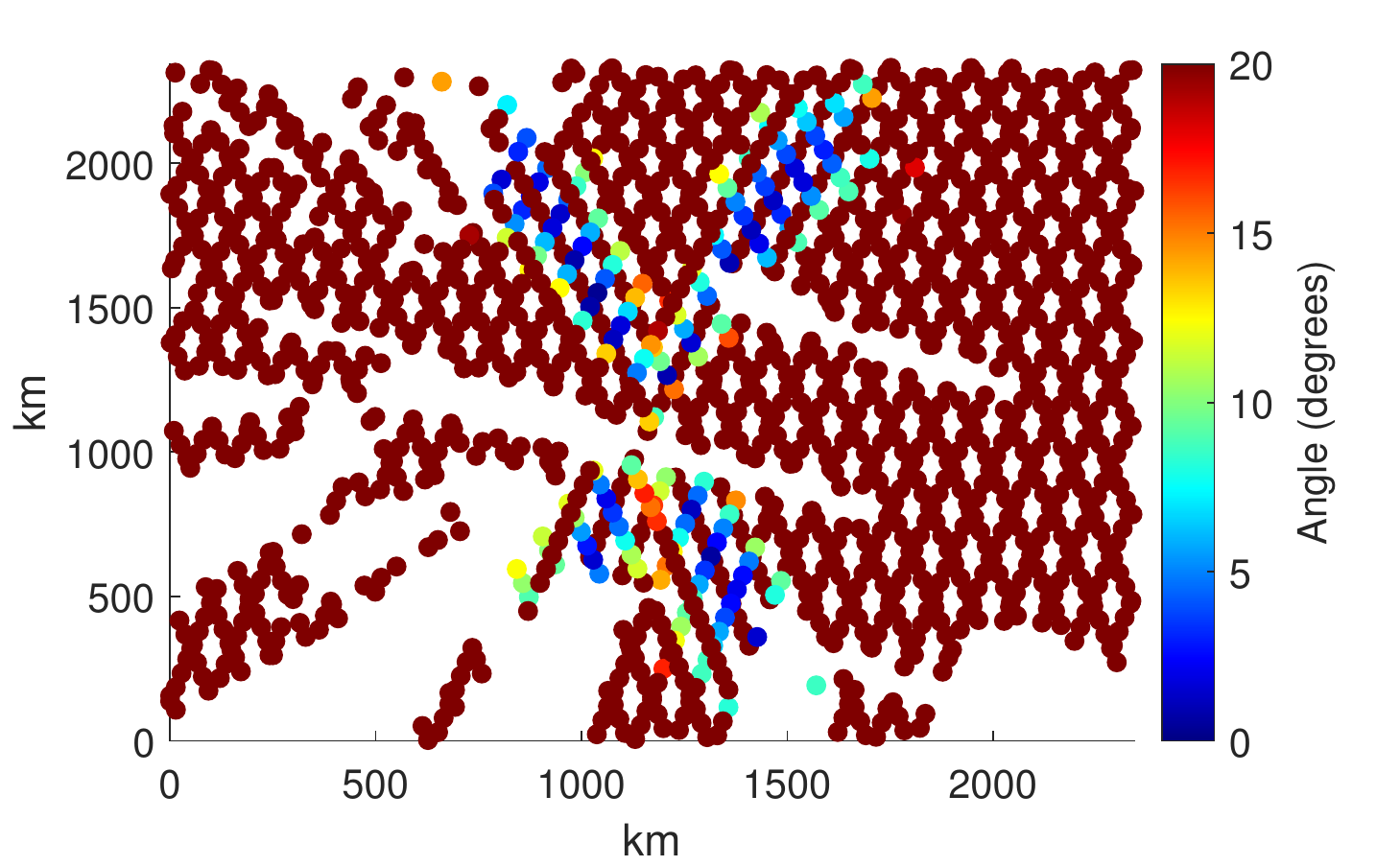} 
 \caption{Angular misalignment between the satellite orientation and the orientation of the strongest interfering ray at each satellite position, for the single example Cell~3 UE as in Figs.~\ref{res:uplink_sat_heatmap_1ue}~and~\ref{res:scatter_heatmaps}.}
  \label{fig:angle_sat_tx}
\end{center}
\end{figure}

Fig.~\ref{res:sat_heatmaps} shows results complementary to those in Fig.~\ref{res:ues_elevation_dist}, namely the heatmaps of the maximum received interference at each satellite position, out of all combinations of satellite antenna orientations and interfering UEs from Cell~3. 
The interference caused by a single example UE in Fig.~\ref{res:uplink_sat_heatmap_1ue} does not reach all satellite positions.
However, if the 100~served UE locations are considered, all satellite positions receive interference from at least one of these UEs, as shown in  Fig.~\ref{res:uplink_sat_heatmap_100ue}. As such, in the realistic case of multiple UEs per cell, although interference from individual UEs may not be strong enough to exceed the interference thresholds at all satellite positions, each of these positions receives some interference. This is important especially in view of multiple \emph{simultaneous} transmissions in the \emph{network of interferers scenarios}, since each satellite position is potentially subject to harmful aggregate interference, as will be discussed in detail subsequently in Section~\ref{section:network_inter}.  

Finally, we observe that the maximum received interference in Fig.~\ref{res:sat_heatmaps} has an X-shaped pattern over the space study area. This is due to the geometry of the 3D coexistence scenario and the combination of the UE and satellite directional antenna patterns, as illustrated in Fig.~\ref{res:scatter_heatmaps}. The heatmap of the maximum received interference at each satellite position without applying any antenna pattern in Fig.~\ref{res:scatter_heatmap_no_patterns} shows a higher interference in the center of the space area, as expected due to being closer to the ground study area; the interference is otherwise rather uniform. After applying the UE antenna pattern in Fig.~\ref{res:scatter_heatmap_just_tx} and  
especially after applying also the satellite antenna pattern (for all possible orientations) in Fig.~\ref{res:scatter_heatmap_tx_rx}, the highest interference is increasingly concentrated towards the center,  forming an X shape. 
To explain this effect in more detail, Fig.~\ref{fig:angle_sat_tx} shows the angular misalignment between the scanning direction of the satellite and the direction of the incoming ray causing the maximum interference at each given position.
Consistent with the interference pattern, the small misalignment values also form an X shape above the ground study area. This shows that, as expected, the strongest interference is caused when the satellite is scanning in a direction close to that of a strong interfering ray. Furthermore, this alignment occurs when the satellite crosses (or passes in the proximity of) the ground study area, due to its scanning direction being perpendicular to the path direction. 

The results in this section show overall that uplink out-of-band interference from a \emph{single} UE and within 3GPP leaked power limits is not harmful from the ITU-R perspective, or the satellite sensitivity perspective with interference thresholds down to $\gamma_4$=--161~dBm. Nonetheless, aggregate network interference from simultaneous uplink ground transmissions, as typically expected in practice, could become harmful; we discuss this subsequently in Section~\ref{section:network_inter}. We also showed that the UEs causing the highest interference levels and also affecting many satellite positions are those with a high antenna elevation. 

\subsubsection{Single Downlink Interferer Scenario}
\label{section:single_inter_dl}

Let us consider the interference distribution in Fig.~\ref{res:downlink_single_interferer}. The interference caused by a single BS in the downlink varies overall between \mbox{--300}~dBm and \mbox{--120}~dBm and is thus higher than that for the \emph{single uplink interferer scenario} in Fig.~\ref{res:uplink_single_interferer}. This is expected, due to the higher leaked EIRP of the BS compared with that of a UE, as specified in Section~\ref{section:sysmodel}.
Importantly, unlike for the uplink, all harmful interference thresholds are exceeded for the downlink, although the likelihood is higher than 0.01\% only for $\gamma_4$=\mbox{--161}~dBm and not for the ITU-R threshold $\gamma_1$=\mbox{--136}~dBm, or the other thresholds (\emph{cf.} Table~\ref{tab:likelihood_singleInterferer}).  
This shows that downlink interference from a single BS can be harmful for the weather satellite sensitivity, if the weather prediction software has poor capabilities to process and eliminate interference. Nonetheless,  interference would not be deemed as harmful for weather prediction software with better processing algorithms (i.e. tolerating interference between \mbox{--151} and \mbox{--136}~dBm), or the \mbox{ITU-R} criteria.

As an insight, the interference caused in the downlink by Cells~2 and~5 is very similar and overall higher than that from other cells. This is due to these two cells being placed in open areas. Thus, for BS transmissions, where the transmit antenna is oriented downwards, the ground reflects most of the rays, allowing more reflections and, in turn, causing more interference to the weather satellite. 
However,  unlike Cells~2 and~5, Cell~6 causes overall the lowest interference,  despite also being placed in an open area. This effect is due to the low-height buildings facing the BS in Cell~6, so that even though many rays are reflected off the ground, there are few subsequent building reflections, so that not many interference rays reach the satellite.

Consistent with our analysis for the \emph{single uplink interferer scenario}  in Section~\ref{section:single_inter_ul}, let us study the spatial structure of the downlink interference. Fig.~\ref{res:downlink_ue3_heatmap} shows the percentage of satellite positions interfered by the BS when its antenna is oriented to serve each LOS UE position in Cell~3. The largest numbers of interfered satellite positions are observed when the BS is oriented towards closer UEs and thus its antenna has a lower elevation angle $\phi_{BS}$, resulting in more ground reflections. 
Furthermore,  Fig.~\ref{res:downlink_elevation} shows that lower $\phi_{BS}$ values not only cause interference at many satellite positions, but also cause the strongest maximum interference at different positions.
 
We complement these results with Fig.~\ref{res:downlink_sat_heatmap}, which shows the heatmap of the maximum received interference at each satellite position as caused by the BS in Cell~3, over all satellite and BS antenna orientations, for $p$=50\%. 
These results are thus representative of the tail of the distribution in Fig.~\ref{res:downlink_single_interferer}, where interference may exceed the harmful thresholds. 
We observe that the satellite positions receiving interfering rays are overall the same as for the single example UE in the \emph{single uplink interferer scenario} in Fig.~\ref{res:uplink_sat_heatmap_1ue},  where some of the satellite positions are not interfered with at all. However, those positions that are interfered with receive much stronger interference from downlink transmissions than uplink ones.
Comparing the results in Figs.~\ref{res:downlink_sat_heatmap}~and~\ref{res:uplink_sat_heatmap_100ue} thus emphasizes that the interference caused by a given cell with typically multiple UEs is different for the downlink versus the uplink, both quantitatively and qualitatively. Whereas uplink interference can reach all satellite positions due to the variability in the location of the ground interfer UE over the cell, downlink interference reaches only some satellite positions corresponding with the fixed BS location in the cell, but with stronger rays. 

The results in this section show overall that out-of-band downlink interference from a single BS is stronger than uplink interference from a single UE and can thus also harm the satellite sensitivity, but only if the weather prediction software cannot tolerate interference above a low threshold of $\gamma_4$=\mbox{--161}~dBm.  The interference in this scenario is not strong enough to be identified as harmful by the ITU-R criteria.  Furthermore, BSs with a low antenna elevation cause the highest interference levels and also affect many satellite positions, due to the ground reflecting many interfering rays.

\subsection{Results for the Network of Uplink and Downlink Interferers Scenarios}
\label{section:network_inter}

\begin{figure}
\begin{center}
  \includegraphics[width=1\linewidth]{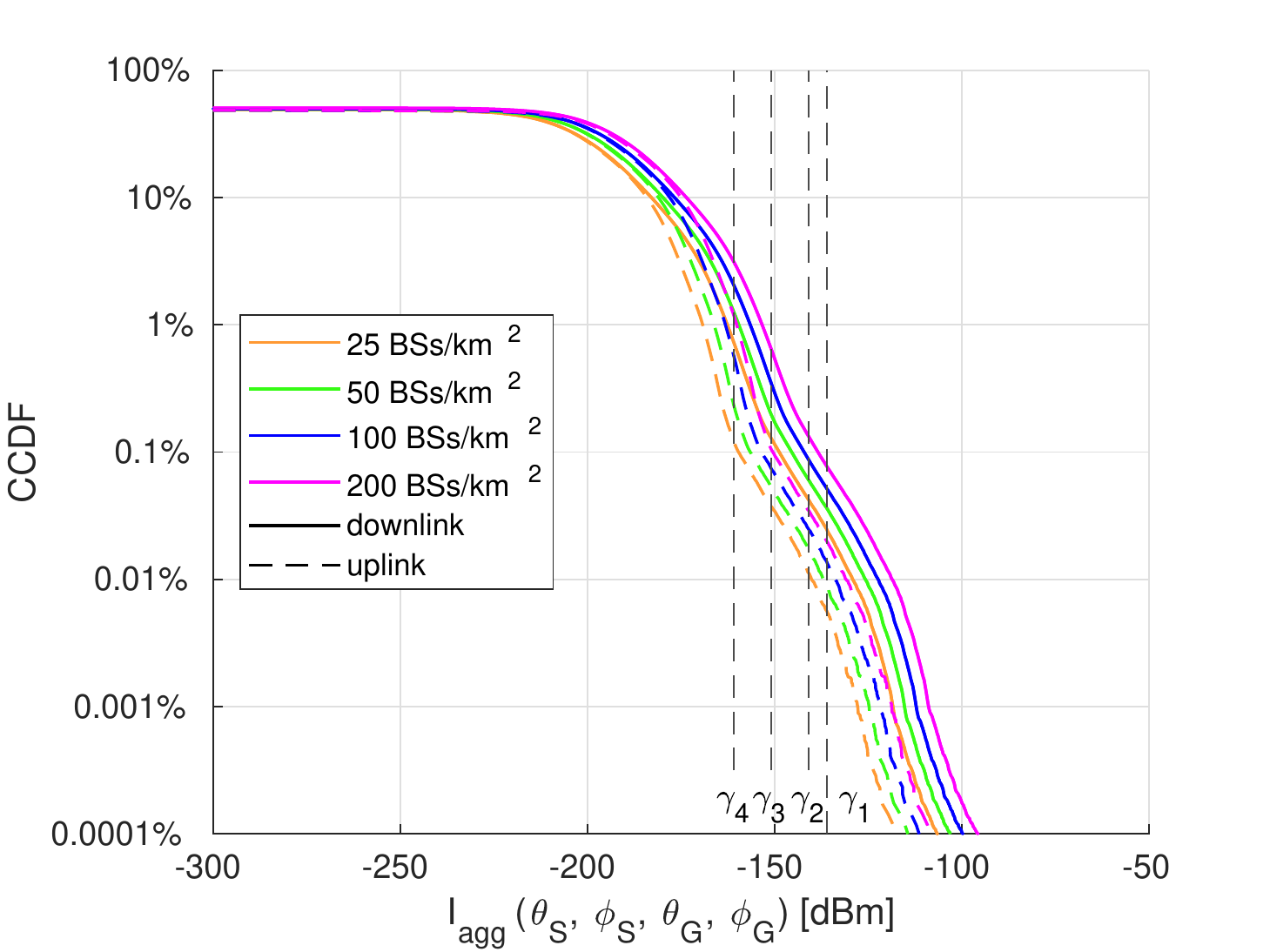} 
  \caption{Distribution of the aggregate interference $I_{agg}(\theta_S, \phi_S, \theta_G, \phi_G)$ for the \emph{network of uplink} and \emph{downlink interferers scenarios} over all possible ground transmitter orientations and all satellite positions and scan angles, for different network densities, $p$=50\%,  and assuming that all $N$ cells cause the same interference as Cell~3.}
  
  \label{fig:cdf_ue3_network}
\end{center}
\end{figure}

In order to study the aggregate interference $I_{agg}(\theta_S, \phi_S, \theta_G, \phi_G)$ caused by multiple cells operating at the same time in the \emph{network of uplink} and \emph{downlink interferers scenarios}, we consider $N$ homogeneous cells as detailed in Section~\ref{section:inter_scenario}. Let us first consider the case where the $N$ cells each cause the same interference levels as the example Cell~3. 
Fig.~\ref{fig:cdf_ue3_network} shows the distribution of $I_{agg}(\theta_S, \phi_S, \theta_G, \phi_G)$ for the \emph{network of uplink} and \emph{downlink interferers scenarios}, for different network densities $\lambda_{cell}$ and $p$=50\%. 
The interference distributions for the uplink and downlink are similar, although the interference levels are higher for the downlink, consistent with the results for the \emph{single uplink} and \emph{downlink interferer scenarios} in Section~\ref{section:single_inter}. 
Furthermore, the aggregate interference increases with the network density, as expected, since the number of simultaneous interferers increases. 
Importantly, for both the uplink and downlink scenarios and most network densities, all harmful interference thresholds $\gamma_1$, $\gamma_2$, $\gamma_3$, and $\gamma_4$ are exceeded with likelihoods higher than 0.01\%. As an exception,  the uplink interference for at most 50~BSs/km\textsuperscript{2} exceeds $\gamma_1$ with a likelihood smaller than 0.01\%.
These are important results that confirm overall that the weather satellite would suffer from harmful aggregate interference in the typical case of multiple ground cells deployed with realistic 5G network densities, not only from the satellite sensitivity perspective, but also from the ITU-R perspective. 
Consequently, although we showed in Section~\ref{section:single_inter} that the 3GPP limit on the leaked transmit power was sufficient to protect the weather satellite from harmful interference from a \emph{single} device according to ITU-R criteria, this limit does not protect the satellite against \emph{aggregate network} interference.

\begin{figure*}[t!]
\centering
\captionsetup[subfigure]{width=0.23\linewidth}
\subfloat[uplink, $\lambda_{cell}=25$ BSs/km\textsuperscript{2}]
          {\includegraphics[width=0.23\linewidth]{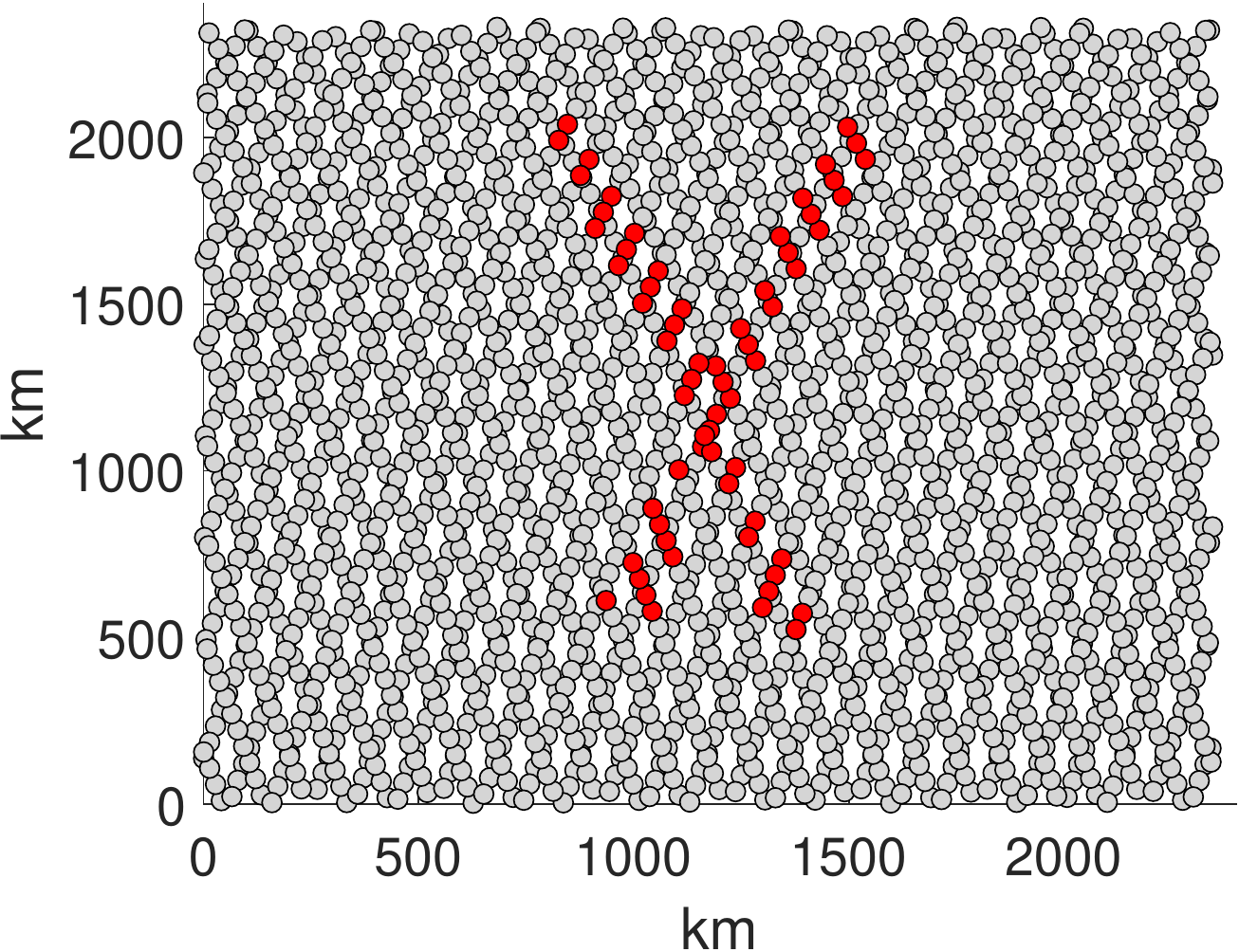} \label{res:cell3_uplink_23_pos_dens25_136}}
\
\subfloat[uplink, $\lambda_{cell}=200$ BSs/km\textsuperscript{2}]
          {\includegraphics[width=0.23\linewidth]{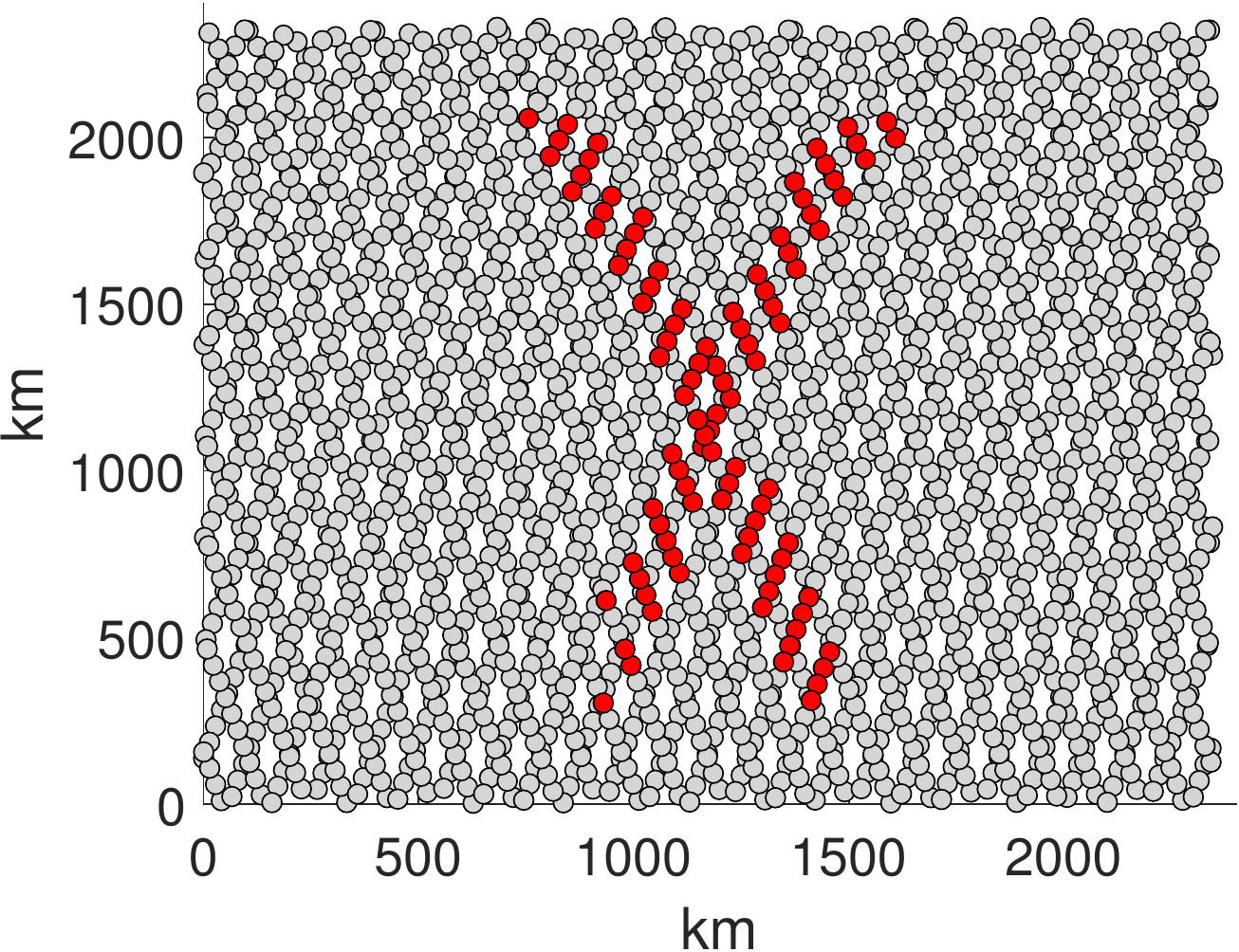} \label{res:cell3_uplink_23_pos_dens50_136}}
\
\subfloat[downlink, $\lambda_{cell}=25$ BSs/km\textsuperscript{2}]
          {\includegraphics[width=0.23\linewidth]{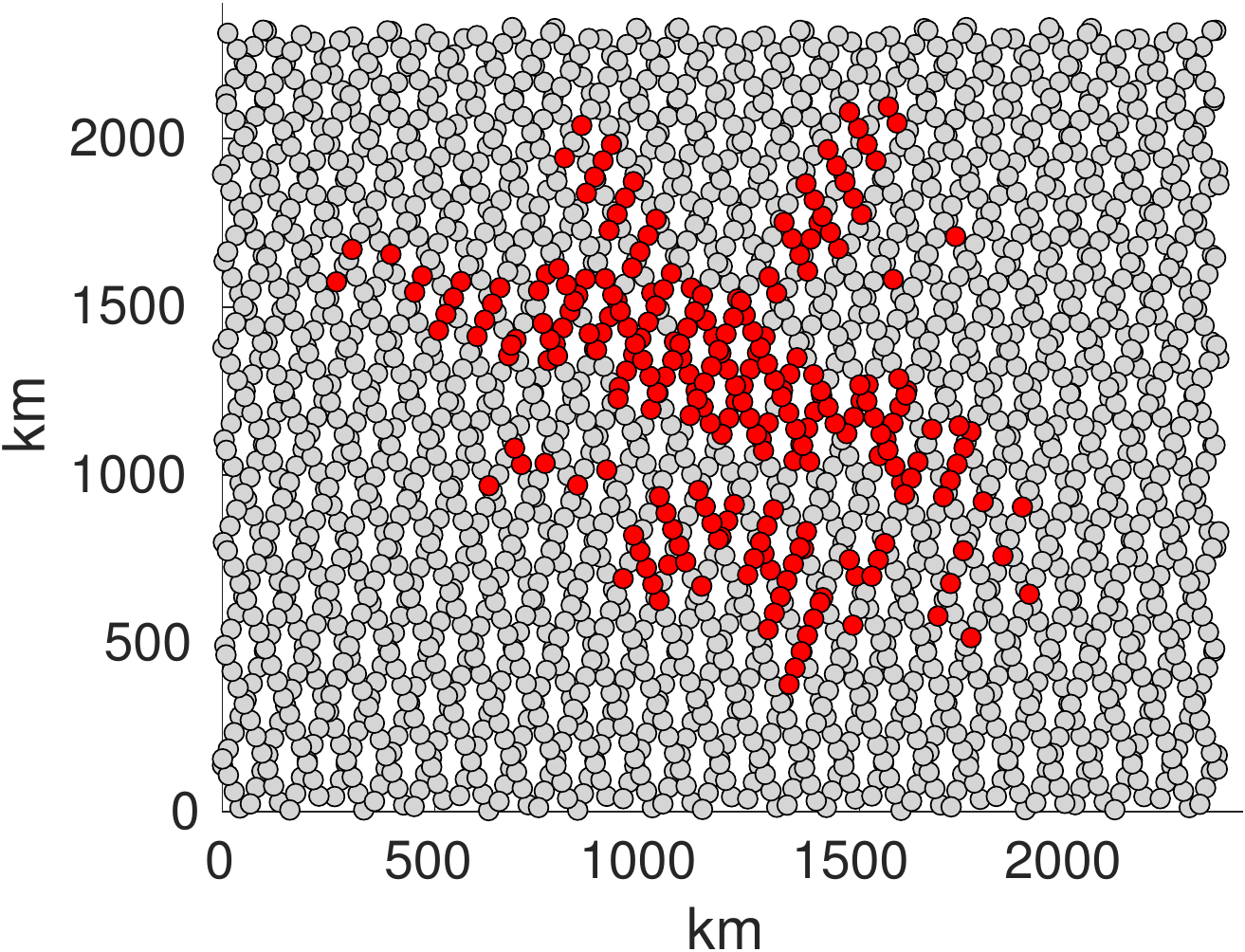} \label{res:cell3_uplink_23_pos_dens100_136}}
\
\subfloat[downlink, $\lambda_{cell}=200$ BSs/km\textsuperscript{2}]
          {\includegraphics[width=0.23\linewidth]{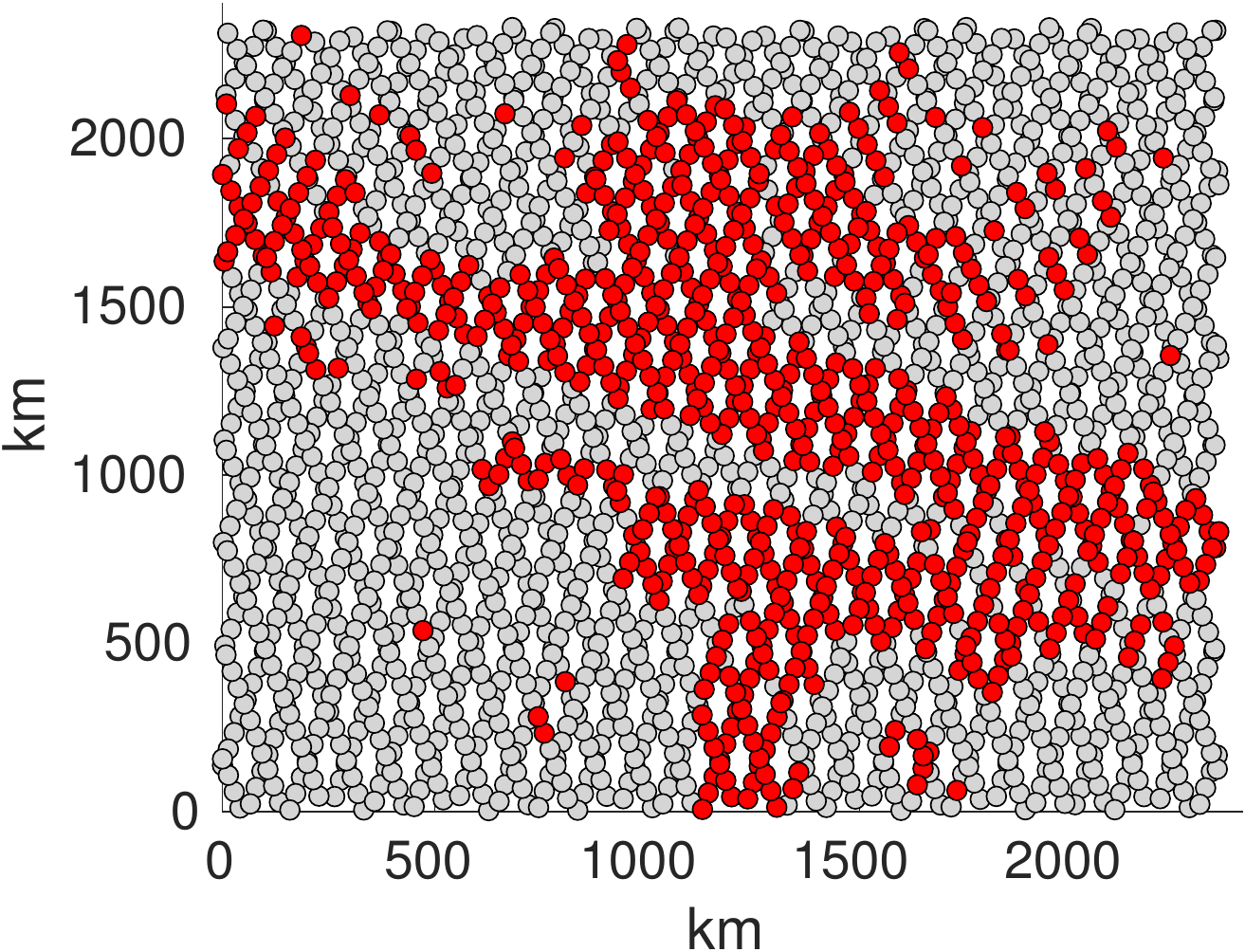} \label{res:cell3_uplink_23_pos_dens200_136}}
          \caption{Satellite positions that receive a maximum $I_{agg}(\theta_S, \phi_S, \theta_G, \phi_G)$ greater than $\gamma_1$=\mbox{--136}~dBm, for the \emph{network of uplink} and \emph{downlink interferers scenarios}, different network densities, $p$=50\%, and assuming that all $N$ cells cause the same interference as Cell~3. The maximum $I_{agg}(\theta_S, \phi_S, \theta_G, \phi_G)$ is taken over the interference estimated for all considered satellite and ground transmitter antenna orientations.}
\label{res:uplink_cell_3_thresh_136}
\end{figure*}

\begin{figure*}[t!]
\centering
\captionsetup[subfigure]{width=0.32\linewidth}
\subfloat[uplink, $\lambda_{cell}=25$ BSs/km\textsuperscript{2}]
          {\includegraphics[width=0.23\linewidth]{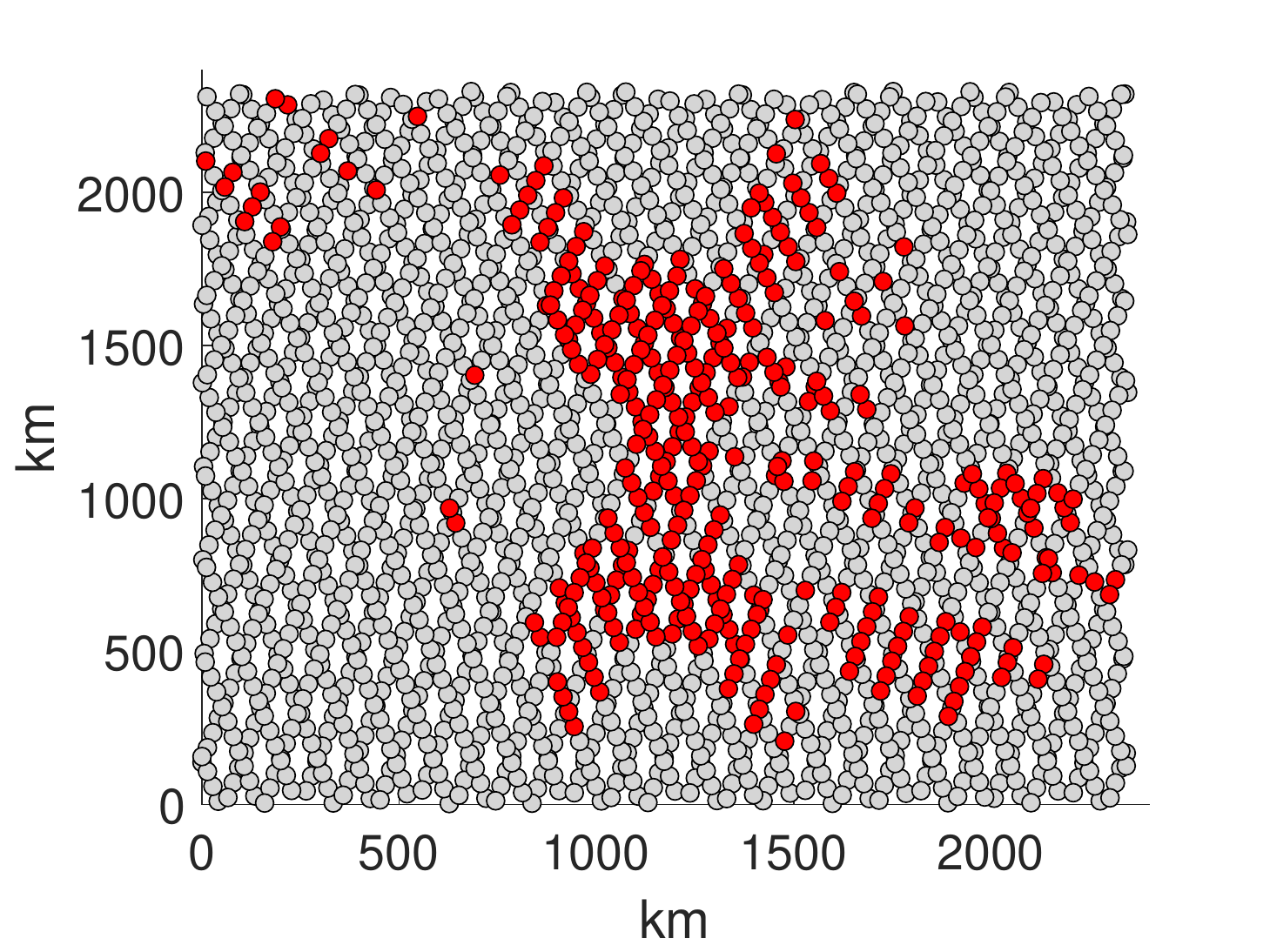} \label{res:cell3_uplink_23_pos_dens25_188}}
\
\subfloat[uplink, $\lambda_{cell}=200$ BSs/km\textsuperscript{2}]
          {\includegraphics[width=0.23\linewidth]{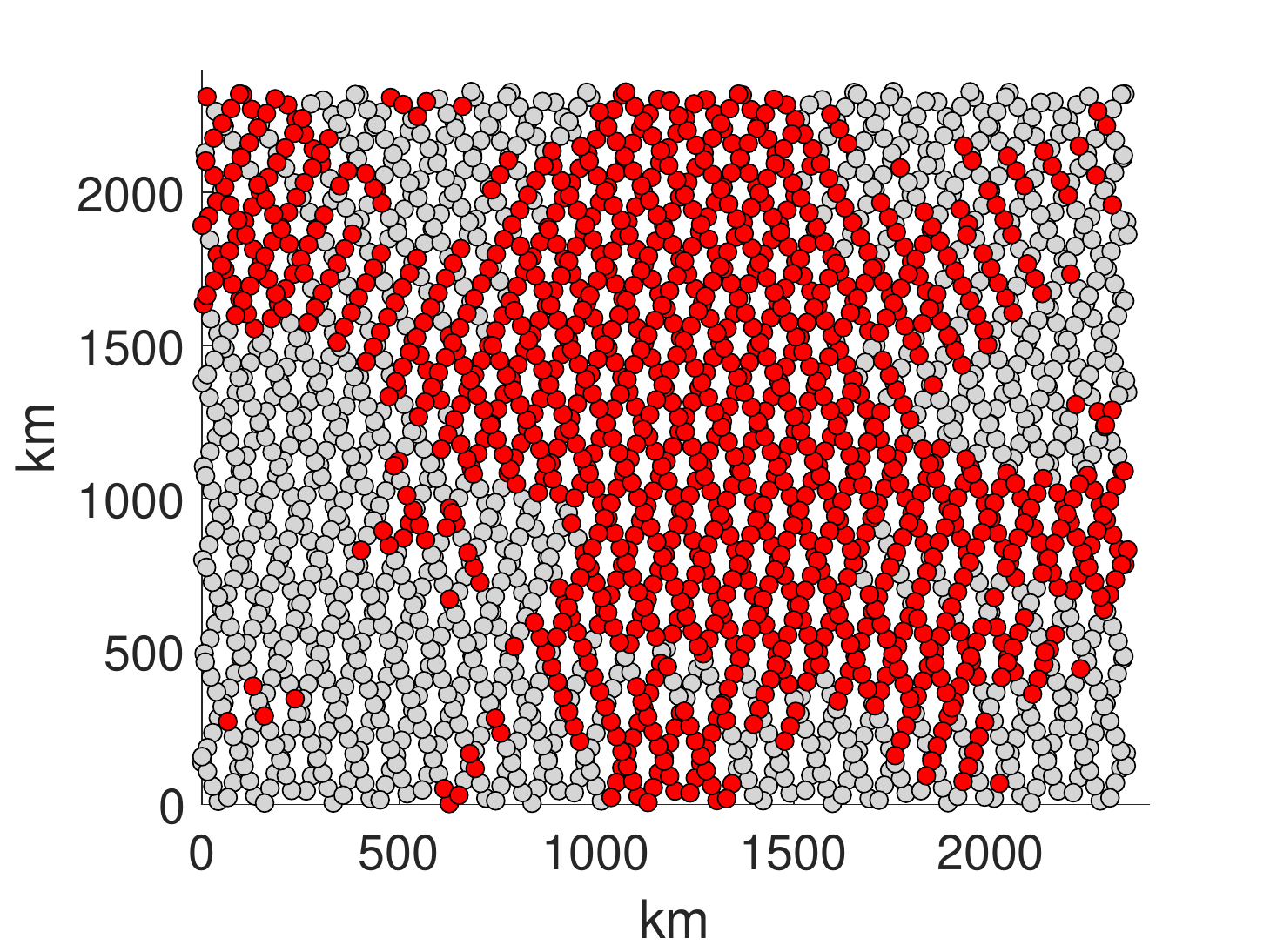} \label{res:cell3_uplink_23_pos_dens50_188}}
\
\subfloat[downlink, $\lambda_{cell}=25$ BSs/km\textsuperscript{2}]
          {\includegraphics[width=0.23\linewidth]{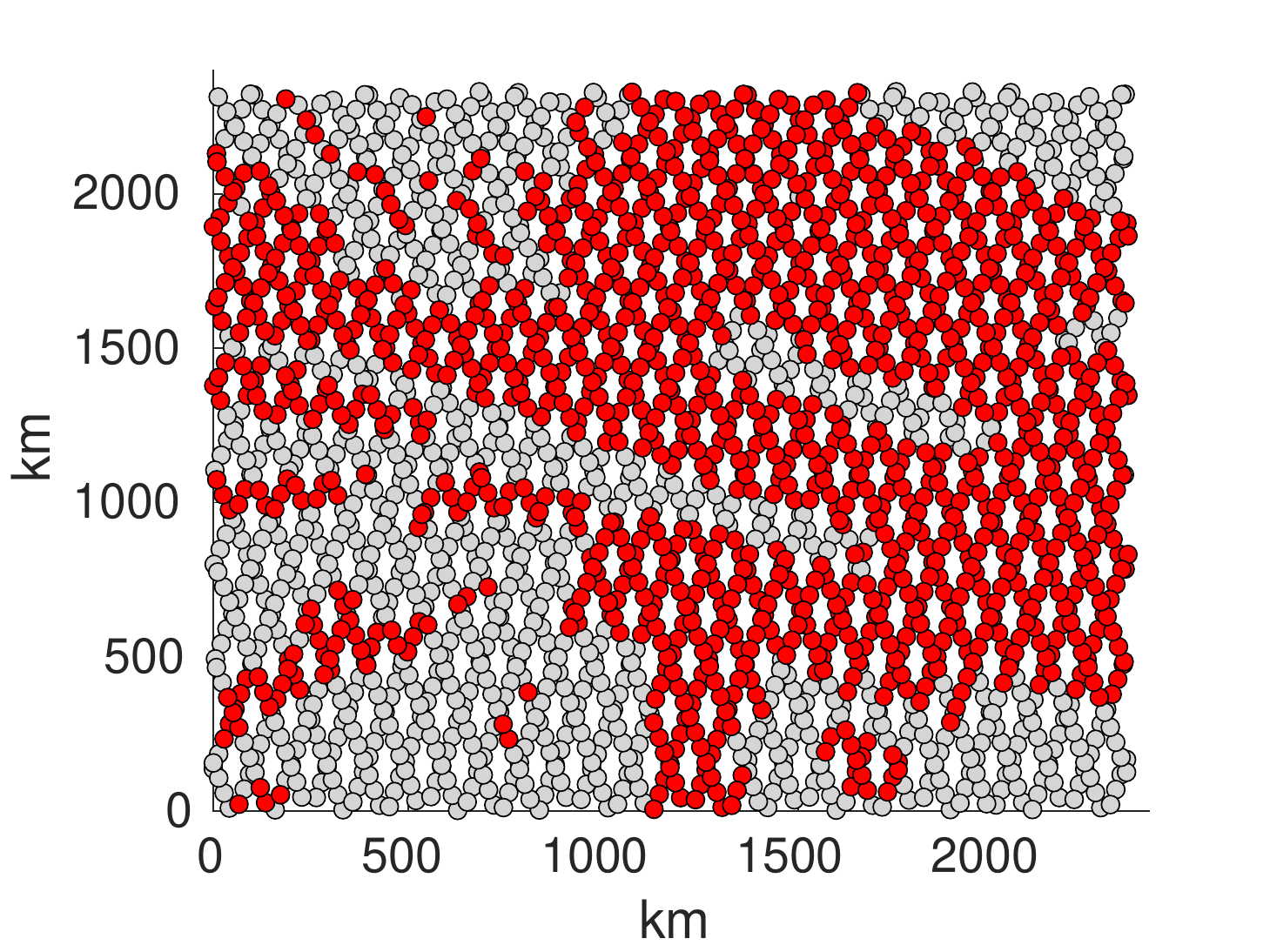} \label{res:cell3_uplink_23_pos_dens100_188}}
\
\subfloat[downlink, $\lambda_{cell}=200$ BSs/km\textsuperscript{2}]
          {\includegraphics[width=0.23\linewidth]{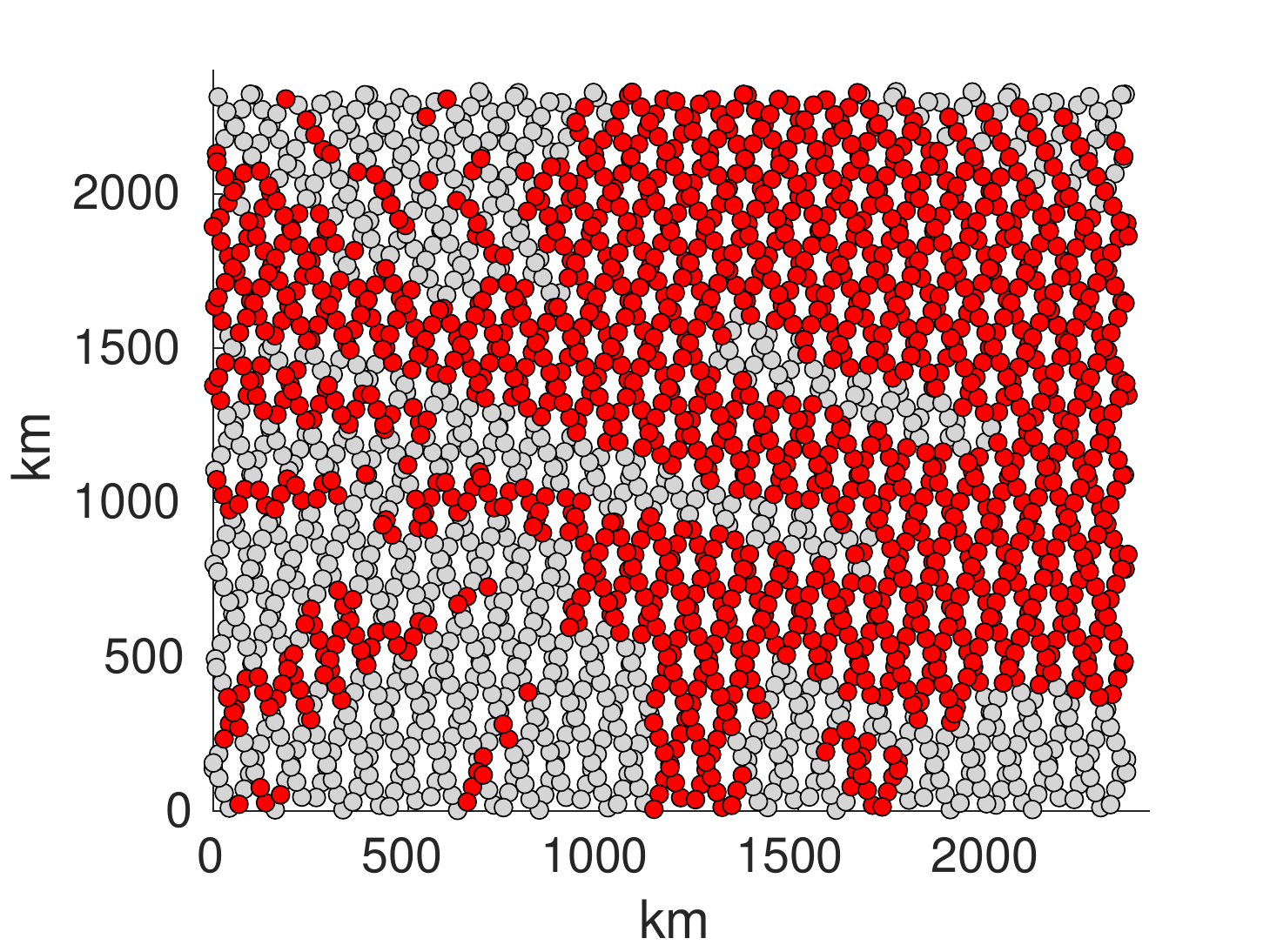} \label{res:cell2_uplink_23_pos_dens200_188}}
\caption{Satellite positions that receive a maximum $I_{agg}(\theta_S, \phi_S, \theta_G, \phi_G)$ greater than $\gamma_4$=\mbox{--161}~dBm, for the \emph{network of uplink} and \emph{downlink interferers scenarios}, different network densities, $p$=50\%, and assuming that all $N$ cells cause the same interference as Cell~3. The maximum $I_{agg}(\theta_S, \phi_S, \theta_G, \phi_G)$ is taken over the interference estimated for all considered satellite and ground transmitter antenna orientations.}
\label{res:uplink_cell_3_thresh_188}
\end{figure*}

\begin{figure}[tb!]
\centering
\captionsetup[subfigure]{width=0.48\linewidth}
{\includegraphics[width=0.8\linewidth]{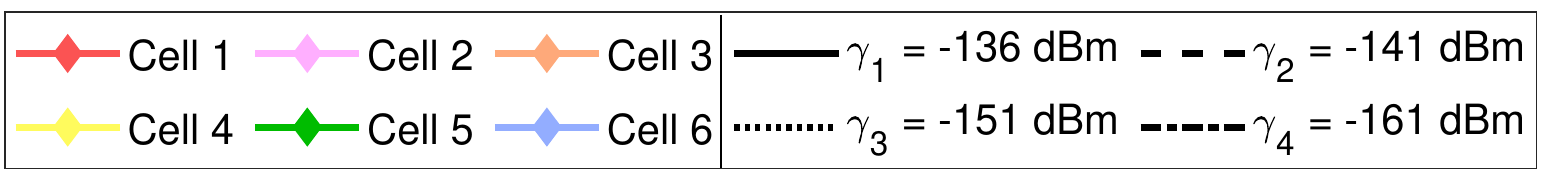}}
\hfill
\subfloat
          {\includegraphics[width=1\linewidth]{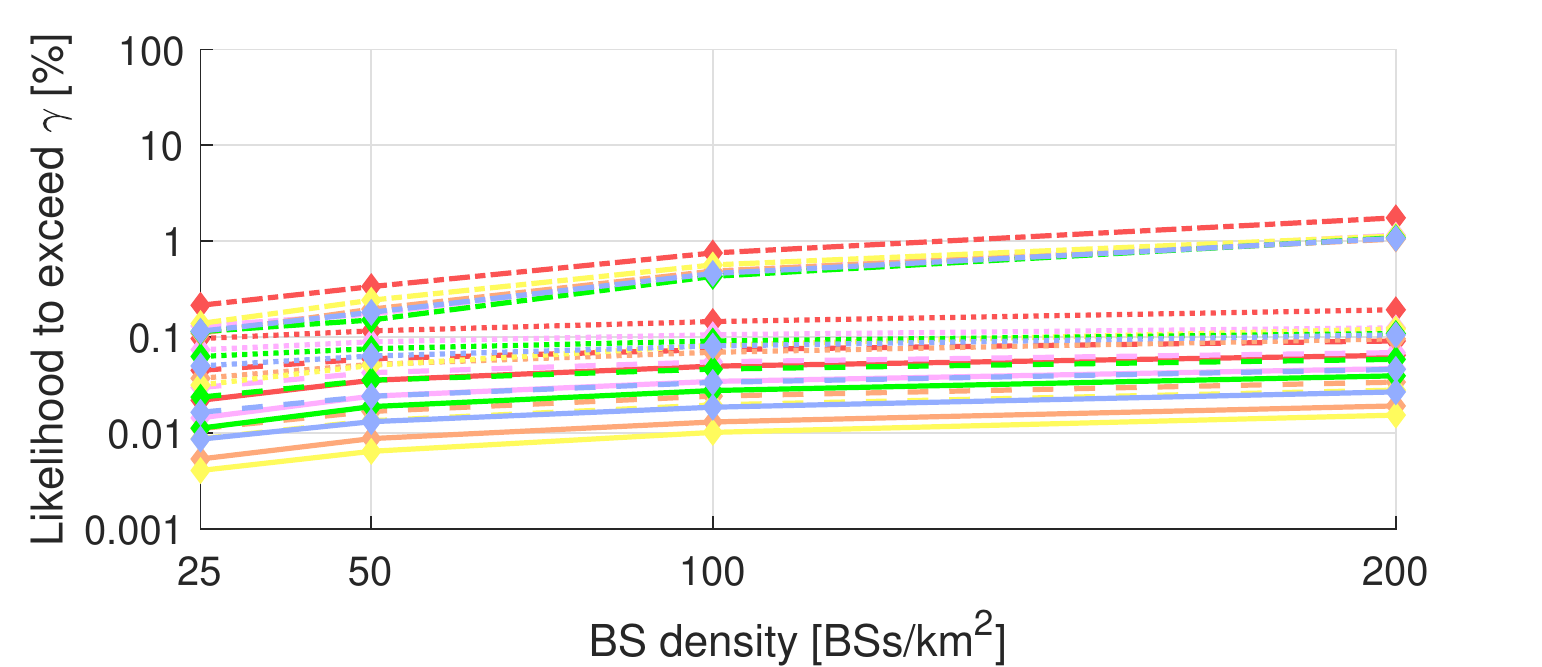} \label{res:uplink_network_interferer_23}}
\caption{Likelihood that $I_{agg}(\theta_S, \phi_S, \theta_G, \phi_G)$ exceeds $\gamma_1$, $\gamma_2$, $\gamma_3$, and $\gamma_4$ versus network density $\lambda_{cell}$, for the \emph{network of uplink interferers scenario} and $p$=50\%.}
\label{res:uplink_network_interferer_threshs}
\end{figure}

\begin{figure}[tb!]
\centering
\captionsetup[subfigure]{width=0.48\linewidth}
{\includegraphics[width=0.8\linewidth]{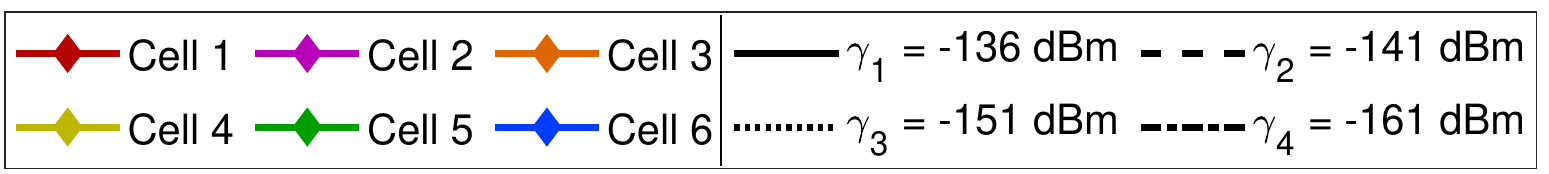}}
\hfill
\subfloat
          {\includegraphics[width=1\linewidth]{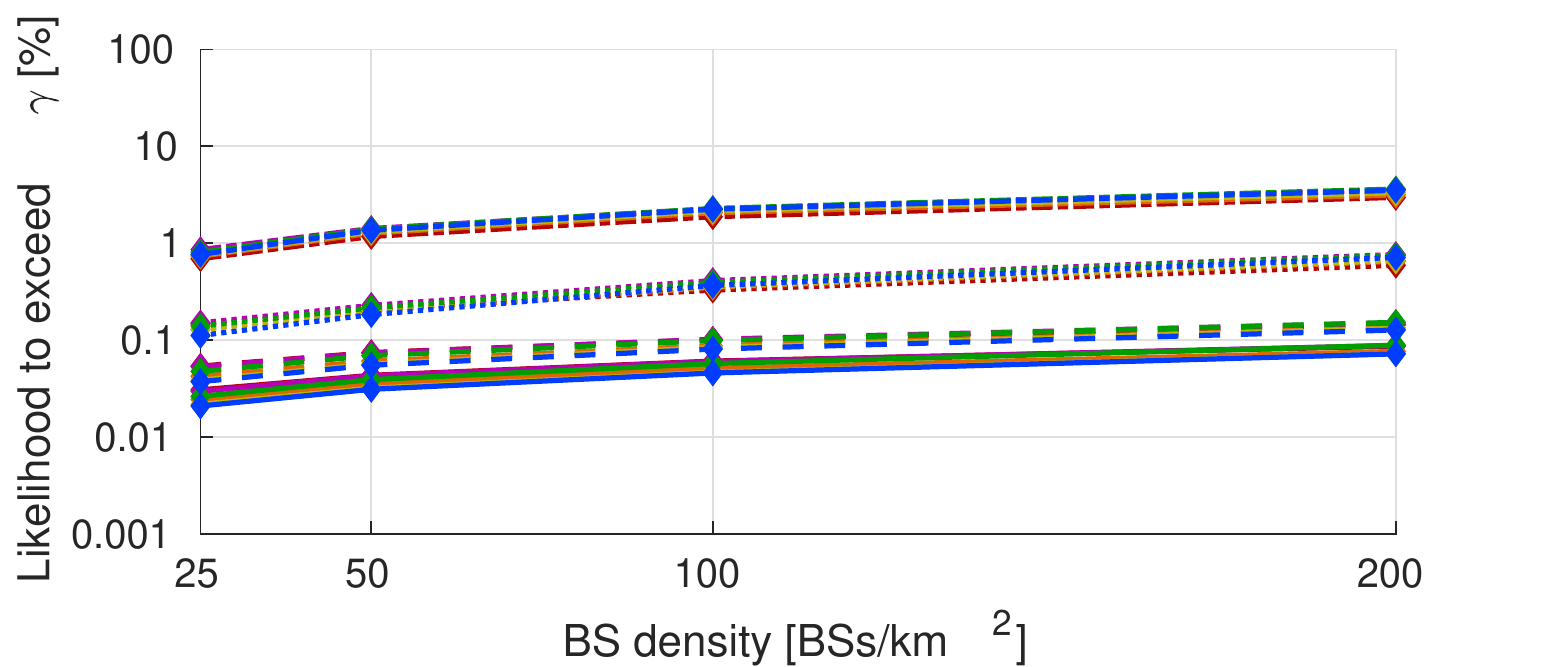} \label{res:downlink_network_interferer_23}}
\caption{Likelihood that $I_{agg}(\theta_S, \phi_S, \theta_G, \phi_G)$ exceeds $\gamma_1$, $\gamma_2$, $\gamma_3$, and $\gamma_4$ versus network density $\lambda_{cell}$, for the \emph{network of downlink interferers scenario} and $p$=50\%.}
\label{res:downlink_network_interferer_threshs}
\end{figure}

We note that at WRC-19 it was decided to decrease the leaked transmit power limit for 3GPP devices by 6~dB starting 2027~\cite{WRC2019}. The impact of this change on the results in Fig.~\ref{fig:cdf_ue3_network} is that the interference curves would all be shifted to the left by 6~dB. In such a case, the aggregate interference would still be identified as harmful according to any of the thresholds and for most combinations of scenario and network density, except for $\gamma_1$ in the uplink (all network densities) and $\gamma_2$ in the uplink and less than 50~BSs/km\textsuperscript{2}. This means that even the reduced future 3GPP leaked power limits set by~\cite{WRC2019} are still not sufficient to protect the weather satellite from downlink cellular network interference, even when considering ITU-R harmful interference threshold which is the most permissive.
          
Let us now discuss in more detail which satellite positions are affected by harmful aggregate interference.
Figs.~\ref{res:uplink_cell_3_thresh_136} and \ref{res:uplink_cell_3_thresh_188} show the heatmaps of the satellite positions that receive a maximum interference $I_{agg}$ higher than $\gamma_1$ and $\gamma_4$, respectively, for the \emph{network of uplink} and \emph{downlink interferers scenarios}. For each satellite position, we consider the maximum interference received over the different satellite and ground transmitter antenna orientations. 
The number of affected satellite positions increases overall with the network density, as expected, due to the larger number of interferers. 
Furthermore, Fig.~\ref{res:uplink_cell_3_thresh_136} shows that rather few satellite positions are affected by uplink  interference above the \mbox{ITU-R} threshold $\gamma_1$=\mbox{--136}~dBm and they follow the X shape discussed for a single UE in Section~\ref{section:single_inter}.\footnote{We note that we expect to observe a similar X shape also for $N$ \emph{heterogeneous} cells within the Manhattan ground network area, since this area is located within a single satellite scan pixel. Thus, the spatial pattern of interference would not change significantly for heterogeneous cells, as long as they are all placed within a ground area corresponding to the same pixel.} Also, the downlink interference is more harmful and affects more positions than the uplink interference.
By contrast, in Fig.~\ref{res:uplink_cell_3_thresh_188} significantly more satellite positions are affected by uplink and downlink interference above $\gamma_4$=\mbox{--161}~dBm.
These results thus show two important aspects: (i)~mm-wave ground deployments can cause harmful aggregate out-of-band interference, from both \mbox{ITU-R} and satellite sensitivity perspectives; and (ii)~despite the interference being overall identified as harmful with any of the considered thresholds, the ITU-R threshold can underestimate by a large extent the impact of out-of-band interference on the satellite sensitivity, especially if only a low interference level such as $\gamma_4$ can be tolerated.

Let us next consider the aggregate interference results for also other types of local urban propagation environments than as for Cell~3. 
Namely, in Figs.~\ref{res:uplink_network_interferer_threshs} and \ref{res:downlink_network_interferer_threshs} we present the likelihoods that $I_{agg}(\theta_S, \phi_S, \theta_G, \phi_G)$ exceeds $\gamma_1$, $\gamma_2$, $\gamma_3$, and $\gamma_4$, when the aggregate interference is caused by $N$ cells identical to each of Cells~1--6, for the \emph{network of uplink} and \emph{downlink interferers scenario}, respectively, for $p$=50\%. We note that these likelihoods are obtained from the $I_{agg}(\theta_S, \phi_S, \theta_G, \phi_G)$ distributions, as shown for Cell~3 in Fig.~\ref{fig:cdf_ue3_network}. 
Fig.~\ref{res:uplink_network_interferer_threshs} shows that the likelihood to exceed $\gamma_4$ due to uplink interference is much higher than 0.01\%, for all network densities and cell types. The likelihoods to exceed $\gamma_2$ and $\gamma_3$ are somewhat lower, but also exceed 0.01\% for all network densities and cells types. This suggests that the satellite sensitivity is always affected by harmful uplink out-of-band interference, regardless of how conservative or relaxed the threshold is. 
By contrast, the likelihood to exceed $\gamma_1$ is overall lower, being sometimes below 0.01\%, i.e. for Cells~3, 4, and 6 and low network densities of at most 50~BSs/km\textsuperscript{2}.  
Furthermore, Fig.~\ref{res:downlink_network_interferer_threshs} shows that, for the \emph{network of downlink interferers scenario}, the likelihood to exceed any of the interference thresholds is higher than 0.01\%, for all network densities and cell types. 

Overall, these results confirm that mm-wave ground deployments with realistic network densities can indeed cause harmful out-of-band interference to the weather satellites, thus strongly supporting the original coexistence concerns raised by weather scientists~\cite{Witze2019, Liu2021, Palmer2020}. This suggests that the current transmit power leakage limits for the UEs and BSs are not sufficient to protect weather satellites from harmful out-of-band-interference, so that further solutions are required to ensure reliable weather predictions when coexisting with emerging mm-wave cellular deployments.

\subsection{Impact of Atmospheric Conditions}
\label{section:atm_att_eff}

\begin{figure}[t!]
\centering
\captionsetup[subfigure]{width=0.48\linewidth}
{\includegraphics[width=0.8\linewidth]{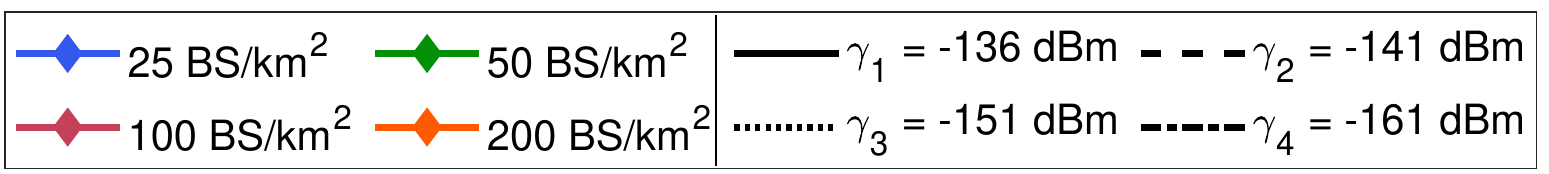}}
\hfill
\subfloat[uplink]
          {\includegraphics[width=1\linewidth]{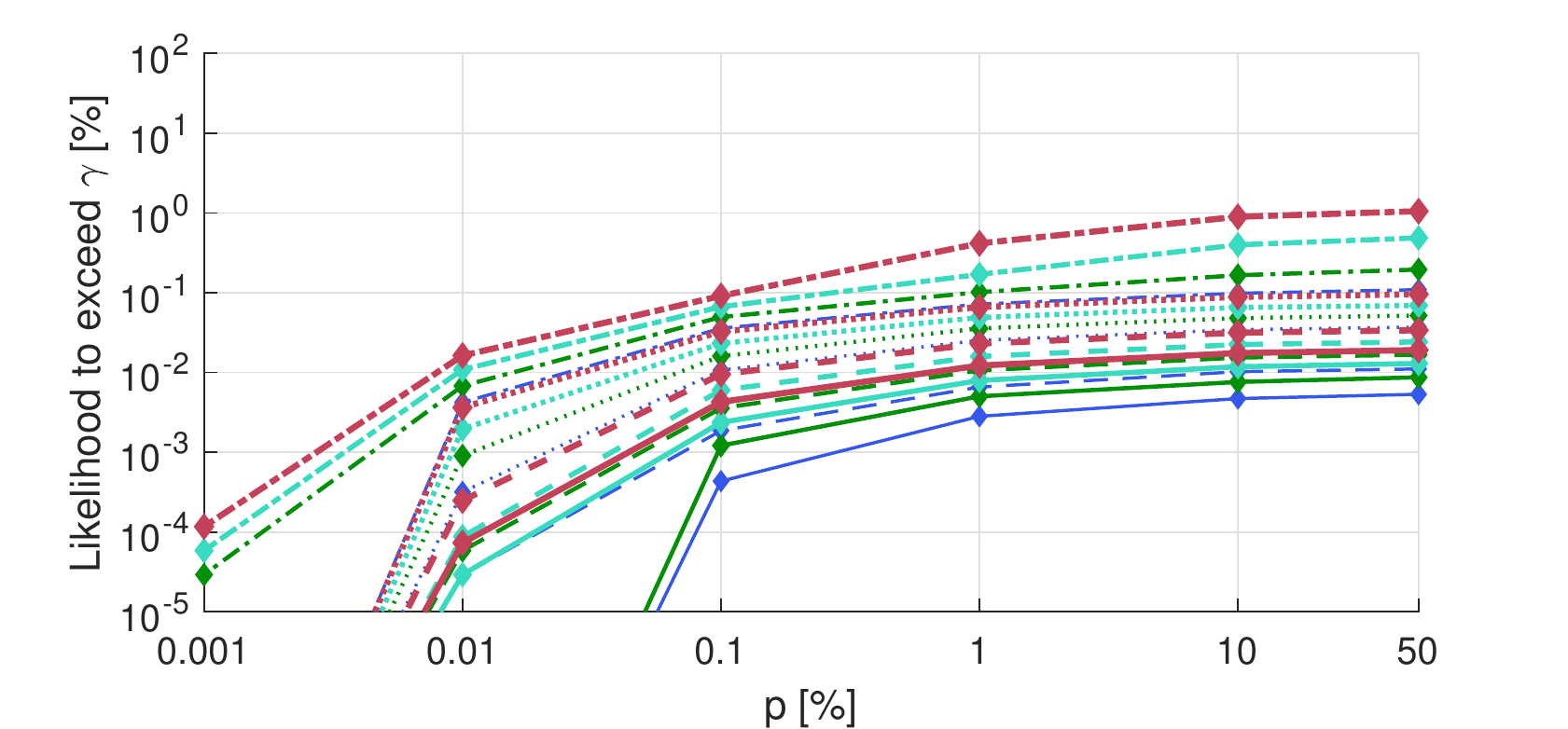} \label{res:uplink_network_interferer_23_probs}}
\hfill
\subfloat[downlink]
          {\includegraphics[width=1\linewidth]{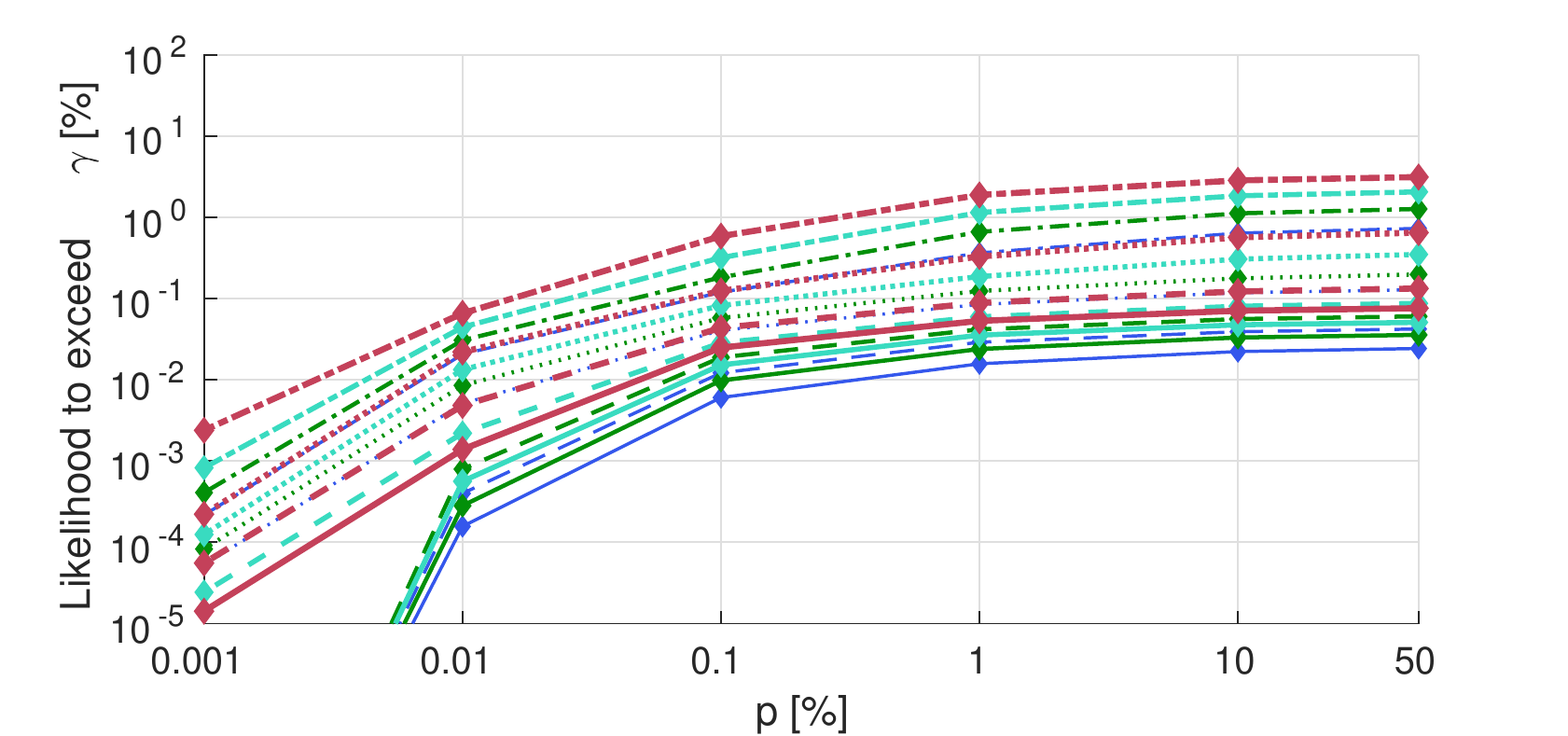} \label{res:downlink_network_interferer_23_probs}}
\caption{Likelihood that $I_{agg}(\theta_S, \phi_S, \theta_G, \phi_G)$ exceeds $\gamma_1$, $\gamma_2$, $\gamma_3$, and $\gamma_4$ versus $p$, for the \emph{network of uplink} and \emph{downlink interferers scenarios}; each of the $N$ cells causes the same interference as Cell~3.}
\label{res:uplink_network_interferer_probs}
\end{figure}
Finally, in this section we study the impact of different atmospheric propagation conditions on the harmful interference at the weather satellite. 
We note that in Sections~\ref{section:single_inter} and~\ref{section:network_inter} we presented results assuming the lowest atmospheric attenuation (i.e. a link unavailability probability $p$=50\%) and thus the worst-case interference.
We now present complementary results, where we vary the atmospheric attenuation by considering $p<50$\%.
We consider again the \emph{network of uplink} and \emph{downlink interferers scenarios} and we focus on the aggregate interference from $N$ cells, each causing an interference equal to that from Cell~3.

Fig.~\ref{res:uplink_network_interferer_probs} shows the likelihood that $I_{agg}(\theta_S, \phi_S, \theta_G, \phi_G)$ exceeds $\gamma_1$, $\gamma_2$, $\gamma_3$, and $\gamma_4$, for different $p$ and network densities. 
The likelihood to exceed any of the harmful interference thresholds is higher than 0.01\% for $p\geq1$\%, for most scenario and network density configurations.  This is an important result and shows that the network interference remains harmful for most atmospheric conditions (i.e. almost the entire range of $p$).
As an exception, the likelihood is lower than 0.01\% for $p$=0.01\%, for most scenario and network density configurations. 
Furthermore, for the \emph{network of uplink interferers scenario} in Fig.~\ref{res:uplink_network_interferer_23_probs}, the likelihood to exceed $\gamma_1$ is lower than 0.01\% also for $p$=1\%, for network densities lower than 200~BSs/km\textsuperscript{2}. Similarly, the likelihood of uplink interference to exceed $\gamma_2$ is lower than 0.01\% for $p$=1\%, for network densities of up to 50~BSs/km\textsuperscript{2}. 
Nonetheless, we consider these cases to be sporadic.
Consequently, the aggregate network interference exceeds overall all harmful interference thresholds with a higher likelihood than that tolerable according to \mbox{ITU-R}, for almost the entire range of atmospheric conditions. We emphasize that we indeed expect these conditions to vary over the entire range of $p\leq 50$\%, thus reflecting the weather changes in practice, which are, in fact, to be measured and predicted by weather satellites.


\section{Conclusions}
\label{section:conclusions}

We presented the most comprehensive coexistence study to date addressing whether and under which circumstances realistic 5G mm-wave ground deployments would cause harmful out-of-band interference to weather satellites sensing in the 23.8~GHz band. We modelled uplink and downlink interference from a single interferer and also a network of interferers consisting of UEs and BSs located in NYC. We used real 3D building data and realistic antenna patterns to perform detailed ray-tracing propagation simulations, also incorporating atmospheric attenuation, and considered real locations of the MetOp-B weather satellite above NYC and its scanning orientation, realistic ground interferer antenna beam orientations from representative urban cell sites, and a range of network densities. We compared the estimated interference not only against the ITU-R threshold of \mbox{--136}~dBm/200~MHz but also a set of proposed alternative harmful interference thresholds \{\mbox{--161}, \mbox{--151}, \mbox{--141}\}~dBm related to the sensitivity of the satellite sensor, as a first step towards directly mapping the impact of 5G ground network interference on weather prediction degradation.

Our results showed that the 3GPP power leakage limits are sufficient to ensure that interference from a \emph{single} 5G-NR device is not harmful from the ITU-R perspective, but can become harmful if the weather prediction software can only tolerate low interference levels. Importantly, our results showed that \emph{aggregate} interference resulting in practice from a network of uplink or downlink interferers with realistic 5G network densities is often harmful, even for the ITU-R threshold which is the least conservative one considered. Our study thus strongly supports the coexistence concerns voiced by weather scientists and suggests that additional engineering and/or regulatory mechanisms will be necessary for harmonious spectrum coexistence between 5G-and-beyond mm-wave networks and passive satellite sensing. Nonetheless, we emphasize that our detailed modelling of urban network deployments and reflecting the interference impact in likelihood-consequence terms was only a first step in fully addressing this important coexistence case. In our future work we plan to work together with weather scientists to more explicitly evaluate the impact of estimated spatial patterns of interference on the resulting weather prediction degradation.

\bibliographystyle{IEEEtran}
\bibliography{IEEEabrv,bibliography}
%



\end{document}